%% file: paper.tex
\documentclass[twocolumn,hyperpdf,amsmath,amssymb,aps,prd,10pt,superscriptaddress,nofootinbib,noeprint,preprintnumbers,floatfix]{revtex4-1}

\input{header}

\begin{document}

\preprint{JLAB-THY-20-3291}
\title{Excited $\bm{J^{--}}$ meson resonances at the $\bm{\mathrm{SU(3)}}$ flavor point from lattice QCD}
\author{Christopher~T.~Johnson}
\email{ctjohnson@email.wm.edu}
\affiliation{Department of Physics, College of William and Mary, Williamsburg, VA 23187, USA}
\affiliation{\lsstyle Thomas Jefferson National Accelerator Facility, 12000 Jefferson Avenue, Newport News, VA 23606, USA}
\author{Jozef~J.~Dudek}
\email{dudek@jlab.org}
\affiliation{Department of Physics, College of William and Mary, Williamsburg, VA 23187, USA}
\affiliation{\lsstyle Thomas Jefferson National Accelerator Facility, 12000 Jefferson Avenue, Newport News, VA 23606, USA}
\collaboration{for the Hadron Spectrum Collaboration}
\date{ \today }
\begin{abstract}
We present the first calculation within lattice QCD of excited light meson resonances with $J^{PC} = 1^{--}$, $2^{--}$ and $3^{--}$. Working with an exact SU(3) flavor symmetry, for the singlet representation of pseudoscalar-vector scattering, we find two $1^{--}$ resonances, a lighter broad state and a heavier narrow state, a broad $2^{--}$ resonance decaying in both $P$-- and $F$--waves, and a narrow $3^{--}$ state. We present connections to experimental $\omega^\star_J, \phi^\star_J$ resonances decaying into $\pi \rho$, $K\kbarSuper{*}$, $\eta \omega$ and other final states. 
\end{abstract}
\maketitle

\section{Introduction}
  \label{Intro}
  \input{1-intro}

\section{Finite-Volume Spectrum}
  \label{fvspectrum}
  \input{2-fvspectrum}

\newpage

\section{Scattering Amplitudes}
	\label{amps}
	\input{3-amps}

\newpage

\section{Resonance Interpretation}
	\label{resonances}
	\input{4-resonances}

\newpage

\section{Summary}
	\label{summary}
	\input{5-summary}

\newpage

\input{acknow}

\appendix

\section{Operator basis}
	\label{ops}
	\input{appA-ops}

\section{Left-hand cut singularities}
	\label{lhc}
	\input{appB-lefthand}

\section{Amplitude parameterizations}
	\label{par}
	\input{appC-params}

\bibliographystyle{apsrev4-1}
\bibliography{bib}


\end{document}

%% file: header.tex
\usepackage{graphicx, color}

\usepackage[letterspace=-10]{microtype} 

\usepackage{bm, amsmath, amsfonts, amssymb,xfrac}
\usepackage{multirow, tabularx, dcolumn}
\usepackage{mathtools,leftidx, braket, slashed, cancel, bigdelim}
\usepackage{blkarray}
\usepackage[figures]{rotating}

\usepackage[utf8]{inputenc} 
\usepackage{hyperref}
\definecolor{jlab_red}{RGB}{192,39,45}
\definecolor{jlab_orange}{RGB}{249,102,0}
\definecolor{jlab_blue}{RGB}{47,122,121}
\definecolor{jlab_green}{RGB}{65,125,10}
\definecolor{jlab_gray}{gray}{0.6}

\newcommand{\SUF}{\ensuremath{\mathrm{SU}(3)_\mathrm{F}}}

\newcommand{\etaOctet}{     \eta^{\scriptscriptstyle\mathbf{8}}}
\newcommand{\etaSinglet}{   \eta^{\scriptscriptstyle\mathbf{1}}}
\newcommand{\omegaOctet}{   \omega^{\scriptscriptstyle\mathbf{8}}}
\newcommand{\omegaSinglet}{ \omega^{\scriptscriptstyle\mathbf{1}}}
\newcommand{\fOneOctet}{    f_1^{\scriptscriptstyle\mathbf{8}}}
\newcommand{\fOneSinglet}{  f_1^{\scriptscriptstyle\mathbf{1}}}

\newcommand{\fZeroSinglet}{  f_0^{\scriptscriptstyle\mathbf{1}}}

\newcommand{\hOneOctet}{    h_1^{\scriptscriptstyle\mathbf{8}}}
\newcommand{\hOneSinglet}{  h_1^{\scriptscriptstyle\mathbf{1}}}

\newcommand{\gSinglet}{   g^{\scriptscriptstyle\mathbf{1}}  }
\newcommand{\gOctet}{     g^{\scriptscriptstyle\mathbf{8}}  }
\newcommand{\hOctet}{     h^{\scriptscriptstyle\mathbf{8}}  }

\newcommand{\threeSone}{\prescript{3\!}{}{S}_1}

\newcommand{\threePzero}{\prescript{3\!}{}{P}_0}
\newcommand{\threePone}{\prescript{3\!}{}{P}_1}
\newcommand{\threePtwo}{\prescript{3\!}{}{P}_2}

\newcommand{\threeDone}{\prescript{3\!}{}{D}_1}
\newcommand{\threeDtwo}{\prescript{3\!}{}{D}_2}
\newcommand{\threeDthree}{\prescript{3\!}{}{D}_3}  
\newcommand{\threeDJ}{\prescript{3\!}{}{D}_J}  
\newcommand{\threeDJs}{\prescript{3\!}{}{D}_{1,2,3}}

\newcommand{\threeFtwo}{\prescript{3\!}{}{F}_2}
\newcommand{\threeFthree}{\prescript{3\!}{}{F}_3}

\newcommand{\kbar}{\,\overline{\!K}}

\newcommand{\kbarSuper}[1]{\,\overline{\!K} \vphantom{K}^{#1} }

\hypersetup{%
pdftitle = {Excited Jmm meson resonances at the SU(3) flavor point from lattice QCD},
pdfsubject = {},
pdfkeywords = {},
pdfauthor = {Hadron Spectrum Collaboration},
colorlinks = {true},
filecolor = {black},
linkcolor = {jlab_blue},
menucolor = {black},
citecolor = {jlab_green},
urlcolor = {jlab_green},
}{}

%% file: 1-intro.tex
 
The lightest vector meson resonances, the $\rho$, $\omega$ and $\phi$, are benchmark states in our understanding of the quark substructure of hadrons~\cite{Shepherd:2016dni}. The near degeneracy of the $\rho$ and the $\omega$, and the preference for $\phi$ to decay to $K\kbar$ even when $\pi\pi\pi$ has a much larger phase-space, leads to the OZI rule and the $u\bar{d}, u\bar{u} + d\bar{d}, s\bar{s}$ assignment for the three states. That such clear conclusions can be drawn comes in part from the fact that these resonances are rather narrow (particularly the $\omega$ and the $\phi$) and that they can be produced in the definitively $J^{PC}=1^{--}$ process of $e^+ e^-$ annihilation, where they appear with essentially no background in simple final states like $ \pi \pi, \pi \pi \pi$ and  $K\kbar$.

In comparison, the spectrum of heavier \emph{excited} vector mesons is far less clear, with proposed experimental candidate states being rather poorly understood~\cite{Clegg:1993mt, Zyla:2020zbs}. Such states lie at or above about 1400 MeV, which is well into the region of coupled-channels, where resonances have multiple possible decay modes. The PDG consensus is for a $\rho(1450)$ with a large total decay width, and a somewhat narrower $\rho(1700)$. The isoscalar states are even less well determined, with preference for an $\omega(1420)$ with a large uncertainty on the width, and an $\omega(1650)$ that is likely to be broad. A relatively narrow $\phi(1680)$ does not appear to have an obvious partner at higher energy\footnote{The next relevant state listed in the PDG is the $\phi(2170)$ observed through its decay to $\phi f_0(980)$ which is too heavy to partner the $\phi(1680)$.}. These assignments of isoscalar resonances to the names $\omega, \phi$ (implying dominantly hidden-light versus hidden-strange $q\bar{q}$ structure) follow from assumptions based upon the OZI rule applied to the decay channels in which the resonances are seen (mostly $\pi\pi\pi$ versus $K \kbarSuper{(*)}$).
  
Within quark models assuming a minimal $q\bar{q}$ structure for mesons, the presence of two $1^{--}$ states in each flavor channel is quite natural, with there being a first radial excitation of the lightest vector states having a quark-antiquark pair in a relative $S$-wave, $q\bar{q}\big[2\threeSone\big]$, and in addition a $D$--wave excitation, $q\bar{q}\big[1\threeDone\big]$. The physical eigenstates can be admixtures of these basis states, although typically the simple model dynamics does not generate a large mixing~\cite{Godfrey:1985xj}. Within these models, one of the $1^{--}$ states would come with spin-orbit partners, $q\bar{q}\big[1\threeDJs\big]$, leading to an expectation of approximately degenerate states with $J^{PC} = 1^{--}, 2^{--}, 3^{--}$. There are experimental candidates for $3^{--}$ states in the form of the narrow resonances $\rho_3(1690), \omega_3(1670)$, and $\phi_3(1850)$, but to date there are no clear signals for the corresponding $2^{--}$ states~\cite{Zyla:2020zbs}.
  
\begin{figure}
\includegraphics[width=0.99\columnwidth]{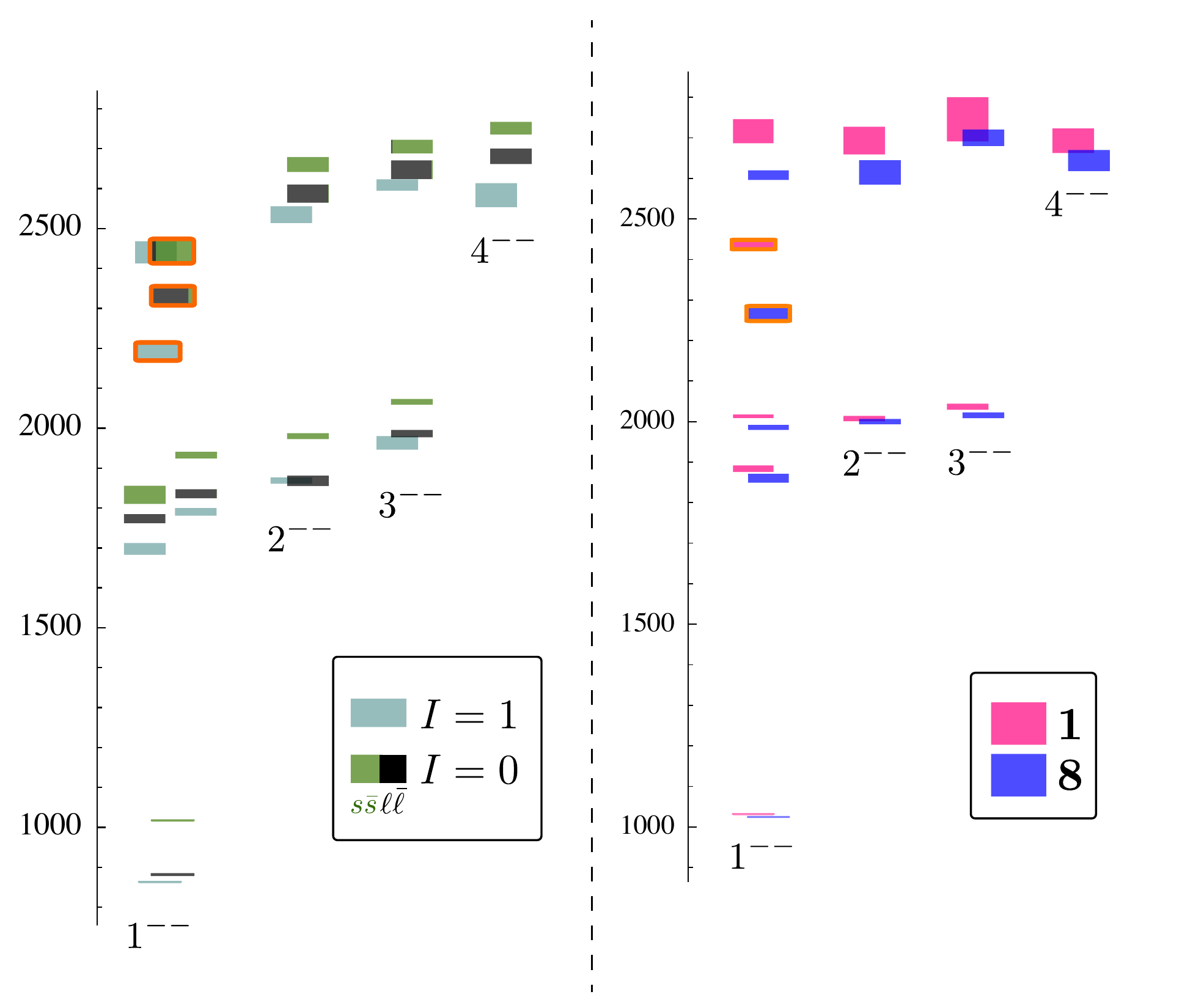}
\caption{
$J^{--}$ meson spectrum (in MeV units) extracted from diagonalization of matrices of fermion bilinear correlation functions. Results taken from Ref~\cite{Dudek:2013yja}. Left panel: Isovector (blue) and isoscalar (green/black) mesons with $m_\pi \sim 391$ MeV. Relative hidden-light and hidden-strange content determined by size of matrix elements $\langle M | \bar{\ell} {\bf \Gamma} \ell | 0 \rangle$, $\langle M | \bar{s} {\bf \Gamma} s | 0 \rangle$. Right Panel: Spectrum in the $\SUF$ limit, in octet ($\bm{8}$, blue) and singlet ($\bm{1}$, pink) representations, with $m_\pi \sim 700$ MeV.
}
\label{singlehadron}
\end{figure}

Recent support for these longstanding quark model expectations comes from lattice QCD calculations of the excited meson spectrum~\cite{Dudek:2009qf, Dudek:2010wm, Dudek:2011tt, Dudek:2013yja}. Lattice QCD is a first-principles numerical approach to QCD in which the quark and gluon fields are discretized on a periodic grid of finite size. By sampling gluon field configurations according to a probability distribution fixed by the QCD action, correlation functions can be computed, and from these physical observables extracted. The simplest calculations of the meson spectrum make use of a large basis of fermion bilinear operators in the construction of matrices of correlation functions, and diagonalisation of these provides a guide to the excited state spectrum of isovector and isoscalar mesons. Figure~\ref{singlehadron}, taken from Ref.~\cite{Dudek:2013yja}, shows the relevant part of the spectrum from two such calculations, one with a heavier than physical light quark mass such that the pion has a mass $\sim 391 \,\mathrm{MeV}$ (left) and another where the light and strange quark masses are degenerate leading to an exact $\SUF$ symmetry and a lightest pseudoscalar of mass $\sim 700$ MeV. The observed spectra support the quark model picture described above, provided it is augmented with $1^{--}$ \emph{hybrid mesons} (highlighted with orange borders) in which a $q\bar{q}$ construction in a color octet is coupled to an excitation of the gluonic field~\cite{Dudek:2011bn}. The lack of significant hidden-light--hidden-strange mixing at the lighter quark mass, and the near degeneracy of the isovector and hidden-light isoscalar states supports the OZI ``rule'' in which $q\bar{q}$ annihilation within a meson, leading to a ``disconnected'' diagram, is a suppressed process.

While relatively simple lattice QCD calculations like those presented in Figure~\ref{singlehadron} provide guidance as to what excited states we expect to find in QCD, they are clearly incomplete in that they do not resolve that excited states are in fact \emph{resonances} which decay rapidly into lighter stable hadrons. In this paper we seek to resolve this omission.

The $\rho^\star$ and $\omega^\star, \phi^\star$ excited mesons are separated in their decay channels by isospin and $G$-parity. In particular $\rho^\star$ states decay to even numbers of pions $\pi\pi, \pi\pi\pi\pi$, while $\omega^\star$ states decay to $\pi\pi\pi$. The separation of isoscalar states into $\omega$ and $\phi$ assignments is not based upon any fundamental symmetry, but is rather motivated by the OZI rule which suggests that $\phi$ states prefer to decay to $K\kbarSuper{(\!*\!)}$, as the initial $s\bar{s}$ does not have to annihilate, over for example $\pi\pi\pi$ where it does\,\footnote{OZI does not forbid $\omega^*$ decays to pairs of strange hadrons which can proceed by production of an $s\bar{s}$ pair. }. Whether the excited $J^{--}$ states remain ideally flavor mixed like the lightest $\omega, \phi$ is a dynamical question, but the lattice calculation shown in the left panel of Figure~\ref{singlehadron} seems to suggest they do.
Regarding the spin-parity structure of decays, we note that $J^P = 1^{-}$ and $3^{-}$ are in the \emph{natural parity sequence} which means that for example $\rho^\star, \rho_3$ can decay into pairs of pseudoscalars, while $2^-$ is in the \emph{unnatural parity sequence} preventing $\rho_2$ decaying into these simplest of final states.

The state-of-the-art until now for theoretical description of the \emph{decays} of excited $J^{--}$ states has been to supplement a $q\bar{q}$ bound-state quark model with a $q\bar{q}$ pair-creation vertex applied in lowest-order perturbation theory. Within such a model an estimate for the OZI-allowed decays of an excited meson $M$ to a pair of lighter mesons, $AB$, follows from evaluation of the matrix element $\langle A B | \mathcal{O}_{q\bar{q}} | M \rangle$, where $A$ and $B$ do not interact with each other, the calculation of which amounts to computing overlap integrals featuring $q\bar{q}$ bound-state wavefunctions. The most successful approach, in the sense of approximately duplicating several measured hadron decay rates, is to assume the $q\bar{q}$ pair is produced with the quantum numbers of the vacuum, the \mbox{``$\threePzero$-model''}~\cite{Barnes:1996ff, Barnes:2002mu}. 
  
Going beyond this to compute directly within QCD, an approach is available which allows us to access the energy dependence of  \emph{scattering amplitudes}, like $AB \to AB$ in which $M$ appears as a resonance. This method makes use of the discrete spectrum of QCD in the finite volume defined by the periodic lattice used in lattice QCD. Consideration of field theories in a cubic volume~\cite{Luscher:1985dn,Luscher:1986pf,Luscher:1990ux,Luscher:1991cf, Rummukainen:1995vs,He:2005ey,Kim:2005gf,Christ:2005gi,Fu:2011xz,Guo:2012hv,Hansen:2012tf,Briceno:2012yi,Gockeler:2012yj,Leskovec:2012gb,Briceno:2014oea} provides a relationship between the $S$-matrix in multiple partial waves and the finite-volume spectrum such that through lattice QCD spectrum computations we can obtain scattering information. By using parameterizations of scattering amplitudes, resonance information follows in a rigorous way from isolating pole singularities at complex values of the scattering energy. Computation of the $\rho$ resonance in elastic $\pi \pi$ scattering is now common~\cite{Aoki:2007rd,Feng:2010es,Aoki:2011yj,Lang:2011mn,Dudek:2012xn,Pelissier:2012pi,Bali:2015gji,Wilson:2015dqa,Bulava:2016mks,Guo:2016zos,Hu:2016shf,Fu:2016itp,Alexandrou:2017mpi,Andersen:2018mau,Werner:2019hxc,Erben:2019nmx,Fischer:2020fvl}, and the extension into the more complicated coupled-channel sector has been pioneered by the \emph{hadspec} collaboration, with calculations of scattering systems containing resonances resembling the $a_0(980), f_0(980), b_1(1235), f_2(1270)$ and $f_2'(1525)$, amongst others~\cite{Dudek:2014qha,Wilson:2014cna,Dudek:2016cru,Moir:2016srx,Briceno:2017qmb,Woss:2018irj,Woss:2019hse}. 

Given the likely complexity of the coupled-channel scattering systems housing the physical $J^{--}$ resonances\,\footnote{In particular the possibility of three-meson decays, the formalism for which has only recently been developed and initially tested~\cite{Hansen:2019nir}.}, we choose in this first calculation of their properties to work in a simplified version of QCD in which three quark flavors are degenerate, $m_u = m_d = m_s$, and where this single quark mass is tuned to approximately match the value of the physical strange quark mass.   The exact $\SUF$ symmetry present in this version of QCD simplifies the scattering systems in which the $J^{--}$ resonances appear, and the relatively large value of the mass of the lightest pseudoscalar $\sim 700$ MeV makes decays to three-meson and higher multiplicity final states kinematically inaccessible. The spectrum of states obtained at this $\SUF$ point when only fermion bilinear operators are used to form correlation functions is shown in the right panel of Figure~\ref{singlehadron}, where we observe \emph{octet} ($\mathbf{8}$) and \emph{singlet} ($\mathbf{1}$) excitations in good agreement with the $q\bar{q}$ picture described above. The \emph{ideal flavor mixing} (states as $u\bar{u}+d\bar{d}, s\bar{s}$) observed away from the $\SUF$ point of course cannot be present here as $\mathbf{8} \sim u\bar{u} + d\bar{d} - 2 s\bar{s}$ and $\mathbf{1} \sim u\bar{u} + d\bar{d} + s\bar{s}$, but the near degeneracy of the octet and singlet states allows for a strong mixing to ideal flavor upon even a small breaking of the $\SUF$ symmetry.

In the first calculation of $J^{--}$ resonances in lattice QCD presented in this paper, we will focus on the $\SUF$ \emph{singlet} ($\mathbf{1}$) states, and seek to determine if there are indeed two resonances in $1^{--}$, one in $2^{--}$ and one in $3^{--}$. We will determine the decay widths of these resonances, and explore how two \emph{overlapping} resonances might manifest in $1^{--}$ scattering amplitudes. We will initially work in a restricted energy region below the expected location of the $1^{--}$ hybrid meson, to avoid the possibility of three-meson decays becoming relevant. We will find that resonances are present which appear in the $\etaOctet \omegaOctet$ scattering channel with negligible coupling to other kinematically accessible channels -- this active scattering channel will be related to $\omega^\star$ and $\phi^\star$ decays to for example $\pi \rho$ and $K\kbarSuper{*}$ in the $\SUF$-broken case.

%% file: 2-fvspectrum.tex

As indicated in the introduction, resonances can be determined from the energy dependence of scattering amplitudes, which are constrained by finite-volume spectra computed using lattice QCD. The spectra follow from diagonalization of matrices of correlation functions which were computed on five anisotropic lattices with volumes $(L/a_s)^3 \times (T/a_t) = \{ 14^3, 16^3, 18^3, 20^3, 24^3 \} \times 128$, where the spatial and temporal lattice spacings are respectively $a_s \sim 0.12\, \mathrm{fm}$ and $a_t = a_s/\xi \sim (4.7 \, \mathrm{GeV})^{-1}$, with anisotropy $\xi \sim 3.5$. Details of the generation of these dynamical three-flavor lattices where $m_u = m_d = m_s$ and where the lightest pseudoscalar has mass $\sim 700$ MeV can be found in Refs.~\cite{Edwards:2008ja,Lin:2008pr}.

\emph{Distillation}~\cite{Peardon:2009gh} was used to compute correlation functions, allowing all relevant Wick contractions to be computed including those featuring $q\bar{q}$ annihilation, which are common when $\SUF$ singlets are being considered. The rank of the distillation space, the number of time sources and the number of gauge configurations used are provided in Table~\ref{lattices}.

\begin{table}[b]
\renewcommand{\arraystretch}{1.2}
\begin{tabular}{c|ccccc}
$L/a_s$ 			& $14$	& $16$ 	& $18$ 	& $20$	& $24$ \\
\hline
$N_\mathrm{cfgs}$	& $397$	& $490$	& $358$	& $477$	& $499$ \\
$N_\mathrm{vecs}$	& $48$	& $64$	& $96$	& $128$	& $160$ \\
$N_\mathrm{tsrcs}$	& $16$	& $4$	& $4$	& $4$	& $1$ 
\end{tabular}
\caption{
Number of distillation vectors ($N_{\text{vecs}}$), gauge configurations ($N_{\text{cfgs}}$) and time-sources ($N_{\text{tsrcs}}$) used in computation of correlation functions on each lattice volume.
}
\label{lattices}
\end{table}

The spectrum of mesons stable against strong decay in this version of QCD was presented in Ref.~\cite{Woss:2020ayi} and is reproduced in Table~\ref{masses}. The dispersion relations (the energy when at momentum ${\vec{p} = \frac{2\pi}{L} \vec{n}}$) for the low-lying mesons which feature in scattering  were also computed  and found to conform to the relativistic expression,
\begin{equation}\label{disp_relation}
\big( a_t E_{\vec{n}} \big)^2 = \big( a_t m \big)^2 + \frac{1}{\xi^2} \left( \frac{2\pi}{L/a_s} \right)^{\!2} |\vec{n}|^2 \, ,
\end{equation}
with an estimate for the anisotropy that accounts for small variations observed for different mesons being ${\xi = 3.486(43)}$ -- see Ref.~\cite{Woss:2018irj} for further details.

\begin{table}
\centering      
\renewcommand{\arraystretch}{1.2}
\begin{tabular}[t]{ r l @{\hskip 3.0ex}|@{\hskip 3.0ex} r l } 
	$\etaOctet$ & 0.1478(1) 	& $\etaSinglet$ & 0.2017(11)   \\
	$\omegaOctet$ & 0.2154(2) 	& $\omegaSinglet$ & 0.2174(3)  \\
                        	   && $\fZeroSinglet$ & 0.2007(18) \\
    $\fOneOctet$ & 0.3203(6)  	& $\fOneSinglet$ & 0.3364(14)  \\
    $\hOneOctet$ & 0.3272(6) 	& $\hOneSinglet$ & 0.3288(17)  \\    
\end{tabular}
\caption{Relevant stable hadron masses, $a_t m$.}
\label{masses}
\end{table} 

The cubic symmetry of the spatial lattice and its boundary is such that $J^P$ are not in general good quantum numbers, rather we should use the irreducible representations (\emph{irreps}) of the cubic symmetry and of its \emph{little group} for systems with nonzero momentum. The irreps we will consider are presented in Table~\ref{subductions} where we observe the \emph{subduction} of many $J^{PC}$ values into each irrep (we show only $J<4$). It is possible for a single $J^{PC}$ to subduce more than once into an irrep, an example being $3^{--}$ which subduces twice into $[110]\,A_2$ -- one way to understand this is in terms of \emph{helicity}~\cite{Thomas:2011rh} where two linear combinations of the seven possible helicities of $J=3$ end up in this irrep. The (undesired) presence of positive parities like $0^{+-}, 2^{+-}$ in the in-flight irreps is unlikely to pose a problem for our calculation as these $J^{PC}$ quantum numbers are \emph{exotic} (inaccessible to $q\bar{q}$), and there is good evidence that the lightest such resonances appear at much higher energies than we will consider~\cite{Dudek:2013yja}. The possible in-flight irreps not listed in Table~\ref{subductions} are excluded because they include subductions of $1^{+-}$ which is expected to feature axial meson resonances -- we choose to avoid the complication of simultaneously describing such resonances in this first study.

\begin{table}[b]
\renewcommand{\arraystretch}{1.2}
\begin{tabular}{l|lll ll}
$[000]\, T_1^-$ & $1^{--}$ &          & $3^{--}$ \\
$[000]\, E^-$   &          & $2^{--}$ &          \\	
$[000]\, T_2^-$ &          & $2^{--}$ & $3^{--}$ \\
$[000]\, A_2^-$ &          &          & $3^{--}$ \\[1ex]
\hline\\[-2ex]
$[100]\, A_1$   & $1^{--}$ &          & $3^{--}$ & $\mathit{0^{+-}}$ & $\mathit{2^{+-}}$ \\
$[100]\, B_1$   &          & $2^{--}$ & $3^{--}$ &                   & $\mathit{2^{+-}}$ \\
$[100]\, B_2$   &          & $2^{--}$ & $3^{--}$ &                   & $\mathit{2^{+-}}$ \\[1ex]
\hline\\[-2ex]
$[110]\, A_1$   & $1^{--}$ & $2^{--}$ & $\big(3^{--}\big)^2$ & $\mathit{0^{+-}}$ & $\big(\mathit{2^{+-}}\big)^2$ \\[1ex]
\hline\\[-2ex]
$[111]\, A_1$   & $1^{--}$ & $2^{--}$ & $\big(3^{--}\big)^2$ & $\mathit{0^{+-}}$ & $\mathit{2^{+-}}$ \\
\end{tabular}
\caption{Subductions of $J^{PC}$ into cubic irreps, superscripts indicate multiple embeddings. Only $J<4$ shown. }
\label{subductions}
\end{table}

The scattering channels that can contribute to $J^{--}$ in the energy region where we expect to find resonances are $\etaOctet \omegaOctet$ with threshold $0.3632(2)$, $\omegaSinglet \fZeroSinglet $ with threshold $0.4181(19)$, and $\etaSinglet \omegaSinglet$ with threshold $0.4191(12)$. As shown in Table~\ref{partialwaves}, the first and last of these feature in $P$-- and $F$--waves, while the $\omegaSinglet \fZeroSinglet $ channel can contribute in $S$--wave.  We will compute correlation functions using operators which resemble all three of these meson-meson configurations, although in practice we will find that $ \omegaSinglet \fZeroSinglet$ and $\etaSinglet \omegaSinglet$ appear to be decoupled from each other, from $\etaOctet \omegaOctet$, and from resonances. The lowest three-meson channel, $\etaOctet \etaOctet \etaOctet$, has its threshold at $0.4434(2)$, but in order to contribute to $J^{--}$ scattering at least two $P$--waves are required, which will render the channel irrelevant in the energy region we consider.

\begin{table}
\begin{tabular}{l|lll}
$1^{--}$ & $\etaOctet \omegaOctet \big\{\! \threePone \!\big\}$ & $\omegaSinglet \fZeroSinglet \big\{\! \threeSone, \threeDone \!\big\}$ &  $\etaSinglet \omegaSinglet \big\{\! \threePone \!\big\}$ \\[2ex]
$2^{--}$ & $\etaOctet \omegaOctet \big\{\! \threePtwo, \threeFtwo \!\big\}$ & $\omegaSinglet \fZeroSinglet \big\{\! \threeDtwo \!\big\}$ &  $\etaSinglet \omegaSinglet \big\{\! \threePtwo, \threeFtwo \!\big\}$ \\[2ex]
$3^{--}$ & $\etaOctet \omegaOctet \big\{\! \threeFthree \!\big\}$ & $\omegaSinglet \fZeroSinglet \big\{\!  \threeDthree \!\big\}$ &  $\etaSinglet \omegaSinglet \big\{\! \threeFthree \!\big\}$ \\[2ex]
\end{tabular}
\caption{Meson-meson scattering partial-waves for each $J^{PC}$-- only waves with $\ell \leq 3$ shown.}
\label{partialwaves}
\end{table}

The construction of meson-meson-like operators which transform irreducibly under the relevant symmetries of the lattice has been discussed in detail previously (see for example Refs.~\cite{Dudek:2012gj, Woss:2018irj, Woss:2019hse}), but in short they are built as sums of products of definite-momentum operators \emph{optimized} for their overlap onto the relevant scattering meson. The summation runs over possible allowed rotations of the momentum of each meson, keeping the total momentum fixed. For example, an operator labelled $\etaOctet_{\scriptscriptstyle{[100]}} \omegaOctet_{\scriptscriptstyle{[110]}}$ will contribute in the $[100]\,A_1$ irrep, and in the limit in which the $\etaOctet$ and $\omegaOctet$ have no meson-meson interactions, this operator would interpolate an eigenstate with a \emph{non-interacting} energy of $\sqrt{{m^2_{\etaOctet}} + \left( \frac{2\pi}{L} \right)^{\!2} } + \sqrt{{m^2_{\omegaOctet}} + 2 \left( \frac{2\pi}{L} \right)^{\!2} } $. Interactions will move the actual finite-volume energy away from this value, and it is ultimately these volume-dependent shifts which allow us to determine the scattering amplitudes.

Our approach is to include all meson-meson operators which have a non-interacting energy, as measured in the center-of-momentum frame, below roughly $a_t E_\mathsf{cm} \sim 0.46$. These are supplemented with a large basis of fermion bilinear operators (``single-meson operators'') expected to have good overlap onto basis states resembling $q\bar{q}$ and hybrid meson configurations. With this basis we expect to obtain a set of energy eigenstates which constitute the complete finite-volume spectrum below $a_t E_\mathsf{cm} \sim 0.46$. The operator basis for each irrep is provided in Appendix~\ref{ops}.

In each irrep, a spectrum is determined by solving a generalized eigenvalue problem featuring the matrix of correlation functions~\cite{Michael:1985ne,Luscher:1990ck,Dudek:2007wv,Dudek:2010wm,Blossier:2009kd}. The resulting eigenvalues each have a time-dependence controlled dominantly by the energy of one finite-volume eigenstate, and the corresponding eigenvectors can be related to the overlap of that state with each operator in the basis. In order to verify that our finite-volume spectra are not overly sensitive to the specific choice of operator basis, we perform several diagonalizations, varying which single-meson operators we include, and also check that excluding those meson-meson operators with the highest non-interacting energies does not lead to a significantly different low-lying spectrum. Any such sensitivity (which is rare) is included as a systematic error on the finite-volume energy.

\begin{figure}
\includegraphics[width=1.04\columnwidth]{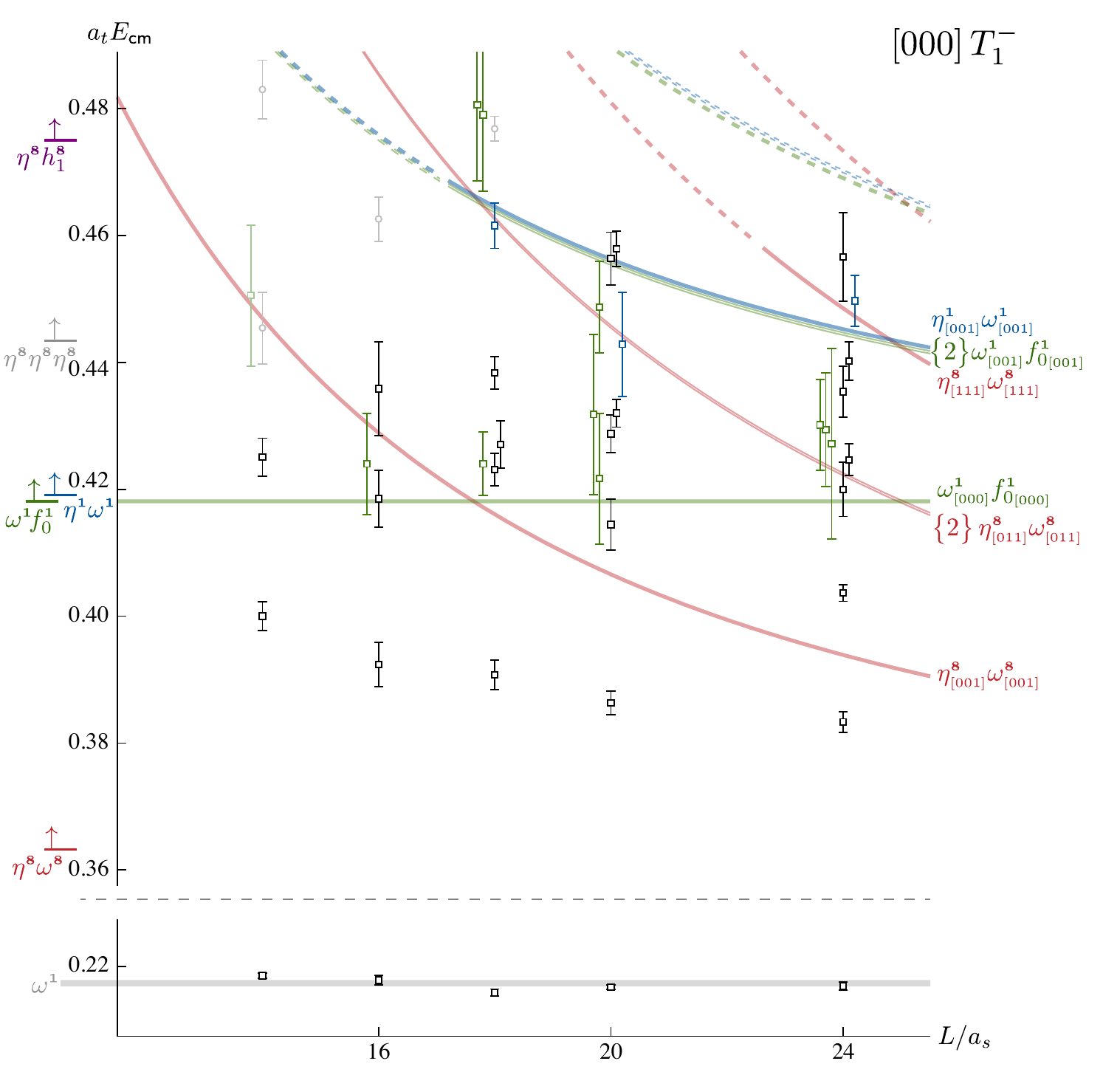}
\caption{
Finite-volume spectra in the $[000]\, T_1^-$ irrep extracted from matrices of correlators built using the operators listed in Table~\ref{ops000}. States color-coded by their dominant operator overlap, as shown in Figure~\ref{T1m_histo}. Curves show non-interacting energies, and when dashed indicate that the corresponding operator(s) were not included in the basis. 
}
\label{T1m}
\end{figure}

An example set of spectra on the five lattice volumes considered is shown in Figure~\ref{T1m} for the case of the $[000]\,T_1^-$ irrep. The lightest state, present at approximately the same energy on each volume, can be identified as the stable $\omegaSinglet$ -- that it shows essentially no volume dependence supports the idea that the lattices used are large enough to avoid significant `polarization' effects, in which a single meson can have an effect on itself around the periodic world. The higher spectra show some large departures from the non-interacting energies (colored curves), and indeed the counting of levels is larger than the number of non-interacting levels, indicating strong meson-meson interactions and likely resonances. The spectra are observed to become dense above the $\omegaSinglet \fZeroSinglet$ threshold, and it is worth examining the overlaps of these finite-volume states onto the set of operators used.

\begin{figure}[b]
\includegraphics[width=0.91\columnwidth]{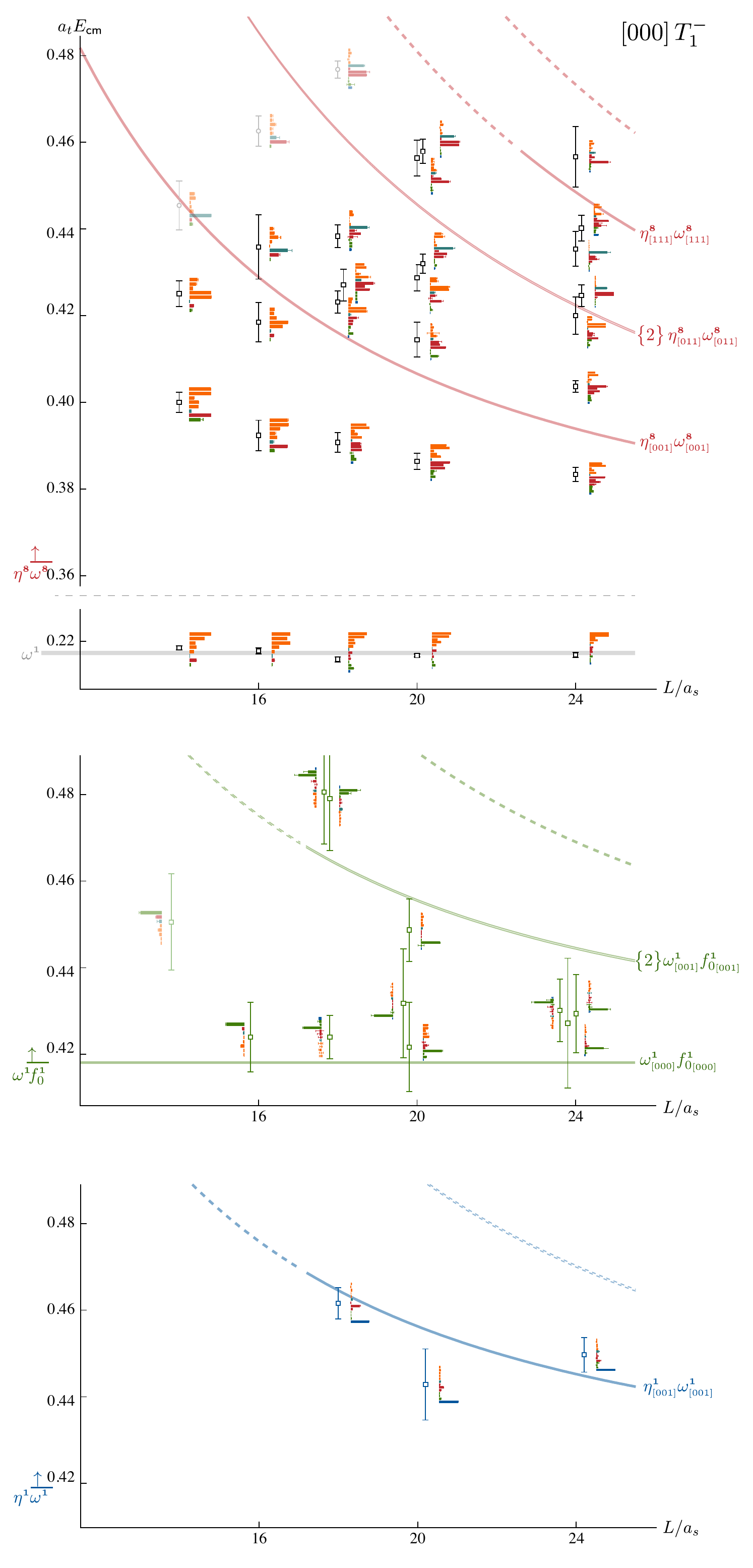}
\caption{
$[000]\, T_1^-$ spectra as in Figure~\ref{T1m} separated by dominant overlap onto $\bar{\psi}{\bf \Gamma}\psi, \etaOctet \omegaOctet$ (top panel), $\omegaSinglet \fZeroSinglet$ (middle panel) or $\etaSinglet \omegaSinglet$ (bottom panel). Histograms show the overlap onto a subset of operators used to build the matrix of correlators. Orange bars (top to bottom): 
$\big(\rho \!\times\! D^{[0]}_{J=0}\big)^{J=1}$, 
$\big(\rho \!\times\! D^{[2]}_{J=0}\big)^{J=1}$, 
$\big(\pi \!\times\! D^{[2]}_{J=1}\big)^{J=1}$, 
$\big(a_0 \!\times\! D^{[1]}_{J=1}\big)^{J=1}$, 
$\big(\rho \!\times\! D^{[2]}_{J=2}\big)^{J=1}$, 
cyan bar: $\big(\rho \!\times\! D^{[2]}_{J=2}\big)^{J=3}$, 
red bars: $\etaOctet \omegaOctet$ (increasing momentum top to bottom), 
green bars: $\omegaSinglet \fZeroSinglet$, 
blue bar: $\etaSinglet \omegaSinglet$.
}
\label{T1m_histo}
\end{figure}

Figure~\ref{T1m_histo} shows the same spectra as Figure~\ref{T1m} with the addition of histograms that illustrate the size of overlaps onto a subset of the operator basis used\,\footnote{The normalization is such that for a given operator, the largest overlap within the complete spectrum of states extracted is given the value 1, and all others are expressed relative to this.}. The five orange bars show overlap onto five single-meson operators with $J=1$ (to be discussed below), the cyan bar shows a single-meson operator with $J=3$, the red bars represent the $\etaOctet \omegaOctet$ operators (ordered top-bottom as lowest-highest non-interacting energy), the green bars the $\omegaSinglet \fZeroSinglet$ operators, and the blue bar the $\etaSinglet \omegaSinglet$ operator. The spectrum has been separated into three panels because it is clear from the overlaps that some states have overlap onto only the $\omegaSinglet \fZeroSinglet$ operators, or onto only the $\etaSinglet \omegaSinglet$ operators, and we notice that these states are statistically compatible with lying on the non-interacting energy curves. This likely indicates that the $\omegaSinglet \fZeroSinglet$ and $\etaSinglet \omegaSinglet$ scattering channels are decoupled from each other, from $\etaOctet \omegaOctet$, and from any resonances.

Examining the upper panel of Figure~\ref{T1m_histo} we have a rather well determined spectrum in which states typically have overlap onto both the single-meson operators (orange, cyan) and the $\etaOctet \omegaOctet$ operators, which may be taken as an indication that there are ``$q\bar{q}$-like'' resonances present which can decay into $\etaOctet \omegaOctet$. The subset of single-meson operators shown are selected for the property that, as discussed in Ref.~\cite{Dudek:2011bn}, certain operators can be characterized by which $q\bar{q}$ constructions they overlap with in the non-relativistic limit. 
The first two orange bars shown represent $\big(\rho \!\times\! D^{[0]}_{J=0}\big)^{J=1}$ and $\big(\rho \!\times\! D^{[2]}_{J=0}\big)^{J=1}$, which have unsuppressed overlap with $q\bar{q}$ in a $\threeSone$ configuration (including radial excitations). 
The third operator, $\big(\pi \!\times\! D^{[2]}_{J=1}\big)^{J=1}$, which features the commutator of two gauge-covariant derivatives, is expected to overlap with hybrid mesons. 
The fourth operator, $\big(a_0 \!\times\! D^{[1]}_{J=1}\big)^{J=1}$, has overlap with both $q\bar{q}[\threeSone]$ and $q\bar{q}[\threeDone]$, 
while the fifth operator, $\big(\rho \!\times\! D^{[2]}_{J=2}\big)^{J=1}$, only overlaps with $q\bar{q}[\threeDone]$. 
We notice that the first excited state, located between $a_t E_\mathsf{cm} = 0.38$ and $0.40$ always has large overlap with the first two orange operators, likely signaling a significant $q\bar{q}[2 \threeSone]$ component. On each volume there is a state near $a_t E_\mathsf{cm} = 0.42$ having large overlap onto the fourth and fifth orange operator corresponding to $q\bar{q}[1 \threeDone]$. There are no states having large overlap with the third orange operator, which matches with our expectation, discussed earlier, that the $1^{--}$ hybrid meson lies at a higher energy than we are considering here\,\footnote{There are finite-volume states at energies larger than we have plotted with overlap onto the third operator.}. 
At least one state near to $a_t E_\mathsf{cm} = 0.44$ at each volume has overlap with the cyan operator, suggesting the presence of a $3^{--}$ resonance.

\begin{figure*}
\includegraphics[width=0.95\textwidth]{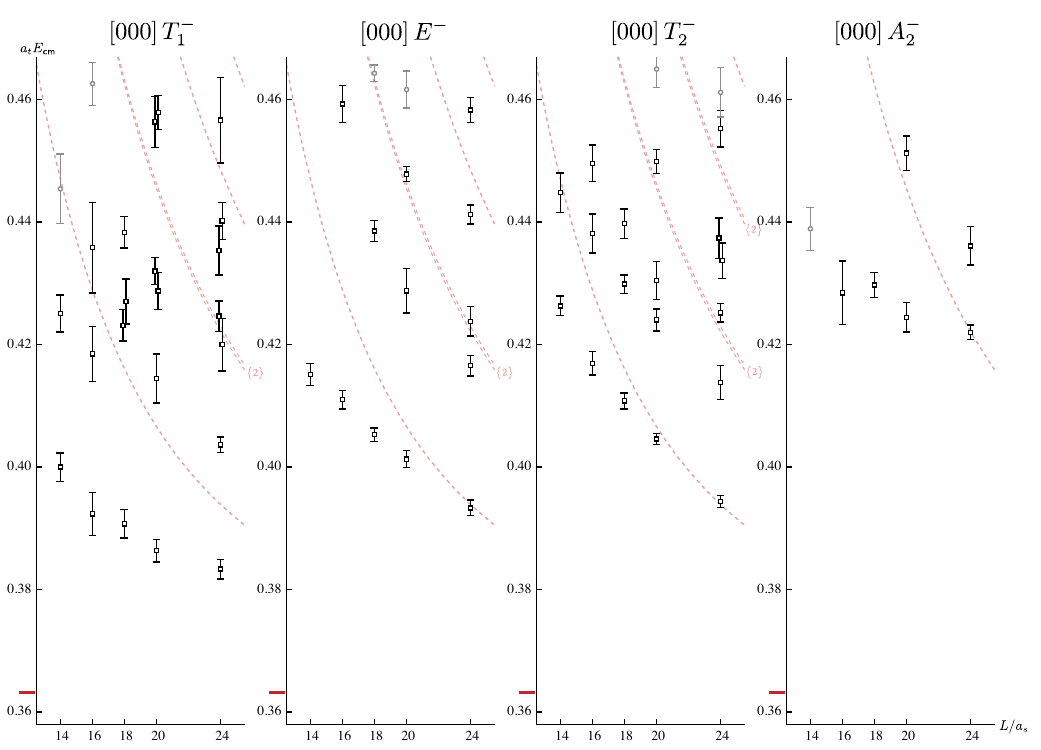}
\includegraphics[width=0.95\textwidth]{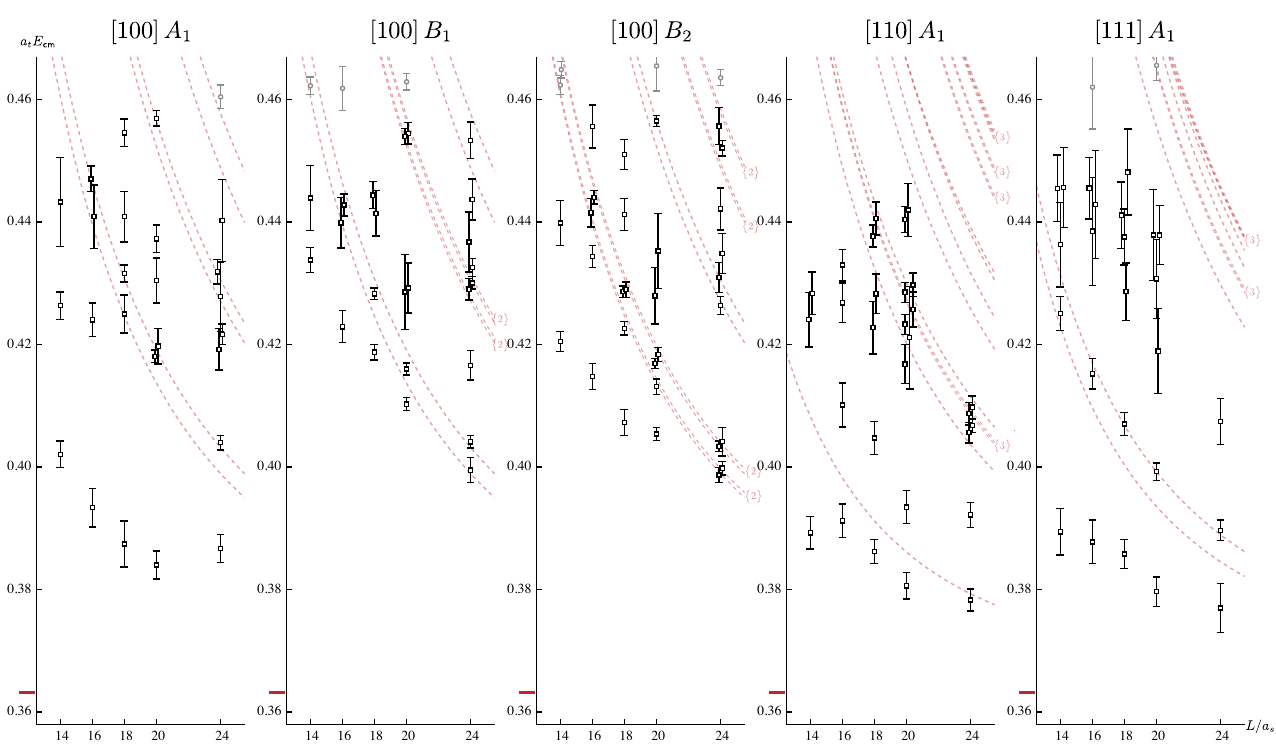}
\caption{
Energy levels with negligible $\etaSinglet \omegaSinglet$, $\omegaSinglet \fZeroSinglet$ overlap, assumed to form part of the $\etaOctet \omegaOctet$ scattering system. These levels will be analysed in terms of elastic $\etaOctet \omegaOctet$ scattering, with gray points not used. 
}
\label{e8o8_spectra}
\end{figure*}

This same procedure of examination of the overlap histograms has been performed for all computed irreps on all volumes, and in every case it appears that $\omegaSinglet \fZeroSinglet$ and $\etaSinglet \omegaSinglet$ are decoupled, and we propose to proceed under the assumption that $\etaOctet \omegaOctet$ can be considered as an elastic scattering system. We will seek to describe all finite-volume energy levels that remain when those levels having overlap onto $\omegaSinglet \fZeroSinglet$ or $\etaSinglet \omegaSinglet$ are excluded, as shown in Figure~\ref{e8o8_spectra}. In total this amounts to nearly 200 energy levels lying below $a_t E_\mathsf{cm} = 0.46$. As we will see later in explicit parameterizations of the relevant scattering amplitudes, the numbers of levels extracted in each energy region matches our expectations of there being two $1^{--}$ resonances, one $2^{--}$ resonance, and one $3^{--}$ resonance. 

\begin{figure}
\includegraphics[width=0.95\columnwidth]{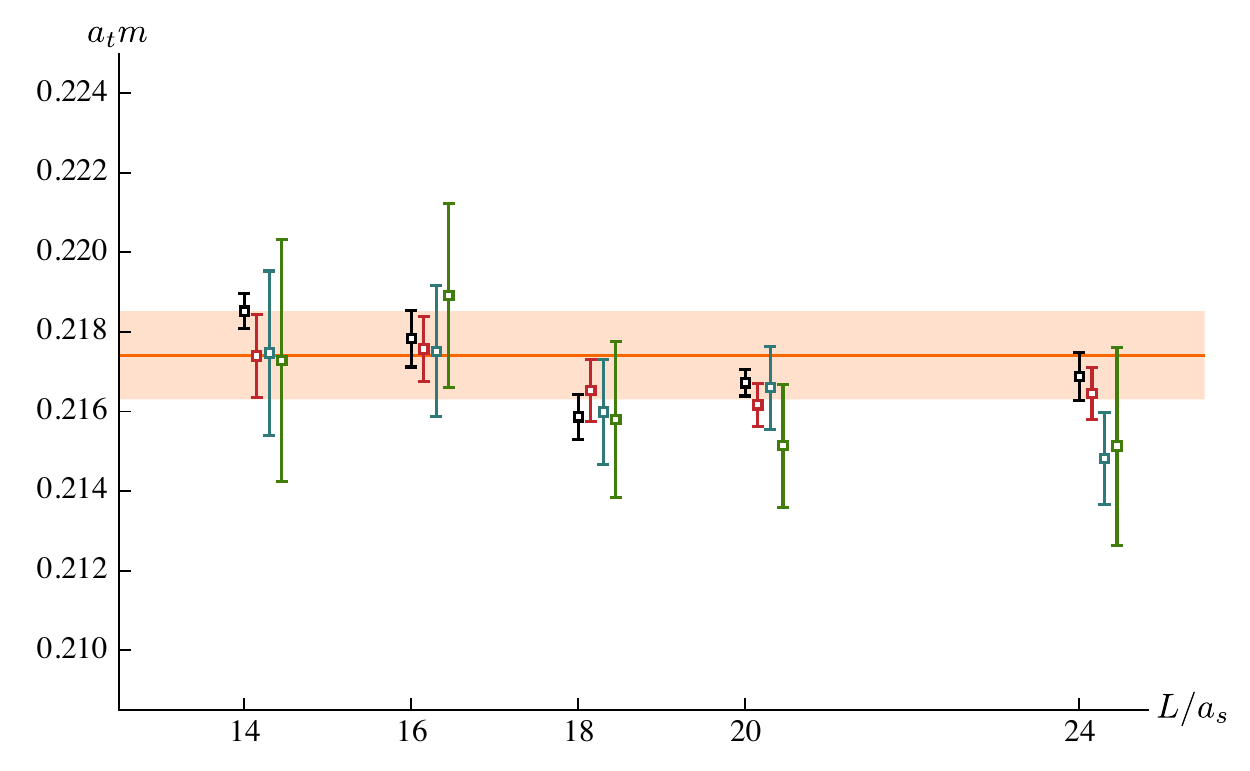}
\caption{ 
$\omegaSinglet$ mass as determining by `boosting' to the $\mathsf{cm}$-frame the lowest energy determined in each of the irreps: $[000]\,T_1^-$ (black), $[100]\,A_1$ (red), $[110]\,A_1$ (cyan), $[111]\,A_1$ (green). The orange band indicates the mass used for this state in scattering analysis.
}
\label{omega1}
\end{figure}

\newpage
The growth in the uncertainty on energy levels plotted as $a_t E_\mathsf{cm}$ as the frame-momentum increases can be traced back to the uncertainty we place on the anisotropy, $\xi$, which is accounted for when we `boost' the calculated energies in the moving-frame back to the center-of-momentum frame. This can be seen clearly in Figure~\ref{omega1} where we show the lowest energy level extracted in the $[000]\,T_1^-$, $[100]\,A_1$, $[110]\,A_1$, and $[111]\,A_1$ irreps, which we expect to be the stable $\omegaSinglet$. We see a consistent mass, but with a growth in uncertainty as the frame momentum increases.

Our hypothesis that the $\etaSinglet \omegaSinglet$ and $\omegaSinglet \fZeroSinglet$ channels are decoupled will be tested explicitly later using a limited set of coupled-channel amplitudes, but we note that should the hypothesis be incorrect, it will likely not be possible to find elastic $\etaOctet \omegaOctet$ amplitudes that are capable of describing all the energy levels in Figure~\ref{e8o8_spectra}. We will find that elastic amplitudes \emph{are} able to describe the spectrum rather well, and we will not find any significant evidence to support channel coupling in this system.

%% file: 3-amps.tex

The relationship between the $t$-matrix describing scattering and the finite-volume spectrum in a periodic $L \!\times\! L \!\times\! L$ box is encoded in the L\"uscher quantization condition,
\begin{equation}
\det \Big[ \bm{1} + i \bm{\rho} \, \bm{t} \, \big( \bm{1} + i \bm{\mathcal{M}} \big)  \Big] = 0, \label{luscher}
\end{equation}
and an extensive discussion of how the relationship can be implemented is presented in Ref.~\cite{Briceno:2017max}. Our approach is to make use of \emph{parameterizations} of the energy dependence of $\bm{t}(E_\mathsf{cm})$ and to attempt to describe as much of an obtained finite-volume lattice QCD spectrum as possible by varying the parameters in the parameterization, solving Eqn.~\ref{luscher} for the finite-volume spectrum for each choice of parameter values, and comparing to the lattice spectrum. An efficient method to solve the above equation, particularly applicable in cases of coupled-channels or coupled partial waves is presented in Ref.~\cite{Woss:2020cmp}.

An important feature of the above quantization condition is that it only has solutions for $t$-matrices which satisfy the unitarity condition that implements the conservation of probability. A straightforward way to ensure this is to make use of $K$-matrices by writing,
\begin{equation}
 \bm{t}^{-1} = \bm{K}^{-1} + \bm{I} \, , \label{tK}
\end{equation}
where $\bm{K}(s = E_\mathsf{cm}^2)$ is a symmetric real matrix in the space of coupled-channels and/or coupled partial waves, and where $\bm{I}(s)$ is a diagonal matrix with imaginary parts of value $\mathrm{Im}\, I_i(s) = - \rho_i(s)$, where the \emph{phase-space}, $\rho = \tfrac{2 k}{\sqrt{s}}$.
The real part of $I_i(s)$ can simply be chosen to be zero, in which case we speak of using the ``naive phase-space'', or we can make the choice to use the result of placing $\rho(s)$ in a dispersive integral, leading to what is often called the ``Chew-Mandelstam phase-space''. The dispersive integral is once-subtracted, and the location of the subtraction can be chosen for our convenience -- our implementation is described in Appendix B of Ref.~\cite{Wilson:2014cna}. 

We have significant freedom to choose parameterization forms for $\mathbf{K}(s)$, and a good approach is to try a range of parameterizations, finding as many as possible that can describe the finite-volume spectrum. If the resulting amplitude has features that are robust under changes of parameterization, we can be confident that they are true features of the actual QCD amplitude. This approach has been used extensively in previous calculations of elastic and coupled-channel scattering by the \emph{hadspec} collaboration~\cite{Dudek:2014qha,Wilson:2014cna,Wilson:2015dqa,Dudek:2016cru,Woss:2018irj,Woss:2019hse}.

In practice, for scattering in partial waves with non-zero orbital angular momentum, $\ell$, we extract from the $K$-matrix the momentum factors needed to get the correct threshold behavior, writing
\begin{equation}
 \left[\bm{t}^{-1}\right]_{\ell \ell'} = \tfrac{1}{(2 k)^\ell} \left[\bm{K}^{-1}\right]_{\ell \ell'} \tfrac{1}{(2 k)^{\ell'}}  + \bm{I}, \label{tKk}
\end{equation}
for the case of two coupled partial waves ($\ell$, $\ell'$). Details of how dynamically coupled partial waves are handled in the $\ell S$ basis can be found in Refs.~\cite{Woss:2018irj, Woss:2019hse, Woss:2020ayi}.

Resonances in a scattering system are associated with \emph{pole singularities} at complex values of $s$ located on \emph{unphysical} Riemann sheets (those where the scattering momentum has a negative imaginary part). The real and imaginary parts of the pole position are commonly given meanings in terms of the mass and total width of the resonance, $\sqrt{s_0} = m \pm \tfrac{i}{2} \Gamma$, where the two signs reflect the fact that these poles always come in complex-conjugate pairs. Couplings to decay channels in each partial wave, $c_i$, can be obtained by factorizing the residue at the pole position,
\begin{equation*}
t_{ij}(s) \sim \frac{c_i \, c_j}{s_0 -s}.
\end{equation*}  

Poles can also lie on the real energy axis below the lowest kinematic threshold -- if they appear on the physical sheet they are associated with stable \emph{bound states} that can appear as asymptotic particles, while if they appear on unphysical sheets they are termed \emph{virtual bound states} which do not have an associated asymptotic particle.

The $t$-matrix can have other singularities, notably \emph{cuts} associated with the dynamics of scattering in crossed-channels. These are known as \emph{left-hand cuts}, and typically their effect on physical scattering is much milder than the effects of narrow resonances, and their net effect above threshold can be modelled by including slowly varying polynomial behavior in the $K$-matrix\,\footnote{But see Refs.~\cite{Briceno:2016mjc, Wilson:2019wfr} for cases where resonances may appear that are very broad and where the physics of the left-hand cut may become relevant.}. They are discussed further in the current context in Appendix~\ref{lhc}.

We now proceed to present descriptions of the finite-volume spectra introduced in the previous section, beginning with the assumed elastic $\etaOctet \omegaOctet$ spectra of Figure~\ref{e8o8_spectra}, using parameterizations of elastic scattering in $J^{PC} = 1^{--}$, $2^{--}$ and $3^{--}$. We will consider several strategies to isolate the amplitudes, firstly considering those irreps which only depend upon scattering with $J^{PC}= 2^{--}$ and/or $3^{--}$, then those irreps which depend only upon $1^{--}$ and $3^{--}$, before finally attempting a global description of all the energy levels.

\subsection{$\bm{J^{PC} = 3^{--}}$ from the $\bm{[000]\, A_2^-}$ irrep}

The only partial wave expected to contribute in the $[000]\, A_2^-$ irrep at the energies we are considering is $\etaOctet \omegaOctet  \{ \threeFthree \}$, and the volume dependence of energy levels in the top right panel of Figure~\ref{e8o8_spectra} appears to be a canonical ``avoided level crossing'' indicating a narrow resonance near $a_t E_\mathsf{cm} \sim 0.43$. 

An elastic $K$-matrix featuring a single pole, ${K(s) = \frac{g^2}{m^2 -s} }$, and either a Chew-Mandelstam or naive phase-space is capable of describing this spectrum with a $\chi^2/N_\mathrm{dof} = \tfrac{5.2}{6-2}= 1.31$. The resulting amplitude has a narrow peak and no other features, and the $t$-matrix has a pole at ${a_t \sqrt{s_0} = 0.4296(16) \pm \tfrac{i}{2} 0.0027(8)}$ with a pole coupling of magnitude $a_t |c_{\etaOctet \omegaOctet}| = 0.047(7)$. Allowing additional freedom in the amplitude by adding a constant to the \mbox{$K$-matrix} leads to a negligible change in the quality of fit, and a consistent resonance pole.

From this analysis of a single irrep, it is clear that there is a narrow $3^{--}$ resonance -- we will delay providing further discussion until we report a more precise determination of its pole parameters using a description of more energy levels.

\subsection{$\bm{J^{PC} = 2^{--}}$ from the $\bm{[000]\, E^-}$ irrep}

Assuming negligible $J=4$ scattering, the $[000]\, E^-$ irrep spectrum is controlled by the coupled partial waves, $\etaOctet \omegaOctet  \{ \threePtwo, \threeFtwo \}$, requiring a two-dimensional $t$-matrix,
\begin{equation*}
\bm{t} = \begin{bmatrix} t_{PP} & t_{PF} \\ t_{PF} & t_{FF}  \end{bmatrix}.
\end{equation*}

 Examining the volume dependence of energy levels in the second panel of Figure~\ref{e8o8_spectra}, we potentially observe an avoided level crossing with the lowest non-interacting $\etaOctet \omegaOctet$ curve, but spread out over a large energy range, which might signal a broad resonance somewhere around $a_t E_\mathsf{cm} \sim 0.42$. A simple amplitude that proves capable of describing this spectrum is given by the $K$-matrix,
\begin{equation}
\bm{ K}(s) = 
\frac{1}{m^2 - s} \begin{bmatrix} g_P^2 && g_P g_F \\ g_P g_F && g_F^2 \end{bmatrix} 
+ \begin{bmatrix} \gamma_{PP} && \gamma_{PF} \\ \gamma_{PF} && \gamma_{FF} \end{bmatrix},
\label{J2_pole_const}
\end{equation}
where when the Chew-Mandelstam phase-space is used (subtracted at $s=m^2$) in Eqn.~\ref{tKk}, the 13 energy levels can be described with a ${\chi^2/N_\mathrm{dof} = \tfrac{5.6}{13-6} = 0.80}$. The $t$-matrix in this case has a pole singularity at 
${a_t \sqrt{s_0} = 0.4181(30) \pm \tfrac{i}{2} 0.0325(77)}$,
and pole couplings to the two partial waves with magnitudes ${a_t |c_{\etaOctet \omegaOctet \{\threePtwo\} }| = 0.145(20)}$ and 
${a_t |c_{\etaOctet \omegaOctet \{\threeFtwo\} }| = 0.030(13)}$. As anticipated this single resonance has a significantly larger width than the one seen in $3^{--}$ and, as we'd expect for a resonance lying only slightly above the relevant decay threshold, the angular momentum barrier ensures an $F$--wave decay coupling that is significantly smaller than the leading $P$--wave.

A more precise determination of this $2^{--}$ amplitude, and of the $3^{--}$ amplitude discussed previously, can be obtained by simultaneously describing both in a description of the 91 finite-volume energy levels in the irreps $[000]\, A_2^-, T_2^-, E^-$ and $ [100]\, B_1, B_2$.


\subsection{$\bm{J^{PC} = 2^{--}, 3^{--}}$ from the $\bm{[000]\, A_2^-, T_2^-, E^-, [100]\, B_1, B_2}$ irreps}

Five irreps, $[000]\, A_2^-, T_2^-, E^-$, and $[100]\, B_1, B_2$, are each sensitive to one or both of the $2^{--}$ and $3^{--}$ scattering amplitudes, and together feature 91 energy levels that we can use to constrain the energy dependence. As an example, using a single $K$-matrix pole for the $3^{--}$ amplitude, $K(s) = \frac{g^2}{m^2 -s}$, and the $K$-matrix presented in Eqn.~\ref{J2_pole_const} for the $2^{--}$ amplitude, with Chew-Mandelstam phase-space (subtracted at $s=m^2$) in both cases, we obtain a best-fit description of the finite-volume spectra with parameters,
\begin{widetext}
\vspace{0.5cm}
	\begin{center}
		\begin{tabular}{crll}
\ldelim\{{8.5}{*}[$J=2\,$]	&		$m =$  & $ 0.4322(15) \cdot a_t^{-1}$   &
			\multirow{9}{*}{ $\begin{bmatrix*}[r] 1 & 0.31 & 0.29  & 0.13  & -0.37 & 0.31  & 0.19  & 0.07 \\[1.1ex]
				                                    & 1    & -0.08 & -0.70 & 0.04  & 0.48  & 0.07  & -0.23 \\[1.1ex]
													&      & 1     & 0.21  & -0.15 & -0.18 & -0.01 & -0.12 \\[1.1ex]
													&      &       & 1     & -0.34 & -0.34 & -0.16 & 0.23 \\[1.1ex]
													&      &       &       & 1     & -0.23 & -0.03 & -0.05 \\[1.1ex]
													&	   &       &       &       & 1     & 0.02  & 0.05 \\[1.5ex]
													&      &       &       &       &       & 1     & -0.04 \\[1.1ex]
													&      &       &       &       &       &       & 1 \\[1.1ex]
													 \end{bmatrix*}$ } \\[1.1ex]
&			$g_P  =$ & $ 0.753(37)  $   & \\[1.1ex]
&			$g_F =$ & $ -4.13(29) \cdot a_t^{2}$   & \\[1.1ex]
&			$\gamma_{PP} = $ & $0.1(33)\cdot a_t^{2}$   & \\[1.1ex]
&			$\gamma_{PF} = $ & $-110(17)\cdot a_t^{4}$   & \\[1.1ex]
&			$\gamma_{FF} = $ & $143(322)\cdot a_t^{6}$   & \\[1.5ex]
\ldelim\{{2.6}{*}[$J=3\, $]	&		$m =$  & $ 0.4341(9) \cdot a_t^{-1}$ \\[1.1ex]
  &			$g  =$ & $ 4.85(28) \cdot a_t^{2} $   & \\[2.5ex]
			\multicolumn{4}{r}{$\chi^2/N_{\text{dof}}=\frac{120.3}{91 - 8}=1.45$\,.}
		\end{tabular}
	\end{center}
	\vspace{-1.1cm}
	\begin{equation} \label{J2J3example}  \end{equation}
\end{widetext}

\begin{figure}[b]
\includegraphics[width=0.98\columnwidth]{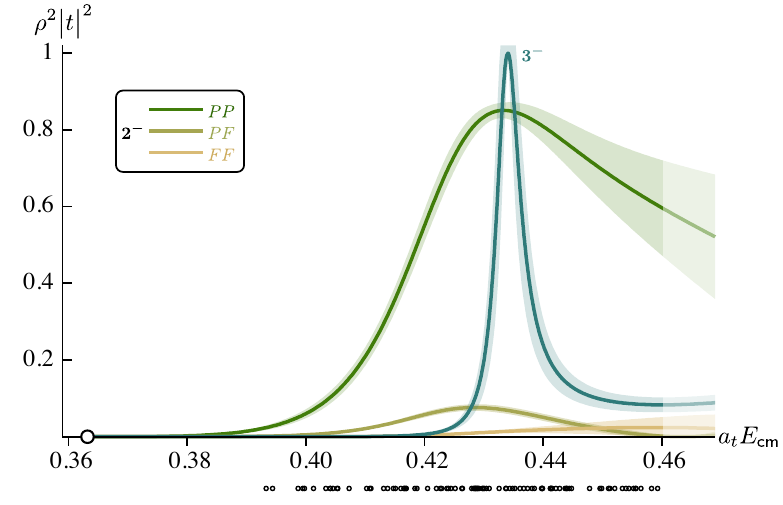}
\caption{
Scattering amplitudes for $J^{PC} = 2^{--}$ (Eqn.~\ref{J2_pole_const}) and $3^{--}$ ($K$-matrix pole), for the best-fit parameters of Eqn.~\ref{J2J3example}. Points below the abscissa show the positions of the finite-volume energy-levels constraining the amplitudes.
}
\label{2m3m_ref_rhotsq}
\end{figure}

The fit quality is quite reasonable, and the parameters show no particularly large correlations. We observe that the constants $\gamma_{PP}, \gamma_{FF}$ are probably redundant, and later we will explore fixing them to zero. The resulting amplitudes are shown in Figure~\ref{2m3m_ref_rhotsq} where the bumps suggest the presence of a narrow resonance in $3^{--}$, and a broader resonance in $2^{--}$. The dominance of $\threePtwo$ over $\threeFtwo$ is obvious, and while the resonance bump is still visible in the off-diagonal element $t_{PF}$, albeit peaking at a slightly lower energy than in the $t_{PP}$ element, there is no clear peak in the weak $t_{FF}$ element.

These best-fit amplitudes feature $2^{--}$ and $3^{--}$ $t$-matrix poles that are compatible with those reported in previous sections in fits to $[000]\, E^-$, $[000]\, A_2^-$ alone, but which now have improved statistical uncertainty:
\begin{align*}
2^{--}: \quad\quad a_t \sqrt{s_0} &= 0.4235(18) \pm \tfrac{i}{2} 0.0375(34) \nonumber \\[1.2ex]
a_t c_{\etaOctet \omegaOctet\{ \threePtwo \}} &= 0.164(12) \, e^{\pm i \pi \, 0.13(2)} \nonumber \\
a_t c_{\etaOctet \omegaOctet\{ \threeFtwo \}} &= 0.057(6) \, e^{\mp i \pi \, 0.88(2)} \nonumber \\[2.2ex]
3^{--}: \quad\quad a_t \sqrt{s_0} &= 0.4338(9) \pm \tfrac{i}{2} 0.0049(6) \nonumber \\[1.2ex]
a_t c_{\etaOctet \omegaOctet\{ \threeFthree \}} &= 0.062(4) \, e^{\pm i \pi \, 0.038(5)} \, .
\end{align*}
We note that the $F$--wave couplings for the $2^{--}$ and $3^{--}$ resonances are of a very similar size. The $2^{--}$ resonance pole (in the lower half plane) has a ratio of $F$--wave to $P$--wave couplings of $0.35(3)\, e^{ i \pi \, 1.01(2)} \approx -0.35(3)$, which is close to being real and negative.

\begin{figure}
\includegraphics[width=0.98\columnwidth]{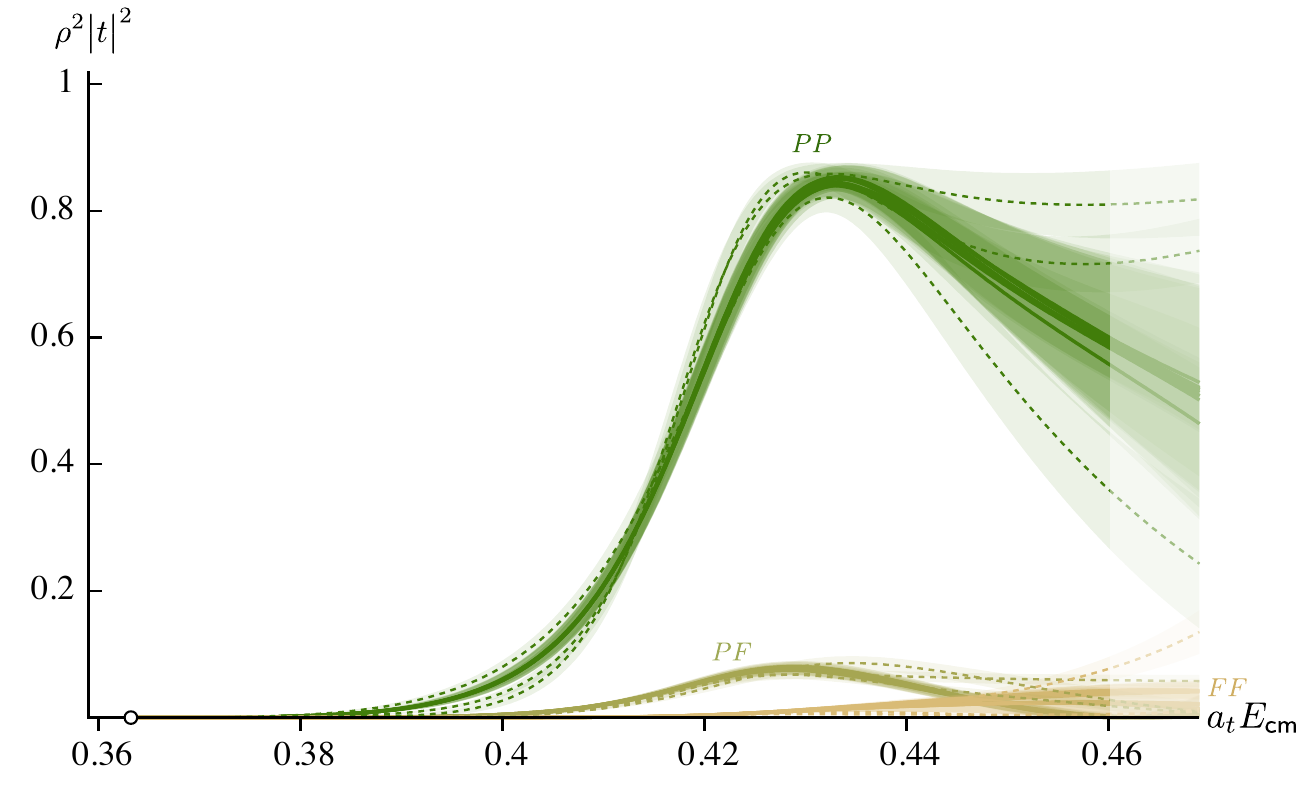}
\caption{
Variation of $2^{--}$ amplitude over parameterization choice. Solid curves and bands show descriptions of the finite-volume spectra with $1.42 < \chi^2/N_\mathrm{dof} < 1.46$, while dashed curves have $1.66 < \chi^2/N_\mathrm{dof} < 1.86$.
}
\label{2m_variation_rhotsq}
\end{figure}

In order to establish that these results are not overly sensitive to details of the particular amplitude parameterization selected, we also explore descriptions of the finite-volume spectra using other choices, which include: replacing the Chew-Mandelstam phase-space with the naive phase space, setting some of the constants added to the $K$-matrix pole in Eqn.~\ref{J2_pole_const} to zero, replacing some of the constants with terms of form $\gamma\! \cdot\! s$, using a second pole in place of the constants, or by writing $\bm{K}^{-1}$ as a matrix of polynomials. The forms used are listed in Table~\ref{2mm_amps} in Appendix~\ref{par}. The resulting $2^{--}$ amplitudes\,\footnote{The $3^{--}$ amplitudes show completely negligible variation.} are shown in Figure~\ref{2m_variation_rhotsq}, and it is quite clear that the amplitude previously presented in Figure~\ref{2m3m_ref_rhotsq} is representative, and in fact it has among the largest statistical uncertainties of those amplitudes considered. While the bulk of the amplitudes tried have a $\chi^2/N_\mathrm{dof}$ very close to the value $1.45$ obtained using Eqn.~\ref{J2_pole_const}, there are three choices that have somewhat larger values: Eqn.~\ref{J2_pole_const} with $\gamma_{PF}$ fixed to zero ($\chi^2/N_\mathrm{dof} = 1.85$), a $K$-matrix built as the sum of two poles ($\chi^2/N_\mathrm{dof} = 1.80$), and a form where $\bm{K}^{-1}$ is parameterized as independent linear polynomials ($a+ b s$) in each element ($\chi^2/N_\mathrm{dof} = 1.67$). These outliers are shown by the dashed lines in Figure~\ref{2m_variation_rhotsq}, where we note that they have a slight difference in $t_{PP}$ peak position, but otherwise only start to show significant deviation from the solid curves above the energy region where constraint is provided by the finite-volume spectra.

\begin{figure}
\includegraphics[width=0.99\columnwidth]{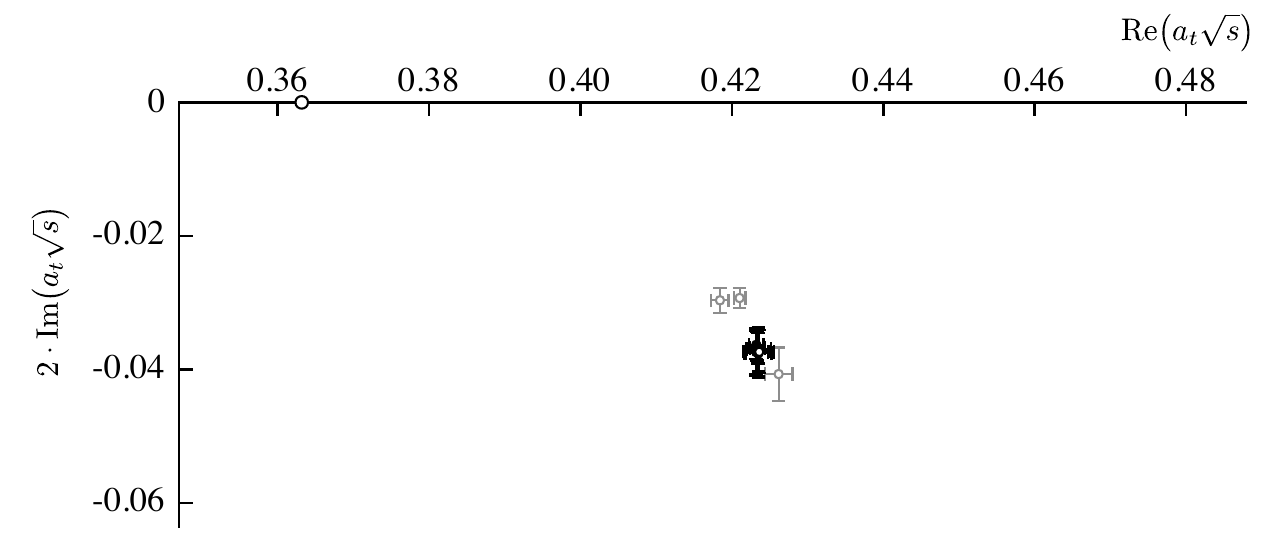}
\includegraphics[width=0.99\columnwidth]{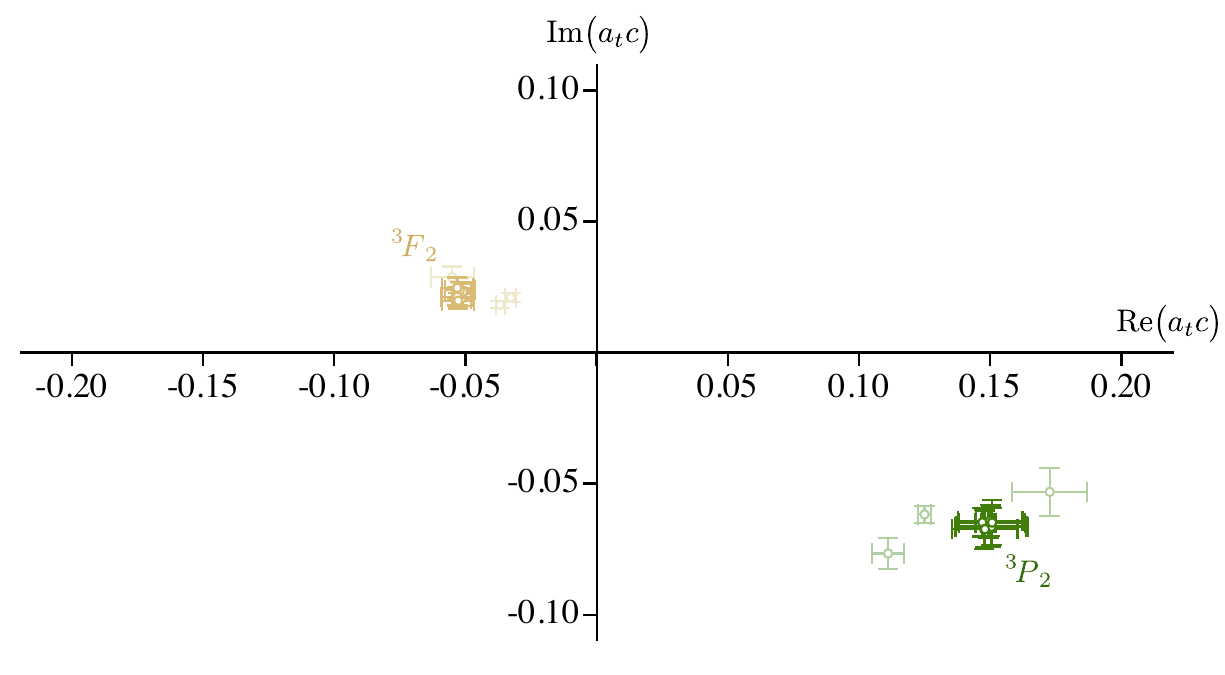}
\includegraphics[width=0.99\columnwidth]{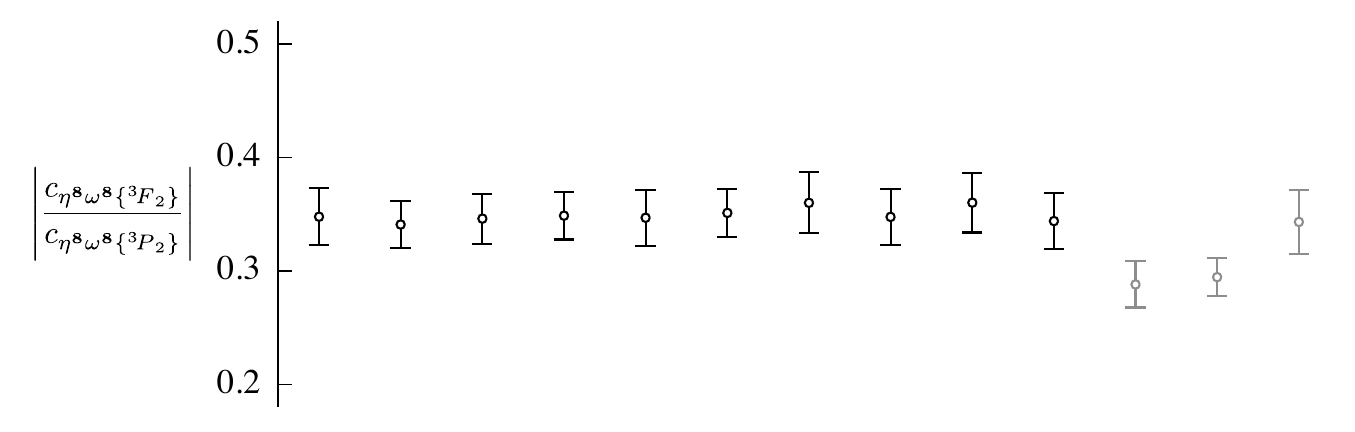}
\caption{
Top panel: $2^{--}$ $t$-matrix pole positions for parameterization variations shown in Figure~\ref{2m_variation_rhotsq} -- black points show the fits with $1.42 < \chi^2/N_\mathrm{dof} < 1.46$ and grey those with $1.66 < \chi^2/N_\mathrm{dof} < 1.86$. Middle panel: the couplings, $c_{\etaOctet \omegaOctet \{\threePtwo\}}$ (green) and $c_{\etaOctet \omegaOctet \{\threeFtwo\}}$ (sand) obtained from factorizing the residue of the $t$-matrix pole in the lower half-plane -- lighter points show those fits with  poorer $\chi^2$. Bottom panel: The magnitude of the ratio of the couplings.
}
\label{2m_variation_pole}
\end{figure}

Figure~\ref{2m_variation_pole} shows the $t$-matrix pole positions and the corresponding pole couplings for the parameterization variations, indicating a clear consensus that agrees with the reference amplitude described previously.
The amplitudes considered do have other pole singularities in addition to the one presented in Figure~\ref{2m_variation_pole}, but they are typically distant from physical scattering and vary with parameterization choice. A typical example, present in the reference amplitude, is a pole on the real axis on the unphysical sheet near $a_t \sqrt{s} \sim 0.23$ -- such a pole is present for many of our amplitudes, although its precise position varies, always remaining far from physical scattering, and as such it remains largely irrelevant to physical scattering. As one might expect given its inferior analytic properties, using the naive phase-space in place of the Chew-Mandelstam function leads to additional singularities, in particular a physical sheet pole on the real energy axis at $a_t \sqrt{s} \sim 0.24$ which is found to have \emph{real-valued} couplings. In odd-$\ell$ scattering, a true bound-state must have imaginary couplings, so the presence of this pole signifies a \emph{ghost} state having negative probability. Such a singularity suggests a flaw in the parameterization, but in practice the ghost pole is so far from physical scattering that it has a negligible impact -- we will further discuss such poles, and their relation to neglected left-hand cuts later, in the context of $1^{--}$ scattering\,\footnote{A ghost pole in $S$-wave is the pathology that causes an amplitude to fail the `sanity check' of Iritani \emph{et.al.}~\cite{Iritani:2017rlk}.}.

The case of an amplitude in which $\bm{K}^{-1}$ is parameterized with linear functions features a different pathology: there are poles \emph{off the real axis on the physical sheet}, albeit fairly deep into the complex plane. Such poles signal a breakdown in \emph{causality} which comes about because we do not place analyticity constraints upon our amplitudes.

We will later return to further discussion of the $2^{--}$ and $3^{--}$ amplitudes in the context of a global analysis of all of our finite-volume energies levels, while now we move to an initial determination of the $J^{PC} = 1^{--}$ amplitude.

\subsection{$\bm{J^{PC} = 1^{--}, 3^{--}}$ from the $\bm{ [000]\, T_1^-,\, [100]\, A_1,\, [111]\, A_1 }$ irreps}

Irreps $[000]\, T_1^-, [100]\, A_1$, and $[111]\, A_1$ all depend upon both the $1^{--}$ and the $3^{--}$ amplitude, but not the $2^{--}$ amplitude. Since we have a well-constrained $3^{--}$ amplitude from the previous subsection, we choose to initially fix this amplitude, and only vary the $1^{--}$ amplitude. We will relax this later when we attempt descriptions of our entire set of finite-volume energy levels.

\begin{figure}
\includegraphics[width=0.98\columnwidth]{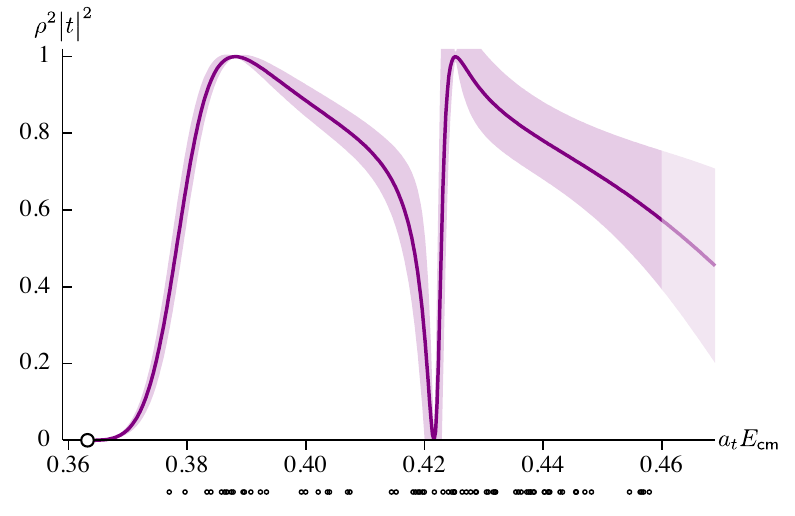}
\caption{
Illustrative two-pole plus constant $\etaOctet \omegaOctet \{\threePone \}$($1^{--}$) elastic scattering amplitude (Eqn.~\ref{J1example}). Points below the abscissa show the positions of the finite-volume energy-levels constraining the amplitudes.
}
\label{1m_ref_rhotsq}
\end{figure}

We will not try to include the very deeply-bound stable $\omegaSinglet$ as an explicit pole in our scattering amplitudes, and hence we exclude the lowest energy level in each irrep on each volume. There are 72 suitable energy levels below $a_t E_\mathsf{cm} = 0.46$ shown in Figure~\ref{e8o8_spectra}, and as discussed in Section~\ref{fvspectrum} and shown in Figure~\ref{T1m_histo}, the spectra and operator overlaps hint at there being two $1^{--}$ resonances present. Narrow resonances are most conveniently parameterized by including explicit poles in the $K$-matrix, and as such a good choice of amplitude to illustrate this case features two poles and a constant, where the constant allows for some flexibility away from a pure superposition of resonances. The Chew-Mandelstam phase-space, subtracted at the lower mass pole ($s=m_\mathsf{a}^2$) is used, and the best fit parameters are found to be,
\vspace{0.1cm}
	\begin{center}
		\begin{tabular}{rll}
 $m_\mathsf{a} =$  & $ 0.3881(14) \cdot a_t^{-1}$   &
			\multirow{5}{*}{ $\begin{bmatrix*}[r] 1 & 0.08  & 0.43  & -0.33  & 0.19  \\[1.1ex]
				                                    & 1     & 0.37  & -0.46  & 0.81  \\[1.1ex]
													&       & 1     & -0.86  & 0.49  \\[1.1ex]
													&       &       & 1      & -0.57 \\[1.1ex]
													&       &       &        & 1     \\[1.1ex]
													 \end{bmatrix*}$ } \\[1.1ex]
			$g_\mathsf{a} = $ &  $ 1.46(10)    $   & \\[1.1ex]
			$m_\mathsf{b} = $ &  $ 0.4242(17) \cdot a_t^{-1} $   & \\[1.1ex]
			$g_\mathsf{b} = $ &  $-0.36(13) $    & \\[1.1ex]
			$\gamma       = $ &  $20.9(86)    \cdot a_t^2$    & \\[2.2ex]
&			\multicolumn{2}{r}{$\chi^2/N_{\text{dof}}=\frac{91.3}{72 - 5}=1.36$\,.}
		\end{tabular}
	\end{center}
	\vspace{-0.5cm}
	\begin{equation} \label{J1example}  \end{equation}
The description is reasonable, and we note that the parameter correlations are modest, with the constant $\gamma$ being statistically significant. The resulting amplitude is shown in Figure~\ref{1m_ref_rhotsq} where we observe a prominent dip with a zero located at $a_t E_\mathsf{cm} = 0.4216(9)$. The points shown beneath the abscissa show the positions of the energy levels used to constrain the amplitude, which cover the entire energy region of interest. When the $t$-matrix is examined in the complex $s$-plane, two pole singularities are found close to the real axis on the unphysical sheet:
\begin{align*}
a_t \sqrt{s_0} &= 0.3787(16) \pm \tfrac{i}{2} 0.0187(13) \nonumber \\[1.1ex]
a_t c_{\etaOctet \omegaOctet\{ \threePone \}} &= 0.144(4) \, e^{\pm i \pi \, 0.202(17)} \, ,
\end{align*}
and 
\begin{align*}
a_t \sqrt{s_0} &= 0.4224(8) \pm \tfrac{i}{2} 0.0030(20) \nonumber \\[1.1ex]
a_t c_{\etaOctet \omegaOctet\{ \threePone \}} &= 0.051(17) \, e^{\pm i \pi \, 0.352(27)} \, .
\end{align*}

\newpage

We interpret these two $t$-matrix poles as being the signal for two $1^{--}$ resonances, a lighter broader state, and a heavier narrow state. The zero on the real energy axis is located close to the second resonance pole\,\footnote{
There is guaranteed to be a zero located between $s= m_\mathsf{a}^2$ and $s= m_\mathsf{b}^2$ whenever a two-pole plus polynomial form is used for an \emph{elastic} $K$-matrix. 
Since $t = \tfrac{K}{1+ I K}$ and $K(s) = \tfrac{g_\mathsf{a}^2}{m_\mathsf{a}^2 - s} + \tfrac{g_\mathsf{b}^2}{m_\mathsf{b}^2 - s} + \gamma(s)$, defining $P(s) = (m_\mathsf{a}^2 - s)(m_\mathsf{b}^2 - s) K(s)$ we have $t(s) = \tfrac{P(s)}{ (m_\mathsf{a}^2 - s)(m_\mathsf{b}^2 - s) + I(s) P(s)} $ and a zero of $t(s)$ will appear when $P(s)=0$. Since $P(m_\mathsf{a}^2) = g_\mathsf{a}^2 ( m_\mathsf{b}^2 - m_\mathsf{a}^2) > 0$ and ${ P(m_\mathsf{b}^2) = - g_\mathsf{b}^2 ( m_\mathsf{b}^2 - m_\mathsf{a}^2) < 0 }$, $P(s)$ must cross zero at least once between $s=m_\mathsf{a}^2$ and $s=m_\mathsf{b}^2$. If $g_\mathsf{b}$ is small, it is clear that $P(m_\mathsf{b}^2)$ will have a value close to zero and hence the zero of $t(s)$ will be close to $s= m_\mathsf{b}^2$.
}.

Elastic unitarity is a strong constraint that significantly restricts the possible behavior of an amplitude like this, and as seen in Figure~\ref{1m_ref_rhotsq}, there is clearly a non-trivial energy dependence, one that does not for example simply consist of two separated bumps as one might anticipate given the resonance content. This is one reason why the use of complex $s$-plane pole positions is advocated as a rigorous identification of resonances -- one could not describe this amplitude as a sum of two Breit-Wigners.

We now move to explore whether the same finite-volume spectrum can be described by other choices of amplitude parameterization, and whether the resulting amplitudes have the same features as just observed. Variations considered include varying the choice for $I(s)$, by changing the subtraction point or by simply using the naive phase-space, and varying what kind of polynomial is added to the two $K$-matrix poles. Table~\ref{1mm_amps} in Appendix~\ref{par} lists the variations, and in Figure~\ref{1m_rhotsq_variation} we show the amplitudes obtained using these parameterization variations, all of which prove capable of describing the finite-volume spectra with ${\chi^2 / N_\mathrm{dof} < 1.44}$. 

\begin{figure}
\includegraphics[width=0.98\columnwidth]{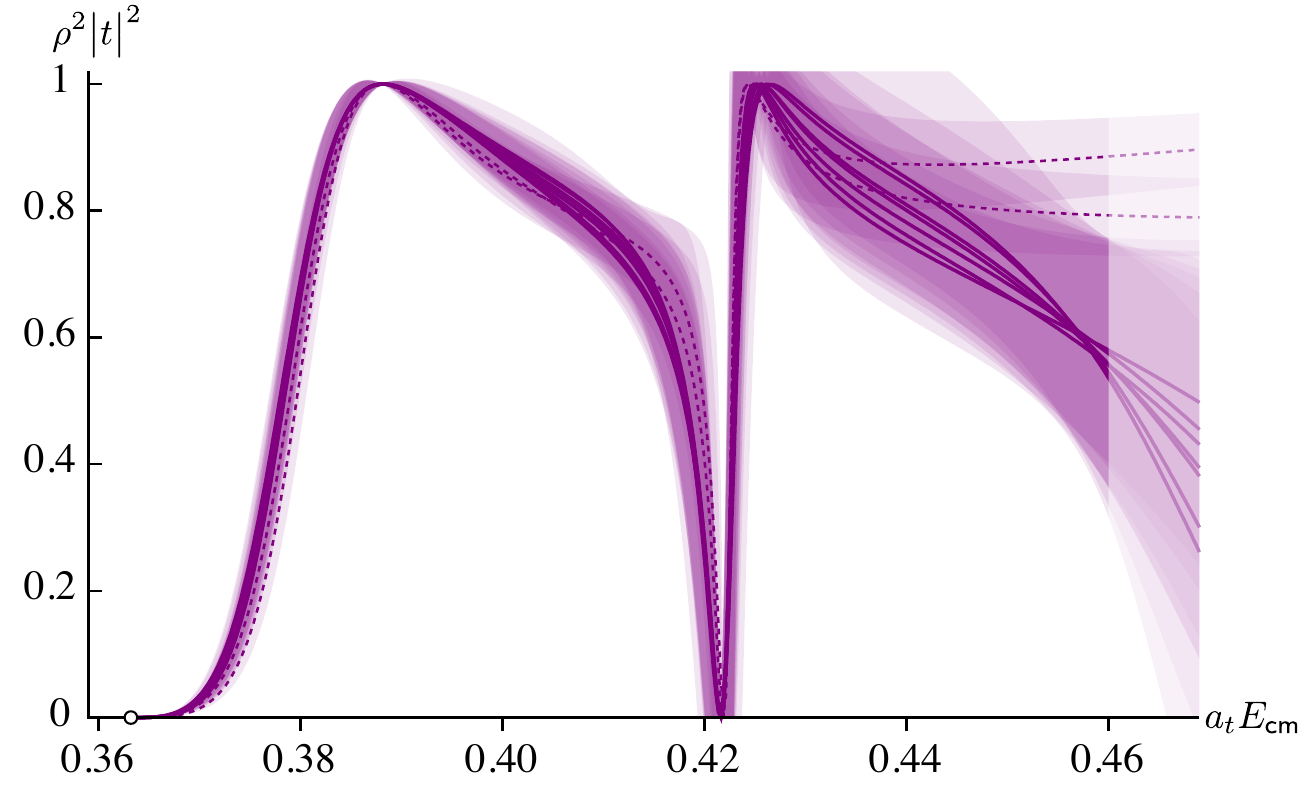}
\caption{
Variation of $1^{--}$ amplitude over parameterization choice. Dashed curves show cases with just two poles in the $K$-matrix and no further freedom.
}
\label{1m_rhotsq_variation}
\end{figure}

\newpage

We note that there is very little observed change in the amplitude except at the highest energies, and in particular the location of the zero in the amplitude appears to be very stable. We display two examples of allowing too little freedom in the amplitude -- the dashed curves show parameterizations featuring only two $K$-matrix poles and no further freedom with either the Chew-Mandelstam phase-space subtracted at the lower pole, or the naive phase-space. We see that they are compatible with the other parameterizations in the region of the resonances, but deviate at high energy. Our conclusion is that some freedom beyond two poles in the $K$-matrix is needed to have the amplitude fall-off at higher energy -- adding a constant seems to be sufficient. 

\begin{figure}
\includegraphics[width=0.98\columnwidth]{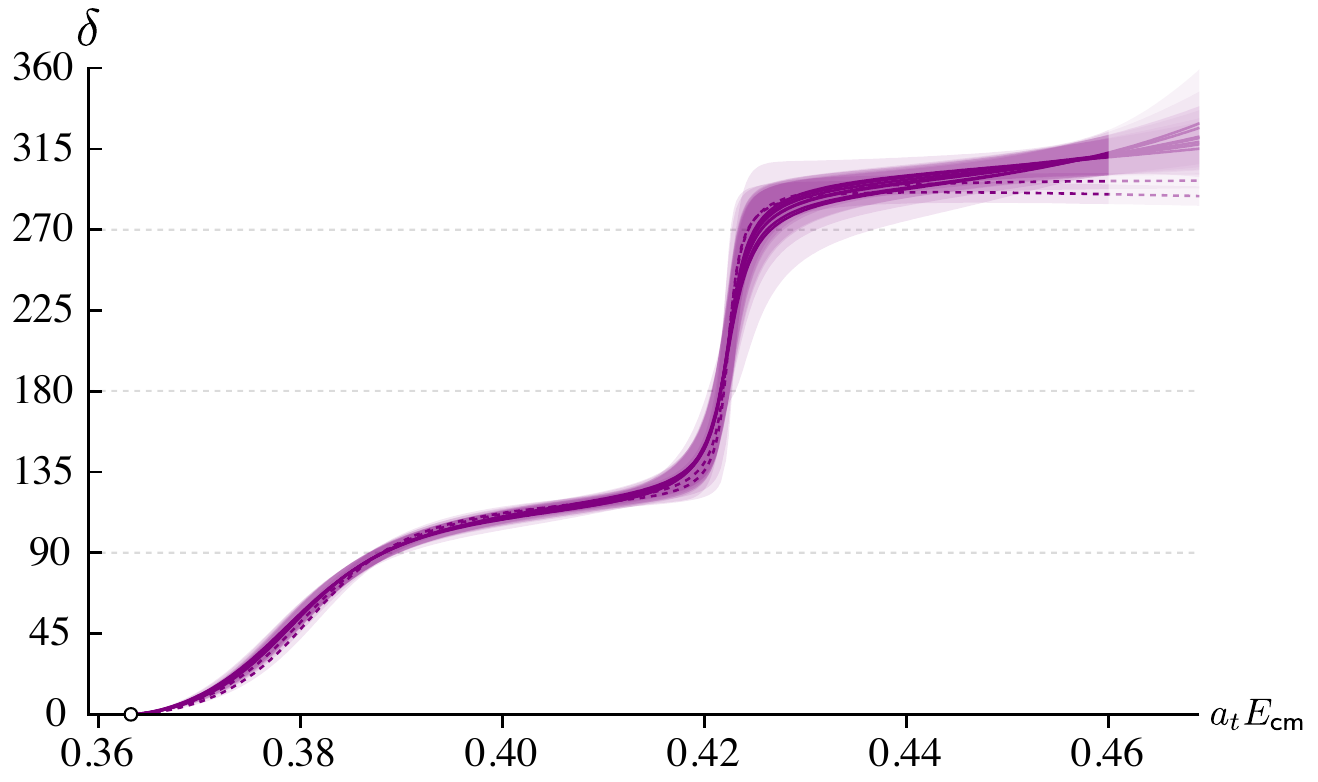}
\caption{Elastic scattering phase-shift for $1^{--}$, variation over parameterization choice. Dashed curves show cases with just two poles in the $K$-matrix and no further freedom.
}
\label{1m_delta_variation}
\end{figure}

Because this process is elastic, we can alternatively display the scattering in terms of an elastic phase shift, $\delta(E_\mathsf{cm})$ , defined by $t = \tfrac{1}{\rho} e^{i \delta} \sin \delta$. The classic signal for an isolated narrow resonance is a rapid rise of $\delta$ passing through $90^\circ$, with the steepness of the rise correlated with the smallness of the resonance width. As observed in Figure~\ref{1m_delta_variation}, the phase shift undergoes two such increases, the first with a low slope and the second being much more rapid. The phase shift passing through $180^\circ$ represents the zero in the amplitude, and the relatively slow approach to $360^\circ$ reflects the slow fall off of the amplitude at high energy.

\begin{figure}
\includegraphics[width=0.98\columnwidth]{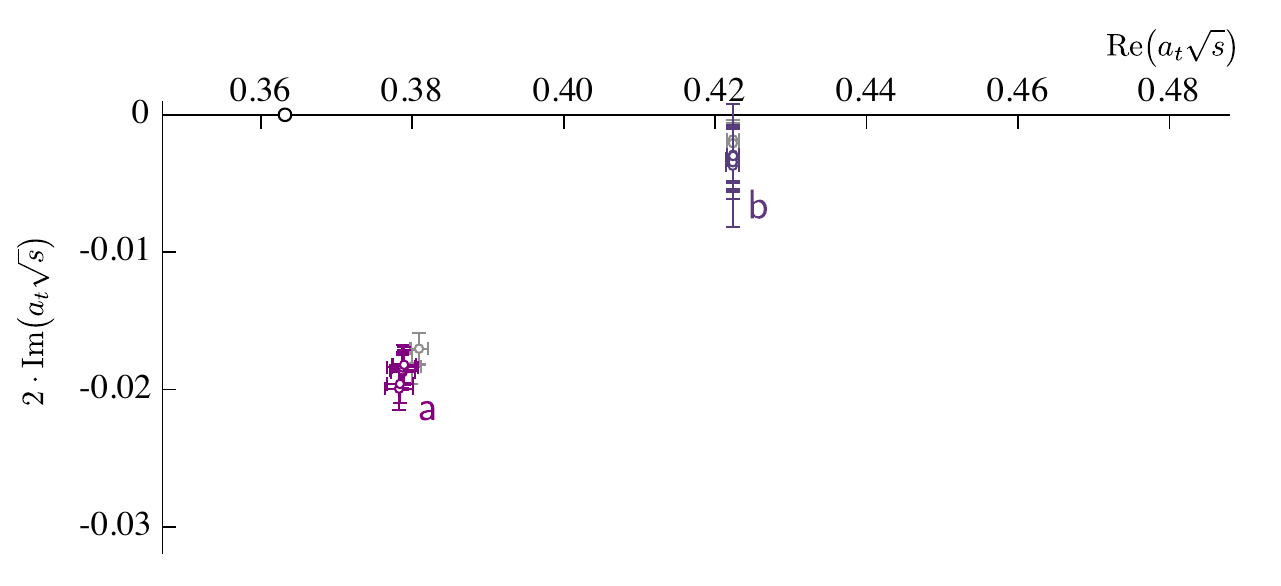}
\includegraphics[width=0.98\columnwidth]{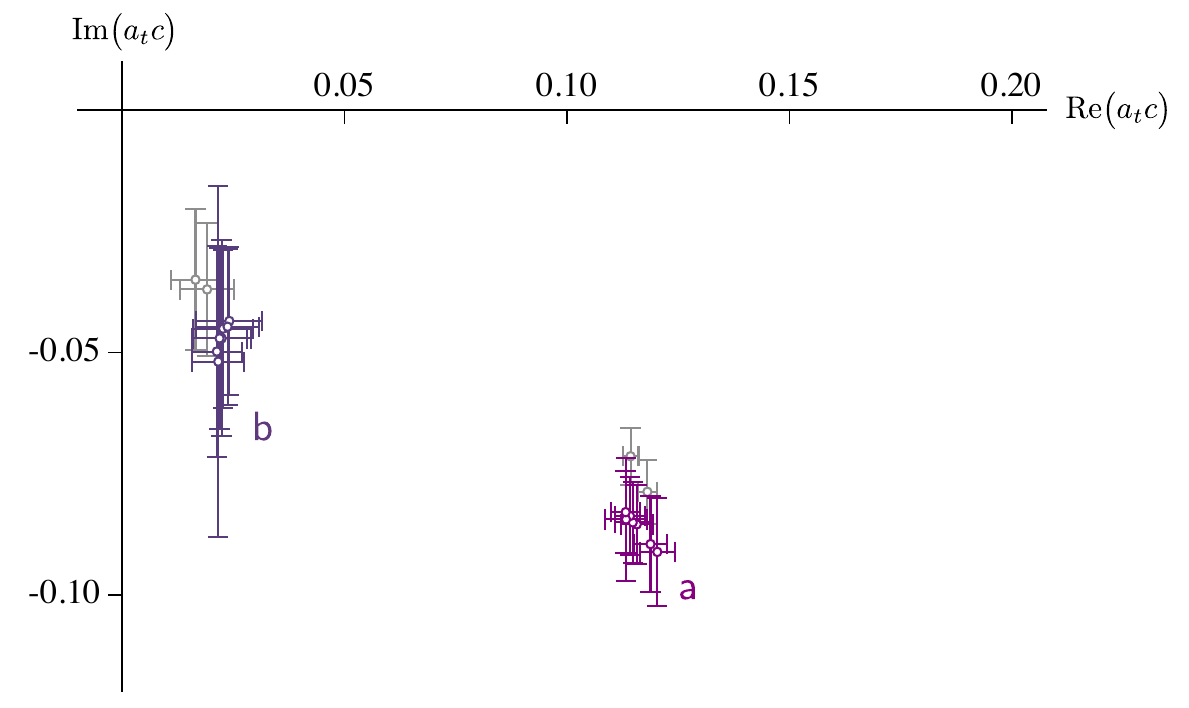}
\caption{
Top panel: $1^{--}$ $t$-matrix pole positions for parameterization variations shown in Figures~\ref{1m_rhotsq_variation},\ref{1m_delta_variation}.  Bottom panel: the coupling, $c_{\etaOctet \omegaOctet \{\threePone\}}$ obtained from factorizing the residue of the $t$-matrix poles in the lower half-plane. Gray points represent the amplitudes having limited freedom shown by the dashed curves in Figures~\ref{1m_rhotsq_variation},\ref{1m_delta_variation}.
}
\label{1m_poles_variation}
\end{figure}

\begin{figure}
\includegraphics[width=0.98\columnwidth]{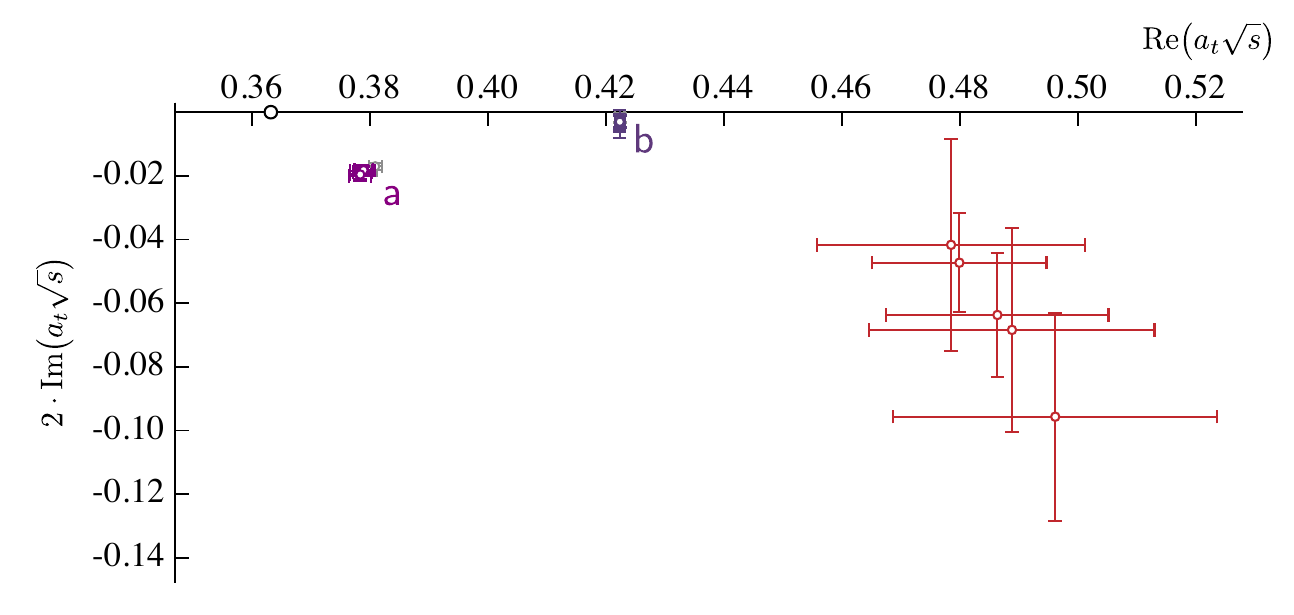}
\caption{
As in Figure~\ref{1m_poles_variation}, the $1^{--}$ $t$-matrix pole positions for parameterization variations shown in Figure~\ref{1m_rhotsq_variation}, including a poorly determined pole at higher energy (red).
}
\label{1m_all_poles_variation}
\end{figure}

Figure~\ref{1m_poles_variation} shows the location of the two $t$-matrix pole singularities and their pole couplings for all the parameterization variations considered. It is clear that there is very little scatter, and that robust conclusions can be drawn about these two resonances appearing in $1^{--}$. Additional pole singularities which lie further from physical scattering are found for some of amplitude variations, in particular several parameterizations feature an extra unphysical sheet pole, lying slightly above the energy region that we have constrained and far into the complex plane. As shown in Figure~\ref{1m_all_poles_variation}, its position is not well determined, and indeed it is not present in all parameterizations, and as such it appears to be an irrelevant artifact\,\footnote{The anticipated $1^{--}$ hybrid meson resonance pole is expected to lie at a somewhat larger energy, and is unlikely to be well constrained without higher-lying energy levels being included in the analysis.}.

Another additional pole singularity is present for several parameterizations, lying on the real energy axis below threshold on the physical sheet. Whenever it appears it is found to have a \emph{real-valued} coupling, indicating that is is a \emph{ghost}. In our illustrative amplitude with two \mbox{$K$-matrix} poles and a constant using Chew-Mandelstam phase-space, it is located at $a_t \sqrt{s} = 0.278(26)$, and it is quite typical across parameterization variations that it lies well below threshold. Such a location is into the left-hand cut region (as discussed in Appendix~\ref{lhc}), and we might interpret the presence of this ghost pole as reflecting the fact that we have made no attempt to parameterize the correct structure of the left-hand cut.

The analyses in the last few sections, considering a subset of all computed irreps each time, has led to a clear picture of the resonance content of the $1^{--}, 2^{--}, 3^{--}$ partial-waves up to an energy $a_t E_\mathsf{cm} \sim 0.46$. We have not so far made any use of the $[110]\, A_1$ irrep which depends upon the scattering amplitudes of \emph{all} the above partial waves. We will now move to consider a global fit of all the $\etaOctet \omegaOctet$ energy levels that will confirm the results seen so far, and lead to reduced statistical uncertainties on some resonance parameters.

\subsection{`Global fit' to all $\bm{ \etaOctet \omegaOctet }$ energy levels} \label{global}

\begin{figure*}
\includegraphics[width=0.95\textwidth]{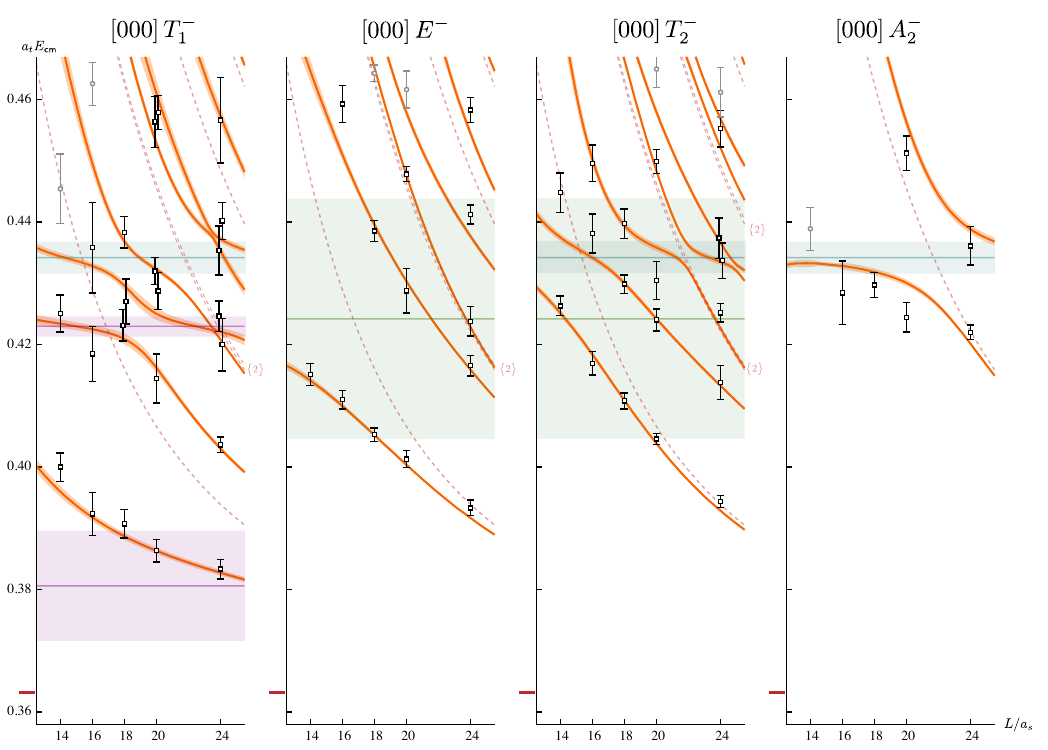}
\includegraphics[width=0.95\textwidth]{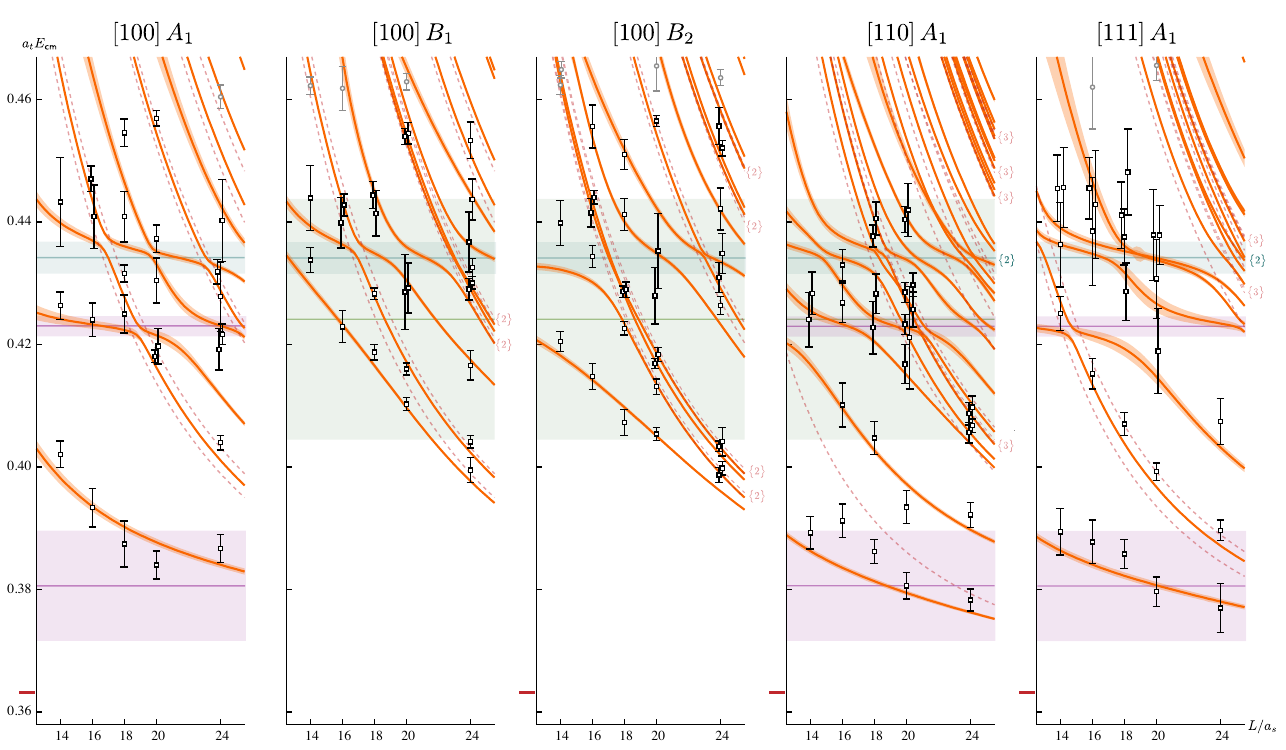}
\caption{
Spectra from Figure~\ref{e8o8_spectra} used to constrain $\etaOctet \omegaOctet$ elastic scattering amplitudes. Orange curves show the finite-volume spectrum corresponding to the best-fit amplitudes. For guidance, the purple ($1^{--}$), green ($2^{--}$) and cyan ($3^{--}$) bands show the resonance masses and widths, allowing avoided level crossings to be observed with the non-interacting $\etaOctet \omegaOctet$ levels (red dashed).
}
\label{e8o8_orange}
\end{figure*}

In this section we attempt a description of the full set of 192 energy levels shown in black in Figure~\ref{e8o8_spectra} using parameterizations of $1^{--}$, $2^{--}$ and $3^{--}$ scattering amplitudes. To illustrate the approach we select amplitudes where $3^{--}$ is described by a single $K$-matrix pole with Chew-Mandelstam phase-space subtracted at the pole, the $2^{--}$ $\threePtwo, \threeFtwo$ coupled system is described by a \mbox{$K$-matrix} pole plus constants in the $PP$ and $PF$ positions with Chew-Mandelstam phase-space subtracted at the pole, and $1^{--}$ is described by two $K$-matrix poles plus a constant with Chew-Mandelstam phase-space subtracted at the lower pole. A description of the complete set of energy levels is found with $\chi^2/N_\mathrm{dof} = 258.3/(192-12) = 1.43$, and the finite-volume spectrum following from the best-fit amplitude is shown by the orange curves in Figure~\ref{e8o8_orange}. As we might guess from the relatively small $\chi^2$, the orange curves are in good agreement with the black points. While the spectrum of `predicted' levels can be dense, it always agrees with our expectations of the number of levels arising from non-interacting levels plus resonances, once multiple subductions are accounted for -- as an example, consider small volumes in the $[110]\,A_1$ irrep, where the six orange curves agrees with an expectation based upon one low-lying $\etaOctet \omegaOctet$ non-interacting level and five resonance contributions ($1^{--}_\mathsf{a}$, $1^{--}_\mathsf{b}$, $2^{--}$, and $3^{--}$ subduced twice). We note that in cases where resonances in different $J^{PC}$ overlap, the ``avoided level crossing'' structure can be somewhat non-trivial.

\begin{figure}
\includegraphics[width=0.98\columnwidth]{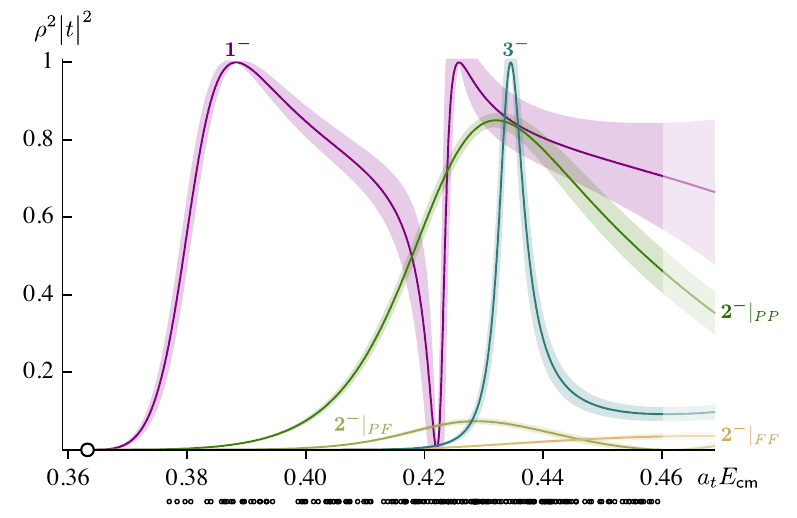}
\caption{
$\etaOctet \omegaOctet$ elastic scattering amplitudes obtained by describing 192 finite-volume energy levels.
}
\label{1m2m3m}
\end{figure}

The amplitudes are shown in Figure~\ref{1m2m3m} where they are seen to be compatible with our previous determinations using subsets of the spectrum. The $t$-matrices are found to feature poles on the unphysical sheet at the following locations (and with pole couplings):
\begin{align}
1^{--}(\mathsf{a}): \quad\quad a_t \sqrt{s_0} &= 0.3806(12) \pm \tfrac{i}{2} 0.0181(12) \nonumber \\[1.2ex]
a_t c_{\etaOctet \omegaOctet\{ \threePone \}} &= 0.141(5) \, e^{\pm i \pi \, 0.18(1)} \nonumber \\[2.2ex]
1^{--}(\mathsf{b}): \quad\quad a_t \sqrt{s_0} &= 0.4230(7) \pm \tfrac{i}{2} 0.0032(16) \nonumber \\[1.2ex]
a_t c_{\etaOctet \omegaOctet\{ \threePone \}} &= 0.052(13) \, e^{\pm i \pi \, 0.34(2)} \nonumber \\[2.2ex]
2^{--}: \quad\quad a_t \sqrt{s_0} &= 0.4242(10) \pm \tfrac{i}{2} 0.0391(24) \nonumber \\[1.2ex]
a_t c_{\etaOctet \omegaOctet\{ \threePtwo \}} &= 0.175(5) \, e^{\pm i \pi \, 0.11(1)} \nonumber \\
a_t c_{\etaOctet \omegaOctet\{ \threeFtwo \}} &= 0.059(4) \, e^{\mp i \pi \, 0.88(2)} \nonumber \\[2.2ex]
3^{--}: \quad\quad a_t \sqrt{s_0} &= 0.4342(6) \pm \tfrac{i}{2} 0.0052(5) \nonumber \\[1.2ex]
a_t c_{\etaOctet \omegaOctet\{ \threeFthree \}} &= 0.064(3) \, e^{\pm i \pi \, 0.040(4)} \, . \label{global_poles}
\end{align}
These are compatible with those found previously. 

Thus far we have not accounted for the effect of the (relatively small) uncertainties on the scattering hadron \mbox{($\etaOctet$, $\omegaOctet$)} masses on the scattering amplitudes, but when considered by varying them by $\pm 1 \sigma$, there is negligible change. The somewhat larger conservative estimate of the uncertainty on the anisotropy, $\xi = 3.486(43)$, has only been accounted for partly, in the boost of moving frame energies back to $a_t E_\mathsf{cm}$. It can also be considered in the computation of $\bm{\mathcal{M}}$ in Eqn.~\ref{luscher}, where varying by $\pm 1 \sigma$ leads to small adjustments in the pole positions given above. The largest effects are observed in the real part of the pole positions which can move by amounts comparable with the statistical error, and in the imaginary part of the $2^{--}$ pole position. In the next section, when we present our best estimates for the resonance pole properties we will include this source of uncertainty in our error estimates.

\subsection{Decoupled $\bm{\etaSinglet \omegaSinglet}$ and $\bm{ \omegaSinglet \fZeroSinglet }$ scattering}

As discussed in Section~\ref{fvspectrum}, the finite-volume spectrum appears to separate into a spectrum due to the resonating $\etaOctet \omegaOctet$ system that we have just considered, and two spectra due to non-resonant systems $\etaSinglet \omegaSinglet$ and $\omegaSinglet \fZeroSinglet$. We will consider these latter sets of energy levels in isolation. 

For $J^{PC} = 1^{--}$ $\etaSinglet \omegaSinglet \{ \threePone \}$ elastic scattering, we use five energy levels as constraint, three levels in $[000]\, T_1^-$ (as shown in the bottom panel of Figure~\ref{T1m_histo}) and two in $[001]\, A_1$, all of which are compatible with lying on the lowest non-interacting $\etaSinglet \omegaSinglet$ curve.

An effective range expansion $k^3  \cot \delta = \tfrac{1}{a} + \tfrac{1}{2} r k^2 + \ldots$ can be used to describe the elastic amplitude. Using only a scattering length $a \neq 0, r = 0$ and no higher terms in the polynomial, the five energy levels can be described with $a = 4.4(40) \times 10^2\, a_t^3$ with a ${\chi^2/N_\mathrm{dof} = 6.2/(5-1) = 1.5 }$. Such a $P$--wave scattering length approximation gives a \mbox{$t$-matrix} pole distribution that is not easily interpreted (three poles evenly spaced around a circle of radius ${k = a^{-1/3}}$), but allowing also a non-zero effective range, which can generate a realistic pole distribution in this case leads to a fit with 100\% correlation between the parameters ($a,r$). 

Alternatively, using a constant $K$-matrix and the Chew-Mandelstam phase-space subtracted at threshold, the energy levels can be described with a ${\chi^2/N_\mathrm{dof} = 6.1/(5-1) = 1.5}$ where the resulting \mbox{$t$-matrix} has a \emph{ghost} pole on the physical sheet at ${  a_t \sqrt{s_0} = 0.315(78)  }$ -- given that the left-hand cut for this process begins at $0.378$ (see Appendix~\ref{lhc}), we can associate this ghost with our lack of control over the crossed-channel physics.

For $J^{PC} = 2^{--}$, assumed to be only in the $\etaSinglet \omegaSinglet \{ \threePtwo \}$ partial wave, we use 8 levels from $[000]\, E^-$, $[000]\, T_2^-$, and $[001]\, B_1$. The $F$--wave $3^{--}$ amplitude is assumed to be negligible. A scattering length description finds ${ a = 4.1(28) \times 10^2\, a_t^3 }$ with a ${ \chi^2/N_\mathrm{dof} = 12.7/(8-1) = 1.8 }$, while a constant $K$-matrix with Chew-Mandelstam subtracted at threshold has a similar $\chi^2$ and a ghost pole at $a_t \sqrt{s_0} = 0.305(73)$.

It is clear that the partial waves $\etaSinglet \omegaSinglet \{ \threePone, \threePtwo \}$ are non-resonant -- the elastic phase-shift reaches only $\sim 20^\circ$ at the largest energies we consider ($a_t E_\mathsf{cm} \sim 0.46$) -- and they appear to have very similar behavior suggesting weak spin-orbit forces in this channel. The large uncertainties on the scattering parameters are only slightly increased if we include the effect of the uncertainty on the $\etaSinglet$ mass.

For $J^{PC} = 1^{--}$ $\omegaSinglet \fZeroSinglet$, we in principle may have a more significant amplitude, owing to the scattering being possible in an $S$--wave ($\threeSone$). In fact a coupled system of partial waves $\{ \threeSone, \threeDone \}$ is required in order to describe the multiplicity of non-interacting energies shown in the middle panel of Figure~\ref{T1m_histo}. We make use of 16 energy levels taken from $[000]\, T_1^-$ and $[001]\, A_1$ irreps, noting that they all have rather large statistical uncertainties and are all compatible with non-interacting $\omegaSinglet \fZeroSinglet$ energies.

An example parameterization uses a diagonal constant $K$-matrix and a Chew-Mandelstam phase-space subtracted at threshold. The resulting constants are statistically compatible with zero in a description of the energy levels with $\chi^2/N_\mathrm{dof} = 21.5/(16-2) = 1.54$. Even larger errors are obtained once the uncertainty on the $\fZeroSinglet$ mass is accounted for, and considering this and variation over parameterizations, the $S$--wave phase-shift remains compatible with zero but with an uncertainty that spreads over at least $\pm 50^\circ$ at $a_t E_\mathsf{cm} = 0.46$.

\subsection{Estimating coupled-channel effects}

The previous sections indicate that the finite-volume spectra can be well described assuming that the $\etaOctet \omegaOctet$, $\etaSinglet \omegaSinglet$, and $\omegaSinglet \fZeroSinglet$ channels are decoupled, with resonances only appearing in $\etaOctet \omegaOctet$. Nevertheless we can attempt a limited study of possible channel coupling.

Using a set of 52 energy levels in irreps sensitive to $J^{PC}=1^{--}$ and $3^{--}$, which includes 4 levels having large $\etaSinglet \omegaSinglet$ overlap, we can try to constrain coupled $(\etaOctet \omegaOctet, \etaSinglet \omegaSinglet)$ $J^{PC} = 1^{--}$ amplitudes parameterized with
\begin{align}
\bm{K}(s) &= 
\frac{1}{m_\mathsf{a}^2 - s} 
\begin{bmatrix} 
(g^{\mathsf{a}}_{\etaOctet \omegaOctet})^2 & g^{\mathsf{a}}_{\etaOctet \omegaOctet}\, g^{\mathsf{a}}_{\etaSinglet \omegaSinglet}\\ 
g^{\mathsf{a}}_{\etaOctet \omegaOctet}\, g^{\mathsf{a}}_{\etaSinglet \omegaSinglet} & (g^{\mathsf{a}}_{\etaSinglet \omegaSinglet})^2 
\end{bmatrix} 
\nonumber \\
&\;\;+ \frac{1}{m_\mathsf{b}^2 - s} 
\begin{bmatrix} 
(g^{\mathsf{b}}_{\etaOctet \omegaOctet})^2 & g^{\mathsf{b}}_{\etaOctet \omegaOctet}\, g^{\mathsf{b}}_{\etaSinglet \omegaSinglet}\\ 
g^{\mathsf{b}}_{\etaOctet \omegaOctet}\, g^{\mathsf{b}}_{\etaSinglet \omegaSinglet} & (g^{\mathsf{b}}_{\etaSinglet \omegaSinglet})^2 
\end{bmatrix} 
\nonumber \\
&\qquad\qquad\;  + \begin{bmatrix} \gamma_{\etaOctet \omegaOctet, \etaOctet \omegaOctet } & \gamma_{\etaOctet \omegaOctet, \etaSinglet \omegaSinglet } \\ \gamma_{\etaOctet \omegaOctet, \etaSinglet \omegaSinglet } & \gamma_{\etaSinglet \omegaSinglet, \etaSinglet \omegaSinglet } \end{bmatrix},
\label{coupled}
\end{align}
and Chew-Mandelstam phase-space subtracted at $s=m_\mathsf{a}^2$. The $J^{PC} = 3^{--}$ amplitude is a single $K$-matrix pole, elastic in $\etaOctet \omegaOctet$ with parameters fixed from previous fits.

\begin{figure}
\includegraphics[width=0.98\columnwidth]{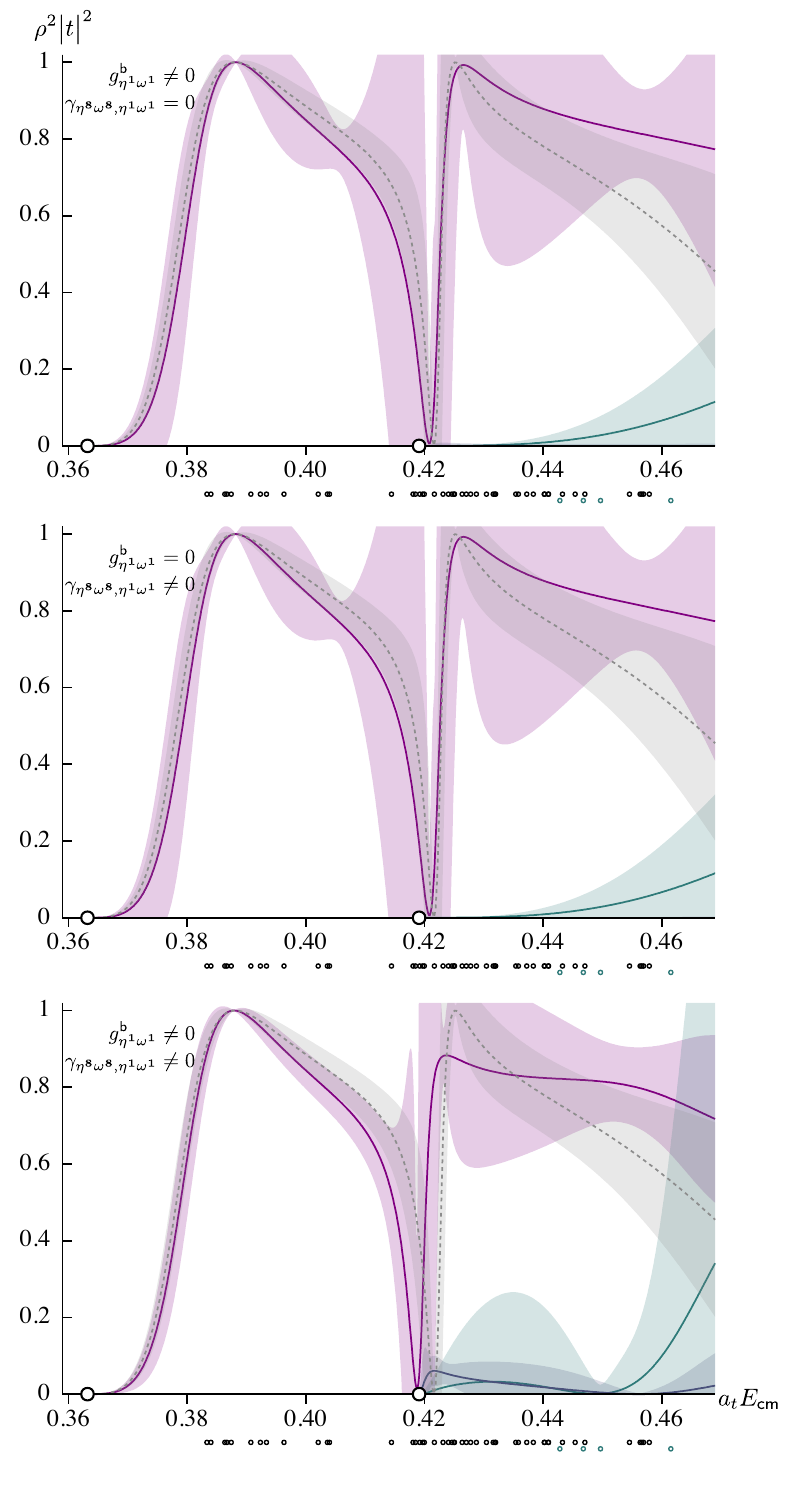}
\caption{
Coupled $(\etaOctet \omegaOctet, \etaSinglet \omegaSinglet)$ $1^{--}$ amplitudes as in Eqns.~\ref{coupled}, \ref{coupled2}. $t_{\etaOctet \omegaOctet, \etaOctet \omegaOctet}$ (purple), $t_{\etaOctet \omegaOctet, \etaSinglet \omegaSinglet}$ (blue), $t_{\etaSinglet \omegaSinglet, \etaSinglet \omegaSinglet}$ (cyan). For comparison the assumed elastic amplitude from Figure~\ref{1m_ref_rhotsq} is shown in grey. The energy levels used to constrain the amplitude are shown below the abscissa, with those having larger overlap onto $\etaSinglet \omegaSinglet$ operators shown in cyan.
}
\label{1m_coupled}
\end{figure}

In practice the $\mathsf{a}$--pole being far below $\etaSinglet \omegaSinglet$ threshold means that the parameter $g^{\mathsf{a}}_{\etaSinglet \omegaSinglet}$ is basically unconstrained so we set it equal to zero. $\gamma_{\etaSinglet \omegaSinglet, \etaSinglet \omegaSinglet }$ is always left free, and we consider three fits:
\begin{align}
g^{\mathsf{b}}_{\etaSinglet \omegaSinglet} \neq 0&, \gamma_{\etaOctet \omegaOctet, \etaSinglet \omegaSinglet } = 0 , \nonumber\\
g^{\mathsf{b}}_{\etaSinglet \omegaSinglet}  = 0&, \gamma_{\etaOctet \omegaOctet, \etaSinglet \omegaSinglet } \neq 0, \nonumber \\
g^{\mathsf{b}}_{\etaSinglet \omegaSinglet} \neq 0&, \gamma_{\etaOctet \omegaOctet, \etaSinglet \omegaSinglet } \neq 0,  
\label{coupled2}
\end{align}
all three of which provide descriptions of the energy levels with $\chi^2 /N_\mathrm{dof} = 1.84$. The resulting amplitudes are shown in Figure~\ref{1m_coupled} where we see that only the $t_{\etaOctet\omegaOctet, \etaOctet \omegaOctet}$ element is significantly non-zero in each case, and that it broadly agrees with the previous elastic analysis. The third fit is somewhat optimistic given the small number of $\etaSinglet \omegaSinglet$ dominated energy levels providing constraint, and indeed it is this amplitude that shows the largest difference with respect to the elastic case, in particular with it having the largest shift in the dip position.

These $t$-matrices have pole singularities on sheets ${\mathsf{III}( \mathrm{Im}\, k_{\etaOctet \omegaOctet} \!<\! 0, \mathrm{Im}\, k_{\etaSinglet \omegaSinglet} \!<\! 0)}$ and ${ \mathsf{II}( \mathrm{Im}\, k_{\etaOctet \omegaOctet} \!<\! 0, \mathrm{Im}\, k_{\etaSinglet \omegaSinglet} \!>\! 0) }$ that are qualitatively unchanged compared to the elastic-only assumption, albeit with larger statistical uncertainties. While it is not well determined, it is possible that the $\mathsf{b}$--pole may have a modest $\etaSinglet \omegaSinglet$ coupling, $ a_t |c_{\etaSinglet \omegaSinglet} | \lesssim 0.04$, that does not change the total width of the heavier resonance because there is so little phase-space for the decay.

\begin{figure}
\includegraphics[width=0.98\columnwidth]{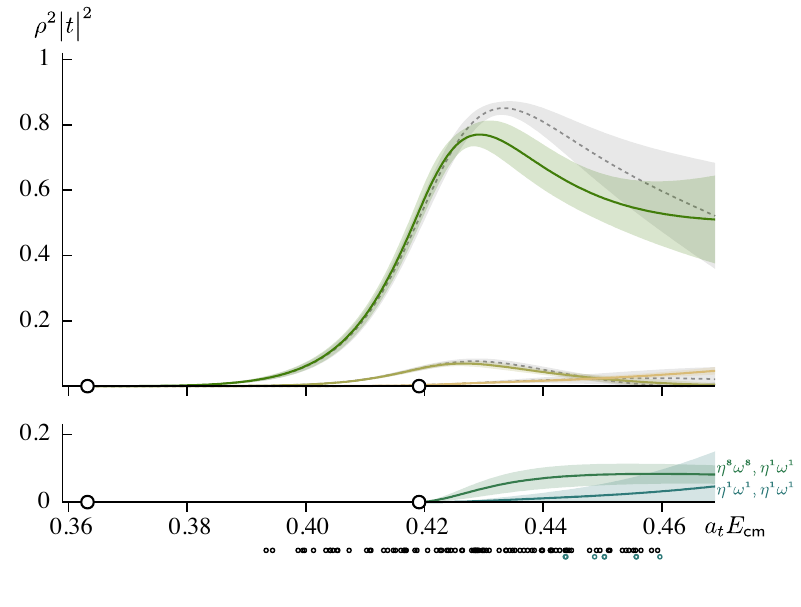}
\caption{
$J^{PC} = 2^{--}$ $t$-matrix, as in Figure~\ref{2m3m_ref_rhotsq} (grey), with the addition of the $\etaSinglet \omegaSinglet \{ \threePtwo \}$ channel.
}
\label{2m_coupled}
\end{figure}

Scattering of $\etaSinglet \omegaSinglet$ in $P$--wave can also impact ${J^{PC} = 2^{--}}$, with the $F$--wave being unlikely to contribute significantly so close to threshold. We augment the pole plus constant $K$-matrix of Eqn.~\ref{J2_pole_const} with an extra $\etaSinglet \omegaSinglet \{ \threePtwo \}$ channel, letting the pole coupling to this new channel and the extra diagonal constant float freely in a description of 96 energy levels, 7 of which have large $\etaSinglet \omegaSinglet$ overlap. The quality of fit, $\chi^2 / N_\mathrm{dof} = \tfrac{125.7}{96 - 9} = 1.45$, is reasonable, and the resulting $t$-matrix elements are shown in Figure~\ref{2m_coupled}. Clearly the additional channel has only a weak effect. The $t$-matrix has a pole (on sheets $\mathsf{III}$ and $\mathsf{II}$) that is in a location compatible with previous estimates. Similarly the $\etaOctet \omegaOctet$ couplings in $P$-- and $F$--waves are not significantly changed. There is a pole coupling to $\etaSinglet \omegaSinglet$ that while small, $a_t | c_{\etaSinglet \omegaSinglet}| \sim 0.07(2)$, is of a comparable size to the $\etaOctet \omegaOctet$ $F$--wave coupling.

It proves to be the case that the large statistical uncertainties on the $\omegaSinglet \fZeroSinglet$ energy levels prevent any meaningful attempt at coupled-channel $(\etaOctet \omegaOctet, \omegaSinglet\! \fZeroSinglet)$ analysis. As such while we cannot rule out non-zero couplings to $\omegaSinglet \fZeroSinglet$ for our resonances, such an outcome seems unlikely given our ability to describe the a huge number of finite-volume energy levels using a set of decoupled amplitudes.

%% file: 4-resonances.tex

In the previous section we presented descriptions of $\etaOctet \omegaOctet$ scattering with $J^{PC} = 1^{--}, 2^{--}$ and $3^{--}$ finding several resonances appearing as poles in the $t$-matrix. We choose to set the lattice scale using the decuplet $\Omega$--baryon mass computed on these lattices, finding ${a_t^{-1} = 4655\, \mathrm{MeV} }$. Our best estimates of the resonance pole properties, with uncertainties which reflect the variations seen in the previous section are,
\begin{align}
1^{--}, \, \omegaSinglet_\mathsf{a}:\quad\quad \sqrt{s_0} &= 1772(7) \pm \tfrac{i}{2} 84(9)\, \mathrm{MeV} \nonumber \\[1.2ex]
\big|c_{\etaOctet \omegaOctet} \big| &= 656(37)\, \mathrm{MeV} \nonumber \\[2.2ex]
1^{--}, \, \omegaSinglet_\mathsf{b}:\quad\quad \sqrt{s_0} &= 1969^{+5}_{-14} \pm \tfrac{i}{2} 15(10)\, \mathrm{MeV} \nonumber \\[1.2ex]
\big|c_{\etaOctet \omegaOctet} \big| &= 242(93)\, \mathrm{MeV} \nonumber \\[2.2ex]
2^{--}, \, \omegaSinglet_2:\quad\quad \sqrt{s_0} &= 1975(10) \pm \tfrac{i}{2} 182(19)\, \mathrm{MeV} \nonumber \\[1.2ex]
\big|c_{\etaOctet \omegaOctet}(\threePtwo) \big| &= 815(30)\, \mathrm{MeV} \nonumber \\
\big|c_{\etaOctet \omegaOctet}(\threeFtwo) \big| &= 275(37)\, \mathrm{MeV} \nonumber \\
\frac{c_{\etaOctet \omegaOctet}(\threeFtwo)}{c_{\etaOctet \omegaOctet}(\threePtwo)} &= -0.35(5) \nonumber \\[2.2ex]
3^{--}, \, \omegaSinglet_3:\quad\quad \sqrt{s_0} &= 2021(8) \pm \tfrac{i}{2} 20^{+2}_{-9}\, \mathrm{MeV} \nonumber \\[1.2ex]
\big|c_{\etaOctet \omegaOctet} \big| &= 298^{+23}_{-51}\, \mathrm{MeV} \, .
\label{best_poles}
\end{align}

The $1^{--}$ amplitude features two resonances: a lighter state with a larger width, and a heavier narrow state which has a tight dip and a zero of the amplitude associated with it. A common parameterization approach in elastic scattering is the \emph{effective range expansion}, in which $k^{2\ell +1} \cot \delta_\ell$ is expanded as a polynomial in $k^2$, truncated at some finite order, with the polynomial coefficients being free parameters, the first two of which are known as the \emph{scattering length} and the \emph{effective range}. The justification for the use of such a series is that it is expected to converge for energies inside a circle centered at threshold which just touches the left-hand cut, the nearest unconsidered singularity. Even with only two terms such a parameterization is capable of describing a single resonance. It is simple to see that our extracted amplitude as shown in e.g. Figure~\ref{1m_ref_rhotsq} \emph{cannot} be described by an effective range expansion, owing to the presence of a \emph{zero} in the amplitude, which would require $k^3 \cot \delta$ to diverge at some positive value of $k^2$, which cannot happen for any finite order polynomial. This appears to present something of a paradox if one takes the view that the left-hand cut represents  the ``potential'' due to particle exchanges in the crossed channels that act to bind the scattering hadrons into a resonance -- such a potential cannot generate the observed zero. Indeed our finding of a lighter broad resonance and a heavier narrow resonance looks quite unnatural in a potential picture where for realistic potential shapes with a centrifugal barrier, one expects the lighter state to have to tunnel through a larger distance than the heavier state in order to decay.

The way out of this is to recognize that relativistic scattering systems have more freedom than those driven by non-relativistic potentials. This can be illustrated by expressing $t(s)$ as a ratio of functions, $t(s) = N(s)/D(s)$, where the numerator houses the left-hand cut, and the denominator has the unitarity cut. In the case of potential scattering, $N(s)$ serves as the potential, and then $D(s)$ is uniquely determined from $N(s)$ by evaluating a dispersive integral. Relativistic scattering differs from this in that there is the freedom to add an arbitrary number of poles to $D(s)$, known as ``CDD poles''~\cite{Castillejo:1955ed}. In elastic scattering, these poles will generate zeros in $t(s)$ at real values of $s$, and nearby $t$-matrix poles at complex values of $s$. Although not a unique interpretation, they are often associated with the idea that the underlying theory (QCD in our case) features particles that would be stable were it not for the presence of pairs of lighter hadrons into which they can decay. This of course matches quite closely with the quark-model picture of $q\bar{q}$ mesons that become stable as the quark mass increases.

Considering the set of resonances as a whole supports an interpretation, bolstered by the overlaps discussed in Section~\ref{fvspectrum}, of the lighter $1^{--}$ state as being dominantly $q\bar{q} [ 2 \threeSone ]$, and the remaining three states as being $q\bar{q} [1 \threeDJ]$ with only small spin-orbit splittings. Which $\etaOctet \omegaOctet$ partial waves are accessible appears to play a role in setting the state decays widths: the $3^{--}$ resonance decays only in $F$--wave and is narrow, while $2^{--}$ also has a $P$--wave decay, and is significantly broader. There is not any obvious explanation for why the lighter $1^{--}$ has a much larger width than the rather narrow heavier $1^{--}$ state. 

The leading method for predicting meson decays prior to this calculation was the $\threePzero$--model. When its assumed form for the $q\bar{q}$ creation vertex is used with harmonic oscillator wavefunctions for the bound $q\bar{q}$ mesons, simple expressions follow for ratios of decay amplitudes of $q\bar{q} [1\threeDJ]$ mesons to pseudoscalar-vector pairs~\cite{Barnes:1996ff} (where we are neglecting the effect of the small mass differences between the decaying mesons). For the $F$--wave decays of the $3^{--}$ and $2^{--}$ states, we have 
$\frac{g_F(3^{--})}{g_F(2^{--})} = \sqrt{\frac{10}{7}} \approx 1.20 $,
and if for comparison we use the ratio of pole couplings presented in Eqn.~\ref{best_poles}, we obtain $\sim 1.1(2)$ in reasonable agreement with the model. For the $P$--wave decays of the $1^{--}[1 \threeDone]$ and $2^{--}$, the model predicts 
$\frac{g_P(1^{--})}{g_P(2^{--})} = \frac{\sqrt{5}}{3} \approx 0.75$,
while, assuming the $\mathsf{b}$--pole is the $q\bar{q}[\threeDone]$ state, Eqn.~\ref{best_poles} suggests $\sim 0.3(1)$ which appears to be in rather poor agreement. The $\threePzero$--model provides an expression for the $2^{--}$ $F/P$ amplitude ratio, one that depends only on the ratio of the decay momentum, $k$, to the harmonic oscillator parameter, $\beta$:
\begin{equation*}
\frac{g_F(2^{--})}{g_P(2^{--})} = - \tfrac{2}{15}\sqrt{\tfrac{2}{3}} 
\left(\tfrac{k}{\beta}\right)^2  \left(1 - \tfrac{2}{15} \left(\tfrac{k}{\beta}\right)^2 \right)^{-1} \, .
\end{equation*}
To describe physical light and strange-quark mesons, it is usual to choose $\beta = 400 \, \mathrm{MeV}$, but since the quarks in our study are somewhat heavier than physical quarks, we might expect the wavefunctions of the mesons to be smaller, and $\beta$ to be larger. When the $\threePzero$--model is applied to charmonium, with still heavier quarks, ${\beta = 500\, \mathrm{MeV}}$ is typical~\cite{Barnes:2005pb}. For our $2^{--}$ resonance, $k \approx 504\,\mathrm{MeV}$, such that taking $400 < \beta < 500 $ MeV, the equation above predicts an $F/P$ ratio between $-0.13$ and $-0.22$, which while the sign agrees, is somewhat smaller in magnitude than our lattice QCD result of $-0.35(5)$.

\subsection{Estimating $\bm{J^{--}}$ meson properties at the physical $\bm{u,d}$ quark mass}

While we have only computed for a single unphysically heavy quark mass, we can attempt to extrapolate consequences at the physical quark mass. This will necessarily be a crude estimate in which we will need to impose additional phenomenological constraints not following directly from our calculation. We begin by expressing the \SUF~representations in terms of more familiar meson states -- the \SUF~singlet can be decomposed~\cite{deSwart:1963pdg, Woss:2020ayi} into states labeled by isospin and strangeness as,
\begin{align*}
\bm{1} = \tfrac{1}{2\sqrt{2}} 
\Big( 
&K^+ \kbarSuper{*-} + K^- \kbarSuper{*+} - K^0 \kbarSuper{*0} - \kbarSuper{0}K^{*0}  \nonumber \\
&+ \pi^+ \rho^- + \pi^- \rho^+ - \pi^0 \rho^0 - \eta_8 \omega_8
\Big) \, ,
\end{align*}
where we use the PDG naming scheme, except for $\eta_8, \omega_8$ by which we mean the neutral flavorless element of the pseudoscalar or vector octet. It is generally accepted that with physical mass quarks, the $\eta$ meson is very close to being $\eta_8$ with only a small admixture of $\eta_1$, while the $\omega$ and $\phi$ are nearly ideally flavor mixed,
\begin{align*}
\omega &= \sqrt{\tfrac{2}{3}} \omega_1 + \tfrac{1}{\sqrt{3}}  \omega_8 \\
\phi   &= \tfrac{1}{\sqrt{3}}  \omega_1 - \sqrt{\tfrac{2}{3}} \omega_8 \, .
\end{align*}
Similar mixing appears in lattice QCD calculations at larger than physical light quark masses, as can be seen in Figure~\ref{singlehadron}, and indeed the ideal flavor mixing appears to be present for excited $\omega^\star_J, \phi^\star_J$ states also. This mixing poses a challenge for us if we wish to estimate decays of these states, as we have only computed the \SUF~singlet component and not the octet. The octet, which for $C=-$ decays to pseudoscalar-vector in the $\bm{8}_1$ representation (see Ref.~\cite{Woss:2020ayi}), has decomposition,
\begin{align*}
\bm{8} = &\tfrac{1}{\sqrt{20}} 
\Big( 
K^+ \kbarSuper{*-} + K^- \kbarSuper{*+} - K^0 \kbarSuper{*0} - \kbarSuper{0}K^{*0} \Big) \nonumber \\
&-\tfrac{1}{\sqrt{5}} \big( \pi^+ \rho^- + \pi^- \rho^+ - \pi^0 \rho^0 - \eta_8 \omega_8
\Big),
\end{align*}
and since we would like to have decays to $\eta \omega$ and $\eta \phi$, we also require the process $\bm{8} \to \bm{8} \otimes \bm{1}$, which has a trivial decomposition $\bm{8} = \eta_8 \omega_1$. We will assume that we can neglect the small admixture of $\eta_1$ in the $\eta$ as a first approximation.

While we have only computed the singlet decays, we can relate the octet decays to these if we implement the OZI rule in a way consistent with the assumed ideal flavor mixing. We define a notation where $\gSinglet$ represents the $\bm{1} \to  \bm{8} \otimes \bm{8}$ decay coupling, $\gOctet$ represents the $\bm{8} \to  \bm{8} \otimes \bm{8}$ decay coupling and $\hOctet$ represents the $\bm{8} \to  \bm{8} \otimes \bm{1}$ decay coupling. A first condition follows from imposing that the decay $\phi^\star \to \pi \rho$ must be zero for exact OZI -- the amplitude for this process is proportional to 
\begin{equation*}
\tfrac{1}{\sqrt{3}} \tfrac{1}{2\sqrt{2}} \, \gSinglet + \Big( \!-\! \sqrt{\tfrac{2}{3}} \Big) \Big( \!- \!\tfrac{1}{\sqrt{5}} \Big) \, \gOctet\, ,
\end{equation*}
where the factors $\tfrac{1}{\sqrt{3}}, -\sqrt{\tfrac{2}{3}}$ are the combination of singlet and octet required to produce the ideally flavor mixed $s\bar{s}$ $\phi^\star$. It follows that exact OZI implies
\begin{equation}
\gOctet = - \frac{\sqrt{5}}{4}\, \gSinglet \, .\label{OZI1}
\end{equation}
We can establish the accuracy of this relation by computing scattering in the \SUF~octet representation, which will be done in the near future. 

A second condition following from OZI can be obtained by insisting that there is zero amplitude for the decay $\phi^\star \to \eta \omega$, which follows since every possible diagram for this process is disconnected. The amplitude is proportional to 
\begin{equation*}
\tfrac{1}{\sqrt{3}} \Big( \!-\! \tfrac{1}{2\sqrt{2}}\Big) \tfrac{1}{\sqrt{3}} \, \gSinglet 
+ \Big( \!-\! \sqrt{\tfrac{2}{3}} \Big) \Big( \!- \!\tfrac{1}{\sqrt{5}} \Big) \tfrac{1}{\sqrt{3}} \, \gOctet
+ \Big( \!-\! \sqrt{\tfrac{2}{3}} \Big) \sqrt{\tfrac{2}{3}} \, \hOctet \, ,
\end{equation*}
where the rightmost factors of $\tfrac{1}{\sqrt{3}}, \sqrt{\tfrac{2}{3}}$ are the combinations of singlet and octet required to produce the ideally flavor mixed $\tfrac{1}{\sqrt{2}} \big( u\bar{u} + d\bar{d} \big)$ $\omega$ in the decay. Using Eqn.~\ref{OZI1}, this amplitude is only zero if 
\begin{equation}
\hOctet = - \frac{1}{2\sqrt{2}}\, \gSinglet \, ,\label{OZI2}
\end{equation}
and again the accuracy of this expression will be tested in future calculations.

Making use of the two OZI conditions, we can write expressions for decays of $\omega^\star_J, \phi^\star_J$ mesons into pseudoscalar-vector final states solely in terms of the computed singlet coupling $\gSinglet$:
\begin{align*}
g \big(\phi^\star \!\to\! K \kbarSuper{*} \big) &= \tfrac{ \sqrt{3} }{4 \sqrt{2} } \, \gSinglet \\
g \big(\phi^\star \!\to\! \eta \phi \big) &= \tfrac{1}{2} \, \gSinglet \\[2.2ex]
g \big(\omega^\star \!\to\! \pi \rho \big) &= \tfrac{ \sqrt{3} }{4} \, \gSinglet \\
g \big(\omega^\star \!\to\! K \kbarSuper{*} \big) &= \tfrac{ \sqrt{3} }{8 } \, \gSinglet \\
g \big(\omega^\star \!\to\! \eta \omega \big) &= -\tfrac{1}{4 } \, \gSinglet \, ,
\end{align*}
where these couplings represent a single charge state. In fact, if we take the OZI relations seriously, they also allow us to use the singlet coupling to predict some decays of the isoscalar members of the octet, the $\rho^\star_J$ mesons, where for these we can use the decomposition of the $I=1, I_z=+1$ member of the octet,
\begin{equation*}
-\sqrt{\tfrac{3}{10}} \Big( K^+ \kbarSuper{*0} + \kbarSuper{0} K^{*+} \Big) + \tfrac{1}{\sqrt{5}} \pi^+ \omega_8 + \tfrac{1}{\sqrt{5}} \eta_8 \rho^+ \, ,
\end{equation*}
so that
\begin{align*}
g \big(\rho^\star \!\to\! \pi \omega \big) &= - \tfrac{ \sqrt{3} }{4} \, \gSinglet \\
g \big(\rho^\star \!\to\! K \kbarSuper{*}  \big) &= \tfrac{ \sqrt{3} }{4 \sqrt{2}} \, \gSinglet \, .
\end{align*}

Using $\Gamma = g^2 \tfrac{\rho}{M}$ for the partial width of a meson of mass $M$ into a final state with coupling $g$, we can obtain
\begin{align*}
\Gamma \big( \omega^\star \to \pi \rho \big)        &= 3\, \tfrac{\rho}{M}  \, \tfrac{3}{16} \big(\gSinglet\big)^2 \\
\Gamma \big( \omega^\star \to K \kbarSuper{*} \big) &= 4\, \tfrac{\rho}{M}  \, \tfrac{3}{64} \big(\gSinglet\big)^2 \\
\Gamma \big( \omega^\star \to \eta \omega \big)     &= 1\, \tfrac{\rho}{M}  \, \tfrac{1}{16} \big(\gSinglet\big)^2 \\[2.2ex]
\Gamma \big( \phi^\star \to K \kbarSuper{*} \big)   &= 4\, \tfrac{\rho}{M}  \, \tfrac{3}{32} \big(\gSinglet\big)^2 \\
\Gamma \big( \phi^\star \to \eta \phi \big)         &= 1\, \tfrac{\rho}{M}  \, \tfrac{1}{4} \big(\gSinglet\big)^2 \\[2.2ex]
\Gamma \big( \rho^\star \to \pi \omega \big)        &= 1\, \tfrac{\rho}{M}  \, \tfrac{3}{16} \big(\gSinglet\big)^2 \\
\Gamma \big( \rho^\star \to K \kbarSuper{*} \big)   &= 2\, \tfrac{\rho}{M}  \, \tfrac{3}{32} \big(\gSinglet\big)^2 \, ,
\end{align*}
where the leftmost integers count the final charge states, and where $K \kbarSuper{*}$ is a shorthand for a sum over all the possible pseudoscalar-vector kaonic final states.

Clearly this combination of \SUF~symmetry and imposition of exact OZI implies there are many relationships that should hold for the experimental states, but unfortunately the lack of a clear experimental picture makes the relationships rather hard to test. Perhaps the simplest is the prediction that, to the extent that an $\omega^\star$ is degenerate with the corresponding $\rho^\star$, the decay width of the former into $\rho \pi$ should be three times larger than the decay width of the latter into $\pi \omega$. For the experimental $\rho_3(1690)$, according to the PDG, the partial width into $\pi \omega$ is $\sim 30(10)\, \mathrm{MeV}$, and while the branching fraction of $\omega_3(1670)$ into $\pi \rho$ is not known, the total width of this state, $168(10)$ MeV, provides an upper limit, so the relation \emph{might} hold provided that decays other than $\pi \rho$ are significant. For the vector states, the analysis of Donnachie and Clegg~\cite{Clegg:1993mt} suggests
\begin{align*}
\omega^\star(1440) \qquad &\Gamma_{\pi \rho} \sim 240 \,\mathrm{MeV} \\
\rho^\star(1463)   \qquad &\Gamma_{\pi \omega} \sim 52 - 78 \, \mathrm{MeV} \, ,
\end{align*}
which is in reasonable agreement with a factor of three, while 
\begin{align*}
\omega^\star(1606) \qquad &\Gamma_{\pi \rho} \sim 84 \,\mathrm{MeV} \\
\rho^\star(1730)   \qquad &\Gamma_{\pi \omega} \sim 0  \, ,
\end{align*}
is less obviously compatible.

We will follow the approach laid out in Ref.~\cite{Woss:2020ayi} to extrapolate our couplings to the physical light quark mass. We interpret the magnitude of the pole couplings $|c_{\etaOctet \omegaOctet}|$ as being suitable for use as $\gSinglet$, and make the simple-minded assumption that there is no dependence on the light-quark mass apart from the scaling of the angular-momentum barrier in a decay with orbital angular momentum $\ell$,
\begin{equation*}
 \gSinglet = \left| \frac{k^{\mathrm{phys}} (M^{\mathrm{phys}}) }{k(M)} \right|^\ell \big| c_{\etaOctet \omegaOctet} \big| \, .
\end{equation*}
This approach breaks \SUF~symmetry only through the masses of the decay hadrons, and requires us to know the relevant resonance masses for physical light quark masses, $M^{\mathrm{phys}}$, which we will take from the PDG when known, or will estimate when not known. 

For $J^{PC}=3^{--}$, using the experimental masses of $\rho_3(1690)$, $\omega_3(1667)$ and $\phi_3(1854)$ we predict
\begin{align*}
\Gamma \big(\rho_3 \to \pi\omega, K\kbarSuper{*} \big)              &= 22, 2 \, \mathrm{MeV} \\
\Gamma \big(\omega_3 \to \pi \rho, K\kbarSuper{*}, \eta\omega \big) &= 62, 2, 1\, \mathrm{MeV} \\
\Gamma \big(\phi_3 \to K\kbarSuper{*}, \eta \phi \big)              &= 20, 3 \, \mathrm{MeV} \, ,
\end{align*}
and we will not quote errors for fear of implying a level of certainty that surely is not present in such a crude extrapolation. There is limited scope for comparison to experiment owing to there being few measured branching ratios. The summed $\omega_3$ estimated partial widths are at least below the measured total width $\sim 168(10)$ MeV, as are the summed $\phi_3$ partial widths compared to $87(25)$ MeV, and in that case there may be a significant contribution from $\phi_3 \to K\kbar$. The $\rho_3$ does have some measured partial widths: $\Gamma_{\pi \omega} \sim 30(10) \,\mathrm{MeV}$, that might be in agreement with our estimate, and $\Gamma_{K\kbar \pi} \sim 7$ MeV which will include $K \kbarSuper{*}$ as a sub-process.

The $\threePzero$--model has been used to predict decays of these states~\cite{Barnes:1996ff, Barnes:2002mu}. It has $\phi_3$ decays to $K\kbarSuper{*}, \eta\phi$ that are in good agreement with our estimates, and in addition predicts larger rates to $K\kbar$ and $K^* \kbarSuper{*}$. The model predictions for $\omega_3$ and $\rho_3$ are also in reasonable agreement with our estimates, with the $\rho_3$ also having significant rates to $\pi\pi$ and $\rho\rho$. To get access to these additional decay modes in the current framework we need to calculate \SUF~octet scattering.

For $J^{PC}=2^{--}$ there are no experimental candidate states, and as such we will proceed assuming masses that are approximately equal to the corresponding $\rho_3$, $\omega_3$ and $\phi_3$ states. In this case there are both $P$--wave and $F$--wave decays and the total partial width for each channel is an incoherent sum of the two. We predict
\begin{align*}
\Gamma \big(\rho_2 \to \pi\omega, K\kbarSuper{*} \big)              &= 125, 36 \, \mathrm{MeV} \\
\Gamma \big(\omega_2 \to \pi \rho, K\kbarSuper{*}, \eta\omega \big) &= 365, 36, 17\, \mathrm{MeV} \\
\Gamma \big(\phi_2 \to K\kbarSuper{*}, \eta \phi \big)              &= 148, 44 \, \mathrm{MeV} \, ,
\end{align*}
which suggests that the $\omega_2$ is likely to have quite a large total width, particularly once decays to final states other than pseudoscalar-vector are added in. The $\rho_2$ and $\omega_2$ might be narrower, particularly given that the largest phase-space modes $\pi \pi$ and $K \kbar$ are not accessible to a $2^{-}$ resonance.

The $\threePzero$--model has $\phi_2$ partial widths that are in good agreement with our estimates, while the $\omega_2$ and $\rho_2$ come out lower in the model. The model predicts a very large $\rho_2 \to a_2 \pi$ rate that leads to a rather large total width for this state.

For $J^{PC}=1^{--}$ we have the problem of associating our two resonances, the lighter broad state $\mathsf{a}$, and the heavier narrow state $\mathsf{b}$, with the physical states. The simplest assumption is that in each flavor channel, the lighter state is purely  $\mathsf{a}$ and the heavier state purely $\mathsf{b}$, with no evolution in a possible basis-state mixing angle with change in light-quark mass. With this assignment we predict
\begin{align*}
\Gamma \big(\rho_\mathsf{a} \to \pi\omega, K\kbarSuper{*} \big)               &= 133, 9 \, \mathrm{MeV} \\
\Gamma \big(\omega_\mathsf{a}  \to \pi \rho, K\kbarSuper{*}, \eta\omega \big) &= 384, 4, 5\, \mathrm{MeV} \\
\Gamma \big(\phi_\mathsf{a}  \to K\kbarSuper{*}, \eta \phi \big)              &= 154, 25 \, \mathrm{MeV} \, ,
\end{align*}
and we can say little more than that these summed partial widths do not over saturate the experimental total widths of the $\rho(1450)$, $\omega(1420)$ and $\phi(1680)$. For the heavier state we predict
\begin{align*}
\Gamma \big(\rho_\mathsf{b} \to \pi\omega, K\kbarSuper{*} \big)               &= 9, 3 \, \mathrm{MeV} \\
\Gamma \big(\omega_\mathsf{b}  \to \pi \rho, K\kbarSuper{*}, \eta\omega \big) &= 25, 3, 1\, \mathrm{MeV} \\
\Gamma \big(\phi_\mathsf{b}  \to K\kbarSuper{*}, \eta \phi \big)              &= 13, 5 \, \mathrm{MeV} \, ,
\end{align*}
which appears to suggest that unless the other allowed decays of the $\rho(1700)$, $\omega(1650)$ and a hypothetical $\phi(1900)$ provide large partial widths, these states should be much narrower than they seem to be in experiment. We do not have a good explanation of this observation, although some degree of basis-state mixing of $\mathsf{a}, \mathsf{b}$ into the physical states might share the decays more evenly and give rise to two moderately broad states.

The $\threePzero$--model, assuming the lighter state is pure $q\bar{q}[2 \threeSone]$ has somewhat larger decay rates for the $\phi$ state, and rates for the $\omega$ and $\rho$ states that are in reasonable agreement with our estimates. Assuming the heavier state is pure $q\bar{q}[1\threeDone]$, the model predicts decays for the hypothetical $\phi^\star$ that are much larger than our estimates, and also has a huge $\sim 500$ MeV branch into $K_1 \kbar$. A similar pattern is observed for the $\omega^\star$ and $\rho^\star$ states, indicating quite poor agreement with our estimates.

%% file: 5-summary.tex

In this paper we have reported on a first lattice QCD study of excited mesons with $J^{--}$ quantum numbers, computing in a version of QCD having exact \SUF~symmetry, and focussing on the singlet representation. We found that the $1^{--}$, $2^{--}$ and $3^{--}$ partial waves at low energies have only a single strongly-interacting channel of pseudoscalar-vector scattering, $\etaOctet \omegaOctet$, with other kinematically open channels being decoupled and weakly interacting. Constraining scattering amplitudes using nearly 200 energy levels across five lattice volumes, we found a unique picture featuring four resonances.

\begin{figure}[b]
\includegraphics[width=0.98\columnwidth]{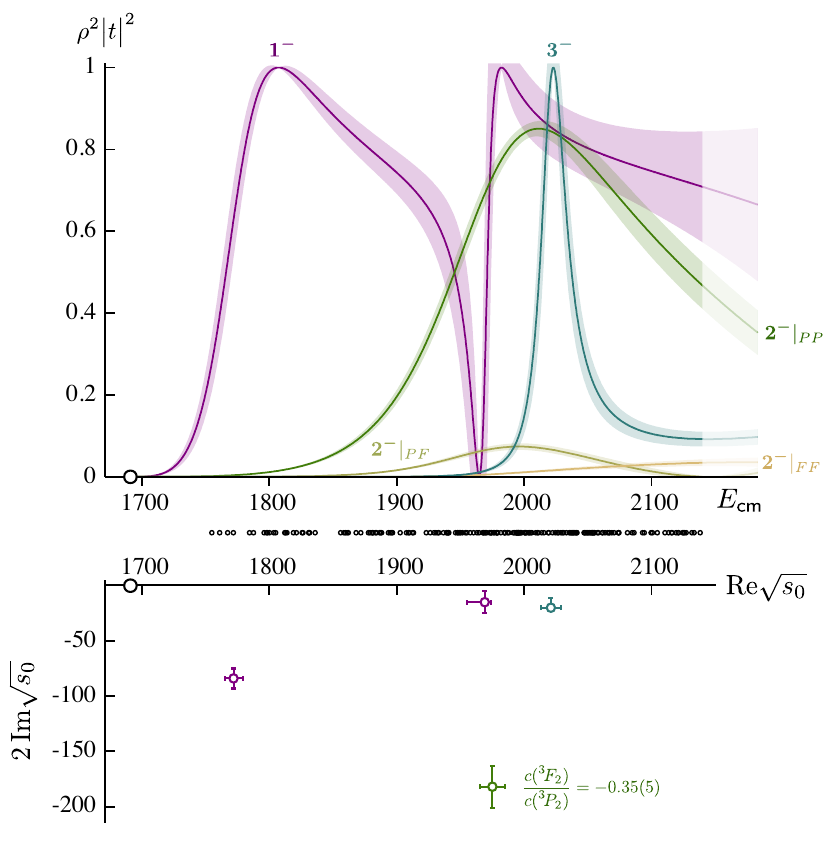}
\caption{
Upper panel: $\etaOctet \omegaOctet$ scattering amplitudes as presented in Section~\ref{global}. Lower panel: Our best estimate for resonance poles from Eqn.~\ref{best_poles}, including variation of amplitude parameterization, scattering meson masses and anisotropy in the error estimates. Scale set to MeV units using the $\Omega$-baryon mass.
}
\label{scale_set}
\end{figure}

A single, isolated narrow resonance with $3^{--}$ appears to match with the well-known experimental states \mbox{($\rho_3$, $\omega_3$, $\phi_3$)}. A first computation within lattice QCD of $2^{--}$ amplitudes, which appear as dynamically coupled $\{\threePtwo, \threeFtwo\}$ partial waves yields a much broader resonance for which there is no experimental evidence to date. The rather novel $1^{--}$ partial wave amplitude features a lighter broad resonance and a heavier narrow resonance. A tight dip and a zero in the amplitude appears on the real energy axis, very close to the heavier resonance pole. We summarize our amplitudes and our best estimates of the resonance poles in Figure~\ref{scale_set}.

In a natural extension of the work reported on in this paper, our next calculation will consider the \SUF~octet system on the same lattices. This will require a first consideration in lattice QCD of coupled pseudoscalar-pseudoscalar and pseudoscalar-vector scattering, but we see no reason why we cannot obtain a comparable number of finite-volume energy levels as in the current study with which to constrain the relevant scattering amplitudes.

Considering higher energy scattering in $1^{--}$ is of particular interest given the suggestion that the next resonance above those we have extracted is expected to be a hybrid meson. The challenge here is the need to implement a so far underdeveloped extension of finite-volume three-body formalism in which two- and three- meson sectors are coupled, but we expect to see progress in this direction in the near future.

Another interesting expansion of scope of the current study would consider the process in which a vector current (describing the virtual photon in $e^+ e^-$ annihilation) produces the $\etaOctet \omegaOctet$ system with $J^{PC} = 1^{--}$. If the quark model picture of the two vector resonances is correct, we'd expect the $q\bar{q}\big[1\threeDone\big]$ state to contribute very little (it has zero wavefunction at the origin, and only appears through the suppressed second derivative), while the $q\bar{q}\big[2\threeSone\big]$ state could be significant. Such a calculation would be a first step towards a first-principles QCD based phenomenology to be used to describe resonance production in $e^+e^-$, a process of primary importance at experiments like BES III.

Developing an understanding of the excited $J^{--}$ resonances in QCD is timely, as we expect a huge new experimental data set in \emph{photoproduction} from the GlueX experiment using which we can obtain better constraint on the properties of these states. It remains to be seen whether an extension of the method presented in Ref.~\cite{Briceno:2015dca, Briceno:2016kkp} for production of the $\rho$ resonance can be practically applied to the current case in order to describe the pion-exchange contribution to photoproduction of excited $J^{--}$ mesons.

The calculation presented in this paper is the first step towards a QCD-based theoretical understanding of the mysterious excited $J^{--}$ resonances.

%% file: acknow.tex

\begin{acknowledgments}
We thank our colleagues within the Hadron Spectrum Collaboration. Special thanks to Antoni Woss and Christopher Thomas for their assistance computing correlation functions with computer resources at Cambridge, and Robert Edwards for his support with using the {\tt Redstar} software system.
JJD and CTJ acknowledge support from the U.S. Department of Energy contract DE-SC0018416 at William \& Mary, and contract DE-AC05-06OR23177, under which Jefferson Science Associates, LLC, manages and operates Jefferson Lab.

The software codes
{\tt Chroma}~\cite{Edwards:2004sx} and {\tt QUDA}~\cite{Clark:2009wm,Babich:2010mu,Clark:2016rdz} were used. 
The authors acknowledge support from the U.S. Department of Energy, Office of Science, Office of Advanced Scientific Computing Research and Office of Nuclear Physics, Scientific Discovery through Advanced Computing (SciDAC) program. 
Also acknowledged is support from the Exascale Computing Project (17-SC-20-SC), a collaborative effort of the U.S. Department of Energy Office of Science and the National Nuclear Security Administration.
This work was performed using the Cambridge Service for Data Driven Discovery (CSD3) operated by the University of Cambridge Research Computing Service (www.hpc.cam.ac.uk), provided by Dell EMC and Intel using Tier-2 funding from the Engineering and Physical Sciences Research Council (capital grant EP/P020259/1), and DiRAC funding from STFC (www.dirac.ac.uk). The DiRAC component of CSD3 was funded by BEIS capital funding via STFC capital grants ST/P002307/1 and ST/R002452/1 and STFC operations grant ST/R00689X/1. DiRAC is part of the National e-Infrastructure.
This work was also performed on clusters at Jefferson Lab under the USQCD Collaboration and the LQCD ARRA Project.
This research was supported in part under an ALCC award, and used resources of the Oak Ridge Leadership Computing Facility at the Oak Ridge National Laboratory, which is supported by the Office of Science of the U.S. Department of Energy under Contract No. DE-AC05-00OR22725.
This research used resources of the National Energy Research Scientific Computing Center (NERSC), a DOE Office of Science User Facility supported by the Office of Science of the U.S. Department of Energy under Contract No. DE-AC02-05CH11231.
The authors acknowledge the Texas Advanced Computing Center (TACC) at The University of Texas at Austin for providing HPC resources.
Gauge configurations were generated using resources awarded from the U.S. Department of Energy INCITE program at the Oak Ridge Leadership Computing Facility, the NERSC, the NSF Teragrid at the TACC and the Pittsburgh Supercomputer Center, as well as at the Cambridge Service for Data Driven Discovery (CSD3) and Jefferson Lab.
This work was performed in part using computing facilities at William \& Mary which were provided by contributions from the National Science Foundation (MRI grant PHY-1626177), and the Commonwealth of Virginia Equipment Trust Fund.
\end{acknowledgments}

%% file: appA-ops.tex

Details of the fermion-bilinear operators, the method for obtaining optimized ``single-meson'' operators, and the construction of meson-meson operators can be found in Refs.~\cite{Dudek:2010wm, Thomas:2011rh, Dudek:2012gj, Wilson:2014cna, Wilson:2015dqa, Woss:2018irj, Woss:2019hse}. The operator basis used in the current calculation is presented in Tables~\ref{ops000}, \ref{ops100}, \ref{ops110} where the meson-meson operators are listed in order of increasing non-interacting energy. Those cases in which more than one construction appears with the same non-interacting energy are indicated by the \emph{multiplicity}, $\{ N \}$.

%
\begin{table*}
{
\renewcommand{\arraystretch}{1.2}
\begin{tabular}{c | lllll}
&\multicolumn{1}{c}{$14^3$} 	
& \multicolumn{1}{c}{$16^3$}
& \multicolumn{1}{c}{$18^3$}
& \multicolumn{1}{c}{$20^3$}
& \multicolumn{1}{c}{$24^3$} \\
\hline\\[-1.3ex]
\parbox[t]{7mm}{\multirow{7}{*}{\rotatebox[origin=c]{90}{$[000]\, T_1^-$ }}} &
$\bar{\psi} {\bf \Gamma} \psi \times 20$ &
$\bar{\psi} {\bf \Gamma} \psi \times 20$ &
$\bar{\psi} {\bf \Gamma} \psi \times 20$ &
$\bar{\psi} {\bf \Gamma} \psi \times 20$ &
$\bar{\psi} {\bf \Gamma} \psi \times 20$ \\[1ex]
&$\etaOctet_{\scriptscriptstyle{[100]}} \omegaOctet_{\scriptscriptstyle{[100]}}$ 
& $\etaOctet_{\scriptscriptstyle{[100]}} \omegaOctet_{\scriptscriptstyle{[100]}}$
& $\etaOctet_{\scriptscriptstyle{[100]}} \omegaOctet_{\scriptscriptstyle{[100]}}$
& $\etaOctet_{\scriptscriptstyle{[100]}} \omegaOctet_{\scriptscriptstyle{[100]}}$
& $\etaOctet_{\scriptscriptstyle{[100]}} \omegaOctet_{\scriptscriptstyle{[100]}}$ \\
& & & $\etaOctet_{\scriptscriptstyle{[110]}} \omegaOctet_{\scriptscriptstyle{[110]}} {\scriptstyle \{2\} }$
& $\etaOctet_{\scriptscriptstyle{[110]}} \omegaOctet_{\scriptscriptstyle{[110]}} {\scriptstyle \{2\} }$
& $\etaOctet_{\scriptscriptstyle{[110]}} \omegaOctet_{\scriptscriptstyle{[110]}} {\scriptstyle \{2\} }$  \\
& & & & & $\etaOctet_{\scriptscriptstyle{[111]}} \omegaOctet_{\scriptscriptstyle{[111]}}$  \\[1ex]
& $\omegaSinglet_{\scriptscriptstyle{[000]}} {\fZeroSinglet}_{\! \scriptscriptstyle{[000]}}$
& $\omegaSinglet_{\scriptscriptstyle{[000]}} {\fZeroSinglet}_{\! \scriptscriptstyle{[000]}}$
& $\omegaSinglet_{\scriptscriptstyle{[000]}} {\fZeroSinglet}_{\! \scriptscriptstyle{[000]}}$
& $\omegaSinglet_{\scriptscriptstyle{[000]}} {\fZeroSinglet}_{\! \scriptscriptstyle{[000]}}$
& $\omegaSinglet_{\scriptscriptstyle{[000]}} {\fZeroSinglet}_{\! \scriptscriptstyle{[000]}}$ \\
& & & $\omegaSinglet_{\scriptscriptstyle{[100]}} {\fZeroSinglet}_{\!\scriptscriptstyle{[100]}} {\scriptstyle \{2\} }$
& $\omegaSinglet_{\scriptscriptstyle{[100]}} {\fZeroSinglet}_{\! \scriptscriptstyle{[100]}} {\scriptstyle \{2\} }$
& $\omegaSinglet_{\scriptscriptstyle{[100]}} {\fZeroSinglet}_{\! \scriptscriptstyle{[100]}} {\scriptstyle \{2\} }$ \\[1ex]
& & & $\etaSinglet_{\scriptscriptstyle{[100]}} \omegaSinglet_{\scriptscriptstyle{[100]}}$
& $\etaSinglet_{\scriptscriptstyle{[100]}} \omegaSinglet_{\scriptscriptstyle{[100]}}$
& $\etaSinglet_{\scriptscriptstyle{[100]}} \omegaSinglet_{\scriptscriptstyle{[100]}}$ \\[1ex]
\hline\\[-1.3ex]
\parbox[t]{7mm}{\multirow{5}{*}{\rotatebox[origin=c]{90}{$[000]\, E^-$ }}} &
$\bar{\psi} {\bf \Gamma} \psi \times 12$ &
$\bar{\psi} {\bf \Gamma} \psi \times 12$ &
$\bar{\psi} {\bf \Gamma} \psi \times 12$ &
$\bar{\psi} {\bf \Gamma} \psi \times 12$ &
$\bar{\psi} {\bf \Gamma} \psi \times 12$ \\[1ex]
&$\etaOctet_{\scriptscriptstyle{[100]}} \omegaOctet_{\scriptscriptstyle{[100]}}$ 
& $\etaOctet_{\scriptscriptstyle{[100]}} \omegaOctet_{\scriptscriptstyle{[100]}}$
& $\etaOctet_{\scriptscriptstyle{[100]}} \omegaOctet_{\scriptscriptstyle{[100]}}$
& $\etaOctet_{\scriptscriptstyle{[100]}} \omegaOctet_{\scriptscriptstyle{[100]}}$
& $\etaOctet_{\scriptscriptstyle{[100]}} \omegaOctet_{\scriptscriptstyle{[100]}}$ \\
& & & $\etaOctet_{\scriptscriptstyle{[110]}} \omegaOctet_{\scriptscriptstyle{[110]}} {\scriptstyle \{2\} }$
& $\etaOctet_{\scriptscriptstyle{[110]}} \omegaOctet_{\scriptscriptstyle{[110]}} {\scriptstyle \{2\} }$
& $\etaOctet_{\scriptscriptstyle{[110]}} \omegaOctet_{\scriptscriptstyle{[110]}} {\scriptstyle \{2\} }$  \\
& & & & & $\etaOctet_{\scriptscriptstyle{[111]}} \omegaOctet_{\scriptscriptstyle{[111]}}$  \\[1ex]
& & & $\etaSinglet_{\scriptscriptstyle{[100]}} \omegaSinglet_{\scriptscriptstyle{[100]}}$
& $\etaSinglet_{\scriptscriptstyle{[100]}} \omegaSinglet_{\scriptscriptstyle{[100]}}$
& $\etaSinglet_{\scriptscriptstyle{[100]}} \omegaSinglet_{\scriptscriptstyle{[100]}}$ \\[1ex]
\hline\\[-1.3ex]
\parbox[t]{7mm}{\multirow{6}{*}{\rotatebox[origin=c]{90}{$[000]\, T_2^-$ }}} &
$\bar{\psi} {\bf \Gamma} \psi \times 18$ &
$\bar{\psi} {\bf \Gamma} \psi \times 18$ &
$\bar{\psi} {\bf \Gamma} \psi \times 18$ &
$\bar{\psi} {\bf \Gamma} \psi \times 18$ &
$\bar{\psi} {\bf \Gamma} \psi \times 18$ \\[1ex]
&$\etaOctet_{\scriptscriptstyle{[100]}} \omegaOctet_{\scriptscriptstyle{[100]}}$ 
& $\etaOctet_{\scriptscriptstyle{[100]}} \omegaOctet_{\scriptscriptstyle{[100]}}$
& $\etaOctet_{\scriptscriptstyle{[100]}} \omegaOctet_{\scriptscriptstyle{[100]}}$
& $\etaOctet_{\scriptscriptstyle{[100]}} \omegaOctet_{\scriptscriptstyle{[100]}}$
& $\etaOctet_{\scriptscriptstyle{[100]}} \omegaOctet_{\scriptscriptstyle{[100]}}$ \\
& & & $\etaOctet_{\scriptscriptstyle{[110]}} \omegaOctet_{\scriptscriptstyle{[110]}} {\scriptstyle \{2\} }$
& $\etaOctet_{\scriptscriptstyle{[110]}} \omegaOctet_{\scriptscriptstyle{[110]}} {\scriptstyle \{2\} }$
& $\etaOctet_{\scriptscriptstyle{[110]}} \omegaOctet_{\scriptscriptstyle{[110]}} {\scriptstyle \{2\} }$  \\
& & & & & $\etaOctet_{\scriptscriptstyle{[111]}} \omegaOctet_{\scriptscriptstyle{[111]}} {\scriptstyle \{2\} }$   \\[1ex]
& & & $\omegaSinglet_{\scriptscriptstyle{[100]}} {\fZeroSinglet}_{\!\scriptscriptstyle{[100]}} $
& $\omegaSinglet_{\scriptscriptstyle{[100]}} {\fZeroSinglet}_{\! \scriptscriptstyle{[100]}} $
& $\omegaSinglet_{\scriptscriptstyle{[100]}} {\fZeroSinglet}_{\! \scriptscriptstyle{[100]}} $ \\[1ex]
& & & $\etaSinglet_{\scriptscriptstyle{[100]}} \omegaSinglet_{\scriptscriptstyle{[100]}}$
& $\etaSinglet_{\scriptscriptstyle{[100]}} \omegaSinglet_{\scriptscriptstyle{[100]}}$
& $\etaSinglet_{\scriptscriptstyle{[100]}} \omegaSinglet_{\scriptscriptstyle{[100]}}$ \\[1ex]
\hline\\[-1.3ex]
\parbox[t]{7mm}{\multirow{2}{*}{\rotatebox[origin=c]{90}{$[000]\, A_2^-$ }}} &
$\bar{\psi} {\bf \Gamma} \psi \times 6$ &
$\bar{\psi} {\bf \Gamma} \psi \times 6$ &
$\bar{\psi} {\bf \Gamma} \psi \times 6$ &
$\bar{\psi} {\bf \Gamma} \psi \times 6$ &
$\bar{\psi} {\bf \Gamma} \psi \times 6$ \\[1ex]
& & & $\etaOctet_{\scriptscriptstyle{[110]}} \omegaOctet_{\scriptscriptstyle{[110]}} $
& $\etaOctet_{\scriptscriptstyle{[110]}} \omegaOctet_{\scriptscriptstyle{[110]}} $
& $\etaOctet_{\scriptscriptstyle{[110]}} \omegaOctet_{\scriptscriptstyle{[110]}} $  \\[3ex]
\end{tabular}
}
\caption{
Operators used to compute matrices of correlations functions in rest-frame irreps. $\bar{\psi} {\bf \Gamma} \psi \times N$ indicates the maximum number of ``single-meson'' operators included in the basis.
}
\label{ops000}
\end{table*}

%
\begin{table*}
{
\renewcommand{\arraystretch}{1.2}
\begin{tabular}{c | lllll}
&\multicolumn{1}{c}{$14^3$} 	
& \multicolumn{1}{c}{$16^3$}
& \multicolumn{1}{c}{$18^3$}
& \multicolumn{1}{c}{$20^3$}
& \multicolumn{1}{c}{$24^3$} \\
\hline\\[-1.3ex]
\parbox[t]{7mm}{\multirow{13}{*}{\rotatebox[origin=c]{90}{$[100]\, A_1$ }}} &
$\bar{\psi} {\bf \Gamma} \psi \times 18$ &
$\bar{\psi} {\bf \Gamma} \psi \times 18$ &
$\bar{\psi} {\bf \Gamma} \psi \times 18$ &
$\bar{\psi} {\bf \Gamma} \psi \times 18$ &
$\bar{\psi} {\bf \Gamma} \psi \times 18$ \\[1ex]
& $\etaOctet_{\scriptscriptstyle{[100]}} \omegaOctet_{\scriptscriptstyle{[110]}}$ 
& $\etaOctet_{\scriptscriptstyle{[100]}} \omegaOctet_{\scriptscriptstyle{[110]}}$
& $\etaOctet_{\scriptscriptstyle{[100]}} \omegaOctet_{\scriptscriptstyle{[110]}}$
& $\etaOctet_{\scriptscriptstyle{[100]}} \omegaOctet_{\scriptscriptstyle{[110]}}$
& $\etaOctet_{\scriptscriptstyle{[100]}} \omegaOctet_{\scriptscriptstyle{[110]}}$ \\
& $\etaOctet_{\scriptscriptstyle{[110]}} \omegaOctet_{\scriptscriptstyle{[100]}}$ 
& $\etaOctet_{\scriptscriptstyle{[110]}} \omegaOctet_{\scriptscriptstyle{[100]}}$
& $\etaOctet_{\scriptscriptstyle{[110]}} \omegaOctet_{\scriptscriptstyle{[100]}}$
& $\etaOctet_{\scriptscriptstyle{[110]}} \omegaOctet_{\scriptscriptstyle{[100]}}$
& $\etaOctet_{\scriptscriptstyle{[110]}} \omegaOctet_{\scriptscriptstyle{[100]}}$ \\
& & & $\etaOctet_{\scriptscriptstyle{[110]}} \omegaOctet_{\scriptscriptstyle{[111]}} $
& $\etaOctet_{\scriptscriptstyle{[110]}} \omegaOctet_{\scriptscriptstyle{[111]}} $
& $\etaOctet_{\scriptscriptstyle{[110]}} \omegaOctet_{\scriptscriptstyle{[111]}} $  \\
& & & & $\etaOctet_{\scriptscriptstyle{[111]}} \omegaOctet_{\scriptscriptstyle{[110]}} $
& $\etaOctet_{\scriptscriptstyle{[111]}} \omegaOctet_{\scriptscriptstyle{[110]}} $  \\
& & & & & $\etaOctet_{\scriptscriptstyle{[110]}} \omegaOctet_{\scriptscriptstyle{[210]}} $ \\
& & & & & $\etaOctet_{\scriptscriptstyle{[210]}} \omegaOctet_{\scriptscriptstyle{[110]}} $ \\[1ex]
& $\omegaSinglet_{\scriptscriptstyle{[100]}} {\fZeroSinglet}_{\!\scriptscriptstyle{[000]}} $
& $\omegaSinglet_{\scriptscriptstyle{[100]}} {\fZeroSinglet}_{\! \scriptscriptstyle{[000]}} $
& $\omegaSinglet_{\scriptscriptstyle{[100]}} {\fZeroSinglet}_{\!\scriptscriptstyle{[000]}} $
& $\omegaSinglet_{\scriptscriptstyle{[100]}} {\fZeroSinglet}_{\! \scriptscriptstyle{[000]}} $
& $\omegaSinglet_{\scriptscriptstyle{[100]}} {\fZeroSinglet}_{\! \scriptscriptstyle{[000]}} $ \\
& $\omegaSinglet_{\scriptscriptstyle{[000]}} {\fZeroSinglet}_{\!\scriptscriptstyle{[100]}} $
& $\omegaSinglet_{\scriptscriptstyle{[000]}} {\fZeroSinglet}_{\! \scriptscriptstyle{[100]}} $
& $\omegaSinglet_{\scriptscriptstyle{[000]}} {\fZeroSinglet}_{\!\scriptscriptstyle{[100]}} $
& $\omegaSinglet_{\scriptscriptstyle{[000]}} {\fZeroSinglet}_{\! \scriptscriptstyle{[100]}} $
& $\omegaSinglet_{\scriptscriptstyle{[000]}} {\fZeroSinglet}_{\! \scriptscriptstyle{[100]}} $ \\
& & & & & $\omegaSinglet_{\scriptscriptstyle{[110]}} {\fZeroSinglet}_{\!\scriptscriptstyle{[100]}} {\scriptstyle \{2\} } $ \\
& & & & & $\omegaSinglet_{\scriptscriptstyle{[100]}} {\fZeroSinglet}_{\!\scriptscriptstyle{[110]}} {\scriptstyle \{2\} } $ \\[1ex]
& & & & & $\etaSinglet_{\scriptscriptstyle{[100]}} \omegaSinglet_{\scriptscriptstyle{[110]}}$ \\
& & & & & $\etaSinglet_{\scriptscriptstyle{[110]}} \omegaSinglet_{\scriptscriptstyle{[100]}}$ \\[1ex]
\hline\\[-1.3ex]
\parbox[t]{7mm}{\multirow{11}{*}{\rotatebox[origin=c]{90}{$[100]\, B_1$ }}} &
$\bar{\psi} {\bf \Gamma} \psi \times 9$ &
$\bar{\psi} {\bf \Gamma} \psi \times 9$ &
$\bar{\psi} {\bf \Gamma} \psi \times 9$ &
$\bar{\psi} {\bf \Gamma} \psi \times 9$ &
$\bar{\psi} {\bf \Gamma} \psi \times 9$ \\[1ex]
& $\etaOctet_{\scriptscriptstyle{[100]}} \omegaOctet_{\scriptscriptstyle{[110]}}$ 
& $\etaOctet_{\scriptscriptstyle{[100]}} \omegaOctet_{\scriptscriptstyle{[110]}}$
& $\etaOctet_{\scriptscriptstyle{[100]}} \omegaOctet_{\scriptscriptstyle{[110]}}$
& $\etaOctet_{\scriptscriptstyle{[100]}} \omegaOctet_{\scriptscriptstyle{[110]}}$
& $\etaOctet_{\scriptscriptstyle{[100]}} \omegaOctet_{\scriptscriptstyle{[110]}}$ \\
& $\etaOctet_{\scriptscriptstyle{[110]}} \omegaOctet_{\scriptscriptstyle{[100]}}$ 
& $\etaOctet_{\scriptscriptstyle{[110]}} \omegaOctet_{\scriptscriptstyle{[100]}}$
& $\etaOctet_{\scriptscriptstyle{[110]}} \omegaOctet_{\scriptscriptstyle{[100]}}$
& $\etaOctet_{\scriptscriptstyle{[110]}} \omegaOctet_{\scriptscriptstyle{[100]}}$
& $\etaOctet_{\scriptscriptstyle{[110]}} \omegaOctet_{\scriptscriptstyle{[100]}}$ \\
& & & $\etaOctet_{\scriptscriptstyle{[110]}} \omegaOctet_{\scriptscriptstyle{[111]}} {\scriptstyle \{2\} }$
& $\etaOctet_{\scriptscriptstyle{[110]}} \omegaOctet_{\scriptscriptstyle{[111]}} {\scriptstyle \{2\} }$
& $\etaOctet_{\scriptscriptstyle{[110]}} \omegaOctet_{\scriptscriptstyle{[111]}} {\scriptstyle \{2\} }$  \\
& & &  $\etaOctet_{\scriptscriptstyle{[111]}} \omegaOctet_{\scriptscriptstyle{[110]}} {\scriptstyle \{2\} }$
& $\etaOctet_{\scriptscriptstyle{[111]}} \omegaOctet_{\scriptscriptstyle{[110]}} {\scriptstyle \{2\} }$
& $\etaOctet_{\scriptscriptstyle{[111]}} \omegaOctet_{\scriptscriptstyle{[110]}} {\scriptstyle \{2\} }$  \\
& & & & & $\etaOctet_{\scriptscriptstyle{[110]}} \omegaOctet_{\scriptscriptstyle{[210]}}$ \\
& & & & & $\etaOctet_{\scriptscriptstyle{[210]}} \omegaOctet_{\scriptscriptstyle{[110]}}$ \\[1ex]
& & & & & $\omegaSinglet_{\scriptscriptstyle{[110]}} {\fZeroSinglet}_{\!\scriptscriptstyle{[100]}} {\scriptstyle \{2\} } $ \\
& & & & & $\omegaSinglet_{\scriptscriptstyle{[100]}} {\fZeroSinglet}_{\!\scriptscriptstyle{[110]}} {\scriptstyle \{2\} } $ \\[1ex]
& & & & & $\etaSinglet_{\scriptscriptstyle{[100]}} \omegaSinglet_{\scriptscriptstyle{[110]}}$ \\
& & & & & $\etaSinglet_{\scriptscriptstyle{[110]}} \omegaSinglet_{\scriptscriptstyle{[100]}}$ \\[1ex]
\hline\\[-1.3ex]
\parbox[t]{7mm}{\multirow{10}{*}{\rotatebox[origin=c]{90}{$[100]\, B_2$ }}} &
$\bar{\psi} {\bf \Gamma} \psi \times 9$ &
$\bar{\psi} {\bf \Gamma} \psi \times 9$ &
$\bar{\psi} {\bf \Gamma} \psi \times 9$ &
$\bar{\psi} {\bf \Gamma} \psi \times 9$ &
$\bar{\psi} {\bf \Gamma} \psi \times 9$ \\[1ex]
& $\etaOctet_{\scriptscriptstyle{[100]}} \omegaOctet_{\scriptscriptstyle{[110]}} {\scriptstyle \{2\} }$ 
& $\etaOctet_{\scriptscriptstyle{[100]}} \omegaOctet_{\scriptscriptstyle{[110]}} {\scriptstyle \{2\} }$
& $\etaOctet_{\scriptscriptstyle{[100]}} \omegaOctet_{\scriptscriptstyle{[110]}} {\scriptstyle \{2\} }$
& $\etaOctet_{\scriptscriptstyle{[100]}} \omegaOctet_{\scriptscriptstyle{[110]}} {\scriptstyle \{2\} }$
& $\etaOctet_{\scriptscriptstyle{[100]}} \omegaOctet_{\scriptscriptstyle{[110]}} {\scriptstyle \{2\} }$ \\
& $\etaOctet_{\scriptscriptstyle{[110]}} \omegaOctet_{\scriptscriptstyle{[100]}} {\scriptstyle \{2\} }$ 
& $\etaOctet_{\scriptscriptstyle{[110]}} \omegaOctet_{\scriptscriptstyle{[100]}} {\scriptstyle \{2\} }$
& $\etaOctet_{\scriptscriptstyle{[110]}} \omegaOctet_{\scriptscriptstyle{[100]}} {\scriptstyle \{2\} }$
& $\etaOctet_{\scriptscriptstyle{[110]}} \omegaOctet_{\scriptscriptstyle{[100]}} {\scriptstyle \{2\} }$
& $\etaOctet_{\scriptscriptstyle{[110]}} \omegaOctet_{\scriptscriptstyle{[100]}} {\scriptstyle \{2\} }$ \\
& & & $\etaOctet_{\scriptscriptstyle{[110]}} \omegaOctet_{\scriptscriptstyle{[111]}} $ 
& $\etaOctet_{\scriptscriptstyle{[110]}} \omegaOctet_{\scriptscriptstyle{[111]}} $ 
& $\etaOctet_{\scriptscriptstyle{[110]}} \omegaOctet_{\scriptscriptstyle{[111]}} $ \\
& & & & $\etaOctet_{\scriptscriptstyle{[111]}} \omegaOctet_{\scriptscriptstyle{[110]}} $
& $\etaOctet_{\scriptscriptstyle{[111]}} \omegaOctet_{\scriptscriptstyle{[110]}} $ \\
& & & & & $\etaOctet_{\scriptscriptstyle{[110]}} \omegaOctet_{\scriptscriptstyle{[210]}} {\scriptstyle \{2\} }$ \\[1ex]
& & & & & $\omegaSinglet_{\scriptscriptstyle{[110]}} {\fZeroSinglet}_{\!\scriptscriptstyle{[100]}}  $ \\
& & & & & $\omegaSinglet_{\scriptscriptstyle{[100]}} {\fZeroSinglet}_{\!\scriptscriptstyle{[110]}} $ \\[1ex]
& & & & & $\etaSinglet_{\scriptscriptstyle{[100]}} \omegaSinglet_{\scriptscriptstyle{[110]}} {\scriptstyle \{2\} }$ \\
& & & & & $\etaSinglet_{\scriptscriptstyle{[110]}} \omegaSinglet_{\scriptscriptstyle{[100]}} {\scriptstyle \{2\} }$ \\[1ex]
\end{tabular}
}
\caption{As Table~\ref{ops000} for irreps with $\vec{P} = [100]$.}
\label{ops100}
\end{table*}

%
\begin{table*}
{
\renewcommand{\arraystretch}{1.2}
\begin{tabular}{c | lllll}
&\multicolumn{1}{c}{$14^3$} 	
& \multicolumn{1}{c}{$16^3$}
& \multicolumn{1}{c}{$18^3$}
& \multicolumn{1}{c}{$20^3$}
& \multicolumn{1}{c}{$24^3$} \\
\hline\\[-1.3ex]
\parbox[t]{7mm}{\multirow{13}{*}{\rotatebox[origin=c]{90}{$[110]\, A_1$ }}} &
$\bar{\psi} {\bf \Gamma} \psi \times 13$ &
$\bar{\psi} {\bf \Gamma} \psi \times 21$ &
$\bar{\psi} {\bf \Gamma} \psi \times 13$ &
$\bar{\psi} {\bf \Gamma} \psi \times 21$ &
$\bar{\psi} {\bf \Gamma} \psi \times 20$ \\[1ex]
& $\etaOctet_{\scriptscriptstyle{[100]}} \omegaOctet_{\scriptscriptstyle{[100]}}$ 
& $\etaOctet_{\scriptscriptstyle{[100]}} \omegaOctet_{\scriptscriptstyle{[100]}}$
& $\etaOctet_{\scriptscriptstyle{[100]}} \omegaOctet_{\scriptscriptstyle{[100]}}$
& $\etaOctet_{\scriptscriptstyle{[100]}} \omegaOctet_{\scriptscriptstyle{[100]}}$
& $\etaOctet_{\scriptscriptstyle{[100]}} \omegaOctet_{\scriptscriptstyle{[100]}}$ \\
& $\etaOctet_{\scriptscriptstyle{[100]}} \omegaOctet_{\scriptscriptstyle{[111]}}$ 
& $\etaOctet_{\scriptscriptstyle{[100]}} \omegaOctet_{\scriptscriptstyle{[111]}}$
& $\etaOctet_{\scriptscriptstyle{[100]}} \omegaOctet_{\scriptscriptstyle{[111]}}$
& $\etaOctet_{\scriptscriptstyle{[100]}} \omegaOctet_{\scriptscriptstyle{[111]}}$
& $\etaOctet_{\scriptscriptstyle{[100]}} \omegaOctet_{\scriptscriptstyle{[111]}}$ \\
& & $\etaOctet_{\scriptscriptstyle{[110]}} \omegaOctet_{\scriptscriptstyle{[110]}} {\scriptstyle \{3\} }$
& $\etaOctet_{\scriptscriptstyle{[110]}} \omegaOctet_{\scriptscriptstyle{[110]}} {\scriptstyle \{3\} }$
& $\etaOctet_{\scriptscriptstyle{[110]}} \omegaOctet_{\scriptscriptstyle{[110]}} {\scriptstyle \{3\} }$
& $\etaOctet_{\scriptscriptstyle{[110]}} \omegaOctet_{\scriptscriptstyle{[110]}} {\scriptstyle \{3\} }$  \\
& & $\etaOctet_{\scriptscriptstyle{[111]}} \omegaOctet_{\scriptscriptstyle{[100]}} $
& $\etaOctet_{\scriptscriptstyle{[111]}} \omegaOctet_{\scriptscriptstyle{[100]}} $
& $\etaOctet_{\scriptscriptstyle{[111]}} \omegaOctet_{\scriptscriptstyle{[100]}} $
& $\etaOctet_{\scriptscriptstyle{[111]}} \omegaOctet_{\scriptscriptstyle{[100]}}$  \\
& & & $\etaOctet_{\scriptscriptstyle{[100]}} \omegaOctet_{\scriptscriptstyle{[210]}} $
& $\etaOctet_{\scriptscriptstyle{[100]}} \omegaOctet_{\scriptscriptstyle{[210]}} $
& $\etaOctet_{\scriptscriptstyle{[100]}} \omegaOctet_{\scriptscriptstyle{[210]}}$  \\
& & & & $\etaOctet_{\scriptscriptstyle{[110]}} \omegaOctet_{\scriptscriptstyle{[200]}} $ \\
& & & & $\etaOctet_{\scriptscriptstyle{[200]}} \omegaOctet_{\scriptscriptstyle{[110]}} $ \\
& & & & $\etaOctet_{\scriptscriptstyle{[210]}} \omegaOctet_{\scriptscriptstyle{[200]}} $\\[1ex]
& $\omegaSinglet_{\scriptscriptstyle{[110]}} {\fZeroSinglet}_{\!\scriptscriptstyle{[000]}} $ 
& $\omegaSinglet_{\scriptscriptstyle{[110]}} {\fZeroSinglet}_{\!\scriptscriptstyle{[000]}} $ 
& $\omegaSinglet_{\scriptscriptstyle{[110]}} {\fZeroSinglet}_{\!\scriptscriptstyle{[000]}} $ 
& $\omegaSinglet_{\scriptscriptstyle{[110]}} {\fZeroSinglet}_{\!\scriptscriptstyle{[000]}} $ 
& $\omegaSinglet_{\scriptscriptstyle{[110]}} {\fZeroSinglet}_{\!\scriptscriptstyle{[000]}} $ \\
& $\omegaSinglet_{\scriptscriptstyle{[000]}} {\fZeroSinglet}_{\!\scriptscriptstyle{[110]}} $ 
& $\omegaSinglet_{\scriptscriptstyle{[000]}} {\fZeroSinglet}_{\!\scriptscriptstyle{[110]}} $ 
& $\omegaSinglet_{\scriptscriptstyle{[000]}} {\fZeroSinglet}_{\!\scriptscriptstyle{[110]}} $ 
& $\omegaSinglet_{\scriptscriptstyle{[000]}} {\fZeroSinglet}_{\!\scriptscriptstyle{[110]}} $ \\
& $\omegaSinglet_{\scriptscriptstyle{[100]}} {\fZeroSinglet}_{\!\scriptscriptstyle{[100]}} {\scriptstyle \{2\} }$ 
& $\omegaSinglet_{\scriptscriptstyle{[100]}} {\fZeroSinglet}_{\!\scriptscriptstyle{[100]}} {\scriptstyle \{2\} }$ 
& $\omegaSinglet_{\scriptscriptstyle{[100]}} {\fZeroSinglet}_{\!\scriptscriptstyle{[100]}} {\scriptstyle \{2\} }$ 
& $\omegaSinglet_{\scriptscriptstyle{[100]}} {\fZeroSinglet}_{\!\scriptscriptstyle{[100]}} {\scriptstyle \{2\} }$ \\[1ex]
& $\etaSinglet_{\scriptscriptstyle{[100]}} \omegaSinglet_{\scriptscriptstyle{[100]}}$ 
& $\etaSinglet_{\scriptscriptstyle{[100]}} \omegaSinglet_{\scriptscriptstyle{[100]}}$ 
& $\etaSinglet_{\scriptscriptstyle{[100]}} \omegaSinglet_{\scriptscriptstyle{[100]}}$ 
& $\etaSinglet_{\scriptscriptstyle{[100]}} \omegaSinglet_{\scriptscriptstyle{[100]}}$ \\[1ex]
\hline\\[-1.3ex]
\parbox[t]{7mm}{\multirow{15}{*}{\rotatebox[origin=c]{90}{$[111]\, A_1$ }}} &
$\bar{\psi} {\bf \Gamma} \psi \times 21$ &
$\bar{\psi} {\bf \Gamma} \psi \times 21$ &
$\bar{\psi} {\bf \Gamma} \psi \times 21$ &
$\bar{\psi} {\bf \Gamma} \psi \times 21$ &
$\bar{\psi} {\bf \Gamma} \psi \times 21$ \\[1ex]
& $\etaOctet_{\scriptscriptstyle{[100]}} \omegaOctet_{\scriptscriptstyle{[110]}}$ 
& $\etaOctet_{\scriptscriptstyle{[100]}} \omegaOctet_{\scriptscriptstyle{[110]}}$
& $\etaOctet_{\scriptscriptstyle{[100]}} \omegaOctet_{\scriptscriptstyle{[110]}}$
& $\etaOctet_{\scriptscriptstyle{[100]}} \omegaOctet_{\scriptscriptstyle{[110]}}$
& $\etaOctet_{\scriptscriptstyle{[100]}} \omegaOctet_{\scriptscriptstyle{[110]}}$ \\
& $\etaOctet_{\scriptscriptstyle{[110]}} \omegaOctet_{\scriptscriptstyle{[100]}}$ 
& $\etaOctet_{\scriptscriptstyle{[110]}} \omegaOctet_{\scriptscriptstyle{[100]}}$
& $\etaOctet_{\scriptscriptstyle{[110]}} \omegaOctet_{\scriptscriptstyle{[100]}}$
& $\etaOctet_{\scriptscriptstyle{[110]}} \omegaOctet_{\scriptscriptstyle{[100]}}$
& $\etaOctet_{\scriptscriptstyle{[110]}} \omegaOctet_{\scriptscriptstyle{[100]}}$ \\
& & & & $\etaOctet_{\scriptscriptstyle{[100]}} \omegaOctet_{\scriptscriptstyle{[211]}}$ 
& $\etaOctet_{\scriptscriptstyle{[100]}} \omegaOctet_{\scriptscriptstyle{[211]}}$  \\
& & & & & $\etaOctet_{\scriptscriptstyle{[200]}} \omegaOctet_{\scriptscriptstyle{[111]}}$  \\
& & & & & $\etaOctet_{\scriptscriptstyle{[100]}} \omegaOctet_{\scriptscriptstyle{[211]}}$  \\
& & & & & $\etaOctet_{\scriptscriptstyle{[110]}} \omegaOctet_{\scriptscriptstyle{[210]}} {\scriptstyle \{3\} }$  \\
& & & & & $\etaOctet_{\scriptscriptstyle{[210]}} \omegaOctet_{\scriptscriptstyle{[110]}} {\scriptstyle \{3\} }$  \\
& & & & & $\etaOctet_{\scriptscriptstyle{[211]}} \omegaOctet_{\scriptscriptstyle{[100]}} $  \\[1ex]
& $\omegaSinglet_{\scriptscriptstyle{[111]}} {\fZeroSinglet}_{\!\scriptscriptstyle{[000]}} $ 
& $\omegaSinglet_{\scriptscriptstyle{[111]}} {\fZeroSinglet}_{\!\scriptscriptstyle{[000]}} $ 
& $\omegaSinglet_{\scriptscriptstyle{[111]}} {\fZeroSinglet}_{\!\scriptscriptstyle{[000]}} $ 
& $\omegaSinglet_{\scriptscriptstyle{[111]}} {\fZeroSinglet}_{\!\scriptscriptstyle{[000]}} $ 
& $\omegaSinglet_{\scriptscriptstyle{[111]}} {\fZeroSinglet}_{\!\scriptscriptstyle{[000]}} $ \\
& $\omegaSinglet_{\scriptscriptstyle{[110]}} {\fZeroSinglet}_{\!\scriptscriptstyle{[100]}} {\scriptstyle \{2\} }$ 
& $\omegaSinglet_{\scriptscriptstyle{[110]}} {\fZeroSinglet}_{\!\scriptscriptstyle{[100]}} {\scriptstyle \{2\} }$ 
& $\omegaSinglet_{\scriptscriptstyle{[110]}} {\fZeroSinglet}_{\!\scriptscriptstyle{[100]}} {\scriptstyle \{2\} }$ 
& $\omegaSinglet_{\scriptscriptstyle{[110]}} {\fZeroSinglet}_{\!\scriptscriptstyle{[100]}} {\scriptstyle \{2\} }$ 
& $\omegaSinglet_{\scriptscriptstyle{[110]}} {\fZeroSinglet}_{\!\scriptscriptstyle{[100]}} {\scriptstyle \{2\} }$ \\
& $\omegaSinglet_{\scriptscriptstyle{[000]}} {\fZeroSinglet}_{\!\scriptscriptstyle{[111]}} $ 
& $\omegaSinglet_{\scriptscriptstyle{[000]}} {\fZeroSinglet}_{\!\scriptscriptstyle{[111]}} $ 
& $\omegaSinglet_{\scriptscriptstyle{[000]}} {\fZeroSinglet}_{\!\scriptscriptstyle{[111]}} $ 
& $\omegaSinglet_{\scriptscriptstyle{[000]}} {\fZeroSinglet}_{\!\scriptscriptstyle{[111]}} $ 
& $\omegaSinglet_{\scriptscriptstyle{[000]}} {\fZeroSinglet}_{\!\scriptscriptstyle{[111]}} $ \\
& $\omegaSinglet_{\scriptscriptstyle{[100]}} {\fZeroSinglet}_{\!\scriptscriptstyle{[110]}} {\scriptstyle \{2\} }$ 
& $\omegaSinglet_{\scriptscriptstyle{[100]}} {\fZeroSinglet}_{\!\scriptscriptstyle{[110]}} {\scriptstyle \{2\} }$ 
& $\omegaSinglet_{\scriptscriptstyle{[100]}} {\fZeroSinglet}_{\!\scriptscriptstyle{[110]}} {\scriptstyle \{2\} }$ 
& $\omegaSinglet_{\scriptscriptstyle{[100]}} {\fZeroSinglet}_{\!\scriptscriptstyle{[110]}} {\scriptstyle \{2\} }$ 
& $\omegaSinglet_{\scriptscriptstyle{[100]}} {\fZeroSinglet}_{\!\scriptscriptstyle{[110]}} {\scriptstyle \{2\} }$ \\[1ex]
& $\etaSinglet_{\scriptscriptstyle{[100]}} \omegaSinglet_{\scriptscriptstyle{[110]}}$ 
& $\etaSinglet_{\scriptscriptstyle{[100]}} \omegaSinglet_{\scriptscriptstyle{[110]}}$ 
& $\etaSinglet_{\scriptscriptstyle{[100]}} \omegaSinglet_{\scriptscriptstyle{[110]}}$ 
& $\etaSinglet_{\scriptscriptstyle{[100]}} \omegaSinglet_{\scriptscriptstyle{[110]}}$ 
& $\etaSinglet_{\scriptscriptstyle{[100]}} \omegaSinglet_{\scriptscriptstyle{[110]}}$ \\
& $\etaSinglet_{\scriptscriptstyle{[110]}} \omegaSinglet_{\scriptscriptstyle{[100]}}$ 
& $\etaSinglet_{\scriptscriptstyle{[110]}} \omegaSinglet_{\scriptscriptstyle{[100]}}$ 
& $\etaSinglet_{\scriptscriptstyle{[110]}} \omegaSinglet_{\scriptscriptstyle{[100]}}$ 
& $\etaSinglet_{\scriptscriptstyle{[110]}} \omegaSinglet_{\scriptscriptstyle{[100]}}$ 
& $\etaSinglet_{\scriptscriptstyle{[110]}} \omegaSinglet_{\scriptscriptstyle{[100]}}$ \\
\end{tabular}
}
\caption{As Table~\ref{ops000} for irreps with $\vec{P} = [110], [111]$.}
\label{ops110}
\end{table*}

%% file: appB-lefthand.tex

The complete scattering amplitude for $\etaOctet \omegaOctet \to \etaOctet \omegaOctet$, $T(s,t)$, has properties which follow from crossing symmetry, the simplest of which is that unitarity should apply not just in the $s$-channel ($\etaOctet \omegaOctet \to \etaOctet \omegaOctet$) but also in the (symmetric) $u$-channel, and in the $t$-channel ($\etaOctet \etaOctet \to \omegaOctet \omegaOctet$). The impact of the required discontinuities across the unitarity branch cut in Mandelstam $t$ and $u$ when the amplitude is projected into $s$-channel partial waves is to generate \emph{left-hand cuts}, i.e. branch cuts which typically lie on the real axis to the left of the $s$-channel threshold in the complex $s$-plane. 

While the discontinuity across these cuts requires knowledge of the scattering dynamics, the \emph{position} of these cuts is simply a function of the masses of the scattering hadrons. Unitarity in the $u$-channel implies a cut running along the real $s$ axis from $-\infty$ to $\big(m(\omegaOctet) - m(\etaOctet) \big)^2$, while unitarity in the $t$-channel provides a cut along the entire negative real $s$ axis, and a circular cut of radius $s = m(\omegaOctet)^2 - m(\etaOctet)^2$. 

In the current case there are additional cuts due to the fact that \emph{stable} mesons appear as bound-state poles in the crossed channels. $\omegaSinglet$ appears in the $u$-channel, and generates an extra ``short-cut'' from $s = \tfrac{ ( m(\omegaOctet)^2 - m(\etaOctet)^2 )^2 }{m(\omegaSinglet)^2}$ to $s = 2\big( m(\omegaOctet)^2 + m(\etaOctet)^2 \big) - m(\omegaSinglet)^2$. $\fZeroSinglet$  appears in the $t$-channel and generates a cut running from $-\infty$ to $s = \left( \sqrt{ m(\etaOctet)^2 - \tfrac{1}{4} m( \fZeroSinglet)^2 } + \sqrt{ m(\omegaOctet)^2 - \tfrac{1}{4} m( \fZeroSinglet)^2 } \right)^2$. Using the hadron masses in Table~\ref{masses}, we find that the rightmost extent of the left-hand cut lies at $a_t \sqrt{s} = 0.299$ and is due to either of the stable exchanges, $\omegaSinglet$, $\fZeroSinglet$. 

For the process $\etaSinglet \omegaSinglet \to \etaSinglet \omegaSinglet$, the nearest left-hand cut is at $a_t \sqrt{s}  = 0.368$, due to $\fZeroSinglet$ exchange in the $t$-channel, and for $\omegaSinglet \fZeroSinglet \to \omegaSinglet \fZeroSinglet $ is at $a_t \sqrt{s}  = 0.367$ due to $\fZeroSinglet$ exchange in the $t$-channel.

%% file: appC-params.tex

Tables~\ref{1mm_amps}, \ref{2mm_amps} list the amplitude parameterization variations discussed in Section~\ref{amps}.

%
\begin{table}
{
\renewcommand{\arraystretch}{1.6}
\begin{tabular}{l   ccc}
\multicolumn{1}{c}{Parameterization} &  
\multicolumn{1}{c}{Phase-space} & 
\multicolumn{1}{c}{$N_\mathrm{pars}$} &
\multicolumn{1}{c}{$\chi^2/N_\mathrm{dof}$} \\
\hline
\multirow{2}{*}{{$K=\frac{g_{\mathsf{a}}^2}{m_{\mathsf{a}}^2-s}+\frac{g_{\mathsf{b}}^2}{m_{\mathsf{b}}^2-s}$}}
 & naive 					& 4 & \textit{1.39} \\ 
 & CM(pole ${\mathsf{a}})$ & 4 & \textit{1.43}\\ 
\hline
\multirow{3}{*}{{$K=\frac{g_{\mathsf{a}}^2}{m_{\mathsf{a}}^2-s}+\frac{g_{\mathsf{b}}^2}{m_{\mathsf{b}}^2-s} +\gamma$}} 
& naive & 5 & 1.38 \\ 
& CM(pole ${\mathsf{a}})$ & 5 & 1.36 \\ 
& CM(pole ${\mathsf{b}})$ & 5 & 1.36 \\ 
\hline
\multirow{3}{*}{{$K=\frac{g_{\mathsf{a}}^2}{m_{\mathsf{a}}^2-s}+\frac{g_b^2}{m_b^2-s} +\gamma \, s$ }}
& naive & 5 & 1.37 \\ 
& CM(pole ${\mathsf{a}}$) & 5 & 1.35 \\ 
& CM(pole ${\mathsf{b}}$) & 5 & 1.35 \\ 
\hline
$K=\frac{g_{\mathsf{a}}^2}{m_{\mathsf{a}}^2-s}+\frac{g_b^2}{m_b^2-s} +\gamma \, s^2 \;\;\;$ 
& CM(pole ${\mathsf{a}}$) & 5 & 1.35 \\ 
\end{tabular}
\caption{
$1^{--}$ amplitude parameterizations, as plotted in Figures~\ref{1m_rhotsq_variation}, \ref{1m_delta_variation}, \ref{1m_poles_variation}, \ref{1m_all_poles_variation}. $\chi^2/N_\mathrm{dof}$ values in italics indicate the amplitudes shown by dashed curves.
}
\label{1mm_amps}
}
\end{table}

\begin{table*}
{
\renewcommand{\arraystretch}{1.6}
\begin{tabular}{lccc}
Parameterization &  Restrictions & $N_\mathrm{pars}$ & $\chi^2/N_\mathrm{dof}$ \\
\hline
\multirow{4}{*}{ $K_{ij}=\frac{g_ig_j}{m^2-s}+\gamma^{(0)}_{ij}$ }  
      & none                                    & 6 & 1.45 \\
      & naive phase-space                       & 6 & 1.45 \\
      & $\gamma^{(0)}_{PP}=\gamma^{(0)}_{FF}=0$ & 4 & 1.42 \\
      & $\gamma^{(0)}_{PF}=0$ & 5 & \textit{1.85} \\[0.8ex]
\hline
\multirow{7}{*}{ $K_{ij}=\frac{g_ig_j}{m^2-s}+\gamma^{(0)}_{ij}+\gamma^{(1)}_{ij}s$ \;\;}  
  & $\gamma^{(0)}_{PP}=\gamma^{(1)}_{PP}=\gamma^{(0)}_{FF}=\gamma^{(1)}_{FF}=0$ & 5 & 1.43\\
  & $\gamma^{(0)}_{PP}=\gamma^{(1)}_{PP}=\gamma^{(0)}_{PF}=\gamma^{(0)}_{FF}=\gamma^{(1)}_{FF}=0$ & 4 & 1.43\\
  & $\gamma^{(1)}_{PP}=\gamma^{(1)}_{PF}=\gamma^{(0)}_{FF}=0$ & 6 & 1.45\\
  & $\gamma^{(1)}_{PP}=\gamma^{(0)}_{PF}=\gamma^{(0)}_{FF} =\gamma^{(1)}_{FF} =0$ & 5 & 1.43\\
  & $\gamma^{(1)}_{PP}=\gamma^{(0)}_{PF}=\gamma^{(1)}_{FF} =0$ & 6 & 1.44\\
  & $\gamma^{(0)}_{PP}=\gamma^{(1)}_{PF}=\gamma^{(1)}_{FF} =0$ & 6 & 1.45\\
  & $\gamma^{(0)}_{PP}=\gamma^{(0)}_{PF}=\gamma^{(1)}_{FF} =0$ & 6 & 1.44\\[0.8ex]
\hline
$K_{ij}=\frac{g_i^\mathsf{a} g_j^\mathsf{a}}{m_\mathsf{a}^2-s}+\frac{g_i^\mathsf{b} g_j^\mathsf{b}}{m_\mathsf{b}^2-s} $& CM(pole $\mathsf{a}$) & 6 & \textit{1.80} \\[0.8ex]
\hline
$\big(K^{-1} \big)_{ij} = c_{ij}^{(0)}+c_{ij}^{(1)}s $ & CM(threshold) & 6 & \textit{1.67} \\
\end{tabular}
\caption{
$2^{--}$ amplitude parameterizations as plotted in Figures~\ref{2m_variation_rhotsq}, \ref{2m_variation_pole}. All use Chew-Mandelstam phase-space subtracted at the $K$-matrix pole unless stated otherwise. $\chi^2/N_\mathrm{dof}$ values in italics indicate the amplitudes shown by dashed curves.
} 
\label{2mm_amps}
}
\end{table*}